\pdfoutput=1
\documentclass[useAMS,usenatbib]{mn2e}
\usepackage[draft]{hyperref}

\usepackage{amsmath}
\usepackage{doi}
\usepackage[authoryear]{natbib}

\usepackage{multirow}
\usepackage{bigdelim}
\usepackage{longtable}
\usepackage{graphicx}
\usepackage{times}

\usepackage{subcaption}

\def\S{SVS13 }

\def \form13{H$_2^{13}$CO }

\def \formd2{D$_2$CO }

\def \met13{$^{13}$CH$_3$OH }

\def \metan2d{CHD$_2$OH }
\def \metan3d{CH$_3$OD }

\newcommand{\araa}{ARA\&A}		
\newcommand{\apj}{ApJ}			
\newcommand{\apjl}{ApJ}		
\newcommand{\apjs}{ApJS}		
\newcommand{\aap}{A\&A}			
\newcommand{\aapr}{A\&A~Rev.}		
\newcommand{\jqsrt}{J. Quant. Spectrosc. Radiat. Trans.}
\newcommand{\mnras}{MNRAS}		
\newcommand{\pasj}{PASJ}		
\newcommand{\ssr}{Space~Sci.~Rev.}	
\newcommand{\nat}{Nature}		
\newcommand{\jcp}{J.~Chem.~Phys.}	

\usepackage{fixltx2e}
\title[SVS13-A iCOMs census]{The census of interstellar complex organic molecules in the Class I hot
  corino of SVS13-A}

\author[E. Bianchi et al.]{E. Bianchi$^{1,2}$\thanks{E-mail:
eleonora.bianchi@univ-grenoble-alpes.fr} , C. Codella$^{2, 1}$, C. Ceccarelli$^{1,3}$, 
F. Vazart$^{1}$, R. Bachiller$^{4}$, N. Balucani$^{5,1,2}$,  
\newauthor M. Bouvier$^{1}$, M. De Simone$^{1}$, J. Enrique-Romero$^{1}$, C. Kahane$^{1}$
 B. Lefloch$^{1,3}$, 
 \newauthor A. L\'opez-Sepulcre$^{1,6}$, J. Ospina-Zamudio$^{1}$, 
 L. Podio$^{2}$, V. Taquet$^{2}$
\\
$^{1}$ Univ. Grenoble Alpes, IPAG, F-38000 Grenoble, France \\
$^{2}$  INAF-Osservatorio Astrofisico di Arcetri, L.go E. Fermi 5, 50125 Firenze, Italy \\
$^{3}$ CNRS, IPAG, F-38000 Grenoble, France \\
$^{4}$ IGN, Observatorio Astron\'omico Nacional, Calle Alfonso XII, 28004 Madrid, Spain\\
$^{5}$ Dipartimento di Chimica, Biologia e Biotecnologie, Universit\'a degli Studi di Perugia, Via Elce di Sotto 8, 06123 Perugia, Italy\\
$^{6}$ Institut de Radioastronomie Millimétrique (IRAM), 300 rue de la Piscine, 38406 Saint-Martin-d'H\'eres, France\\
}

\begin{document}

\date{Accepted date. Received date; in original form date}

\pagerange{\pageref{firstpage}--\pageref{lastpage}} \pubyear{2016}

\maketitle

\label{firstpage}

\begin{abstract}
  We present the first census of the interstellar Complex Organic Molecules
  (iCOMs) in the low-mass Class I protostar SVS13-A, obtained
  by analysing data from the IRAM-30m Large Project ASAI
  (Astrochemical Surveys At IRAM). They consist of an
  high-sensitivity unbiased spectral survey at the 1mm, 2mm and 3mm
  IRAM bands. We detected five iCOMs: acetaldehyde (CH$_3$CHO),
  methyl formate (HCOOCH$_3$), dimethyl ether (CH$_3$OCH$_3$), ethanol
  (CH$_3$CH$_2$OH) and formamide (NH$_2$CHO).
  In addition we searched for other iCOMs and 
  ketene (H$_2$CCO), formic acid (HCOOH) and  
  methoxy (CH$_3$O), whose only ketene was detected.
  The numerous detected lines, from 5 to 37 depending on the species,
  cover a large upper level energy range, between 15 and 254 K.  This
  allowed us to carry out a rotational diagram analysis and derive
  rotational temperatures between 35 and 110 K, and column densities
  between $3\times 10^{15}$ and $1\times 10^{17}$ cm$^{-2}$
  on the 0$\farcs$3 size previously determined by interferometric
  observations of glycolaldehyde. 
  These new observations clearly demonstrate the presence of a rich chemistry in the hot
  corino towards SVS13-A. The measured iCOMs abundances were
  compared to other Class 0 and I hot corinos, as well as comets, previously published in
  the literature. We find evidence that (i) SVS13-A is as chemically rich
  as younger Class 0 protostars, and (ii) the iCOMs relative
  abundances do not substantially evolve during the protostellar
  phase.
\end{abstract}


\section{Introduction}

The measurement of the abundance of interstellar Complex Organic
Molecules (hereinafter iCOMs; C-bearing molecules containing at least
six atoms; \citealt{Herbst2009}; \citealt{Ceccarelli2017}) is one of
the several pieces of the huge puzzle to complete if one wants to
understand or, more humbly, to just shed light on the processes that
led to the emergence of life on Earth and, possibly, on other
planets. Indeed, simple organic molecules formed during the
protostellar and protoplanetary disk phases might survive and be
delivered to the nascent planet, providing seeds for the formation
of the more complex, real pre-biotic species needed to the first
living beings.

In the context of Solar-type protostars and planetary systems, Class 0
hot corinos are the objects where the detection and study of iCOMs has
been traditionally easier. These are compact ($\geq 100$ au), hot
($\geq$ 100 K), dense ($\geq 10^7$ cm$^{-3}$) and iCOMs enriched
regions at the center of the Class 0 protostellar envelopes
(e.g. \citealt{Ceccarelli2007}; \citealt{Caselli2012}). Their
chemical richness is believed to be caused by the sublimation of the
icy mantles that coat the dust grains and the consequent injection
into the gas-phase of species which either are themselves iCOMs
(e.g. \citealt{Garrod2006}) or which react in the gas-phase and form
iCOMs (e.g. \citealt{Charnley1992}; \citealt{Balucani2015}; \citealt{Skouteris2017, Skouteris2018}) or
both.

From an evolutionary point of view, Class I sources represent a bridge
between the Class 0 sources, characterised by a thick collapsing
envelope and an embedded forming disk \citep{Enoch2009, Tobin2015}, and the Class II and III sources, characterised by the
presence of a prominent circumstellar disk, where planets eventually
form \citep{Avenhaus2018, Garufi2017}. However, it is not clear whether this is also valid from a
chemical point of view.  On the one hand, (some) Class 0 sources
possess a hot corino with the detection of several iCOMs: methyl
formate (HCOOCH$_3$), dimethyl ether (CH$_3$OCH$_3$), acetaldehyde
(CH$_3$CHO), ethanol (CH$_3$CH$_2$OH) and formamide (NH$_2$CHO) are
among the most easily detected iCOMs (see references below). On the
other hand, protoplanetary disks only show emission from simpler
iCOMs: methanol (CH$_3$OH), acetic acid (HCOOH), ethyl cyanide
(CH$_3$CN) and cyanoacetylene (HC$_3$N) are the most complex molecules
so far detected (\citealt{Chapillon2012}; \citealt{Dutrey2014}; \citealt{Oberg2015}; 
\citealt{Walsh2016}; \citealt{Favre2018}) .

The question is: are more complex molecules not detected in
protoplanetary disks only because of the current instruments detection
limits or because iCOMs are not present there?  In other words, does the chemical complexity vary with the evolutionary phase?
One can answer this question by starting understanding whether there
is any change in the chemical complexity going from Class 0 to I
sources. This would provide a base to predict the chemistry at work in
the later phase, the protoplanetary disk one.

Thanks to the new powerful facilities developed in the last decade,
such as NOEMA and ALMA, several projects have now focused on the
chemical composition of Class 0 hot corinos
(e.g. \citealt{Jorgensen2012, Jorgensen2016, Jorgensen2018}; \citealt{Maury2014};
\citealt{Taquet2015}; \citealt{Imai2016}; \citealt{Codella2016};
\citealt{Desimone2017};\citealt{Oya2017}; \citealt{Lopez2017};
\citealt{Bianchi2017b}; \citealt{Ospina2018}). On the other hand, very little has been done
so far to study the overall iCOMs composition of Class I hot corinos.
Only a couple of Class I sources have been observed so far in a few
iCOMs with the IRAM-30m (\citealt{Oberg2014}; \citealt{Graninger2016};
\citealt{Bergner2017}), and in none was a full census obtained. 

We present here the first iCOMs census in a Class I source,
SVS13-A. This was obtained by analysing the unbiased spectral survey
of the 3, 2 and 1 mm bands observable with the IRAM-30m, which is part
of the Large Program {\it ASAI} (Astrochemical Surveys At IRAM; {\it
  www.oan.es/asai}; \citealt{Lefloch2018}).

The article is organised as follows.  In Section \ref{Sec:Source}, we
present the source background; in Section \ref{Sec:Observations} the
observations and the line identification procedure; in Section
\ref{Sec:Results} the main results for the analysis of each molecular
species and in Section \ref{Sec:Discussion} we discuss the results.
Finally, Section \ref{Sec:Conclusions} is for the conclusions.

\section{Source background}\label{Sec:Source}

SVS13-A is a Class I protostar located in the SVS13 cluster of the
NGC1333 cloud in Perseus, which lies at a distance of (235$\pm$18) pc
\citep{Hirota2008}.  Next to SVS13-A lies the Class 0 protostar
SVS13-B, about 15$\arcsec$ apart (see e.g. \citealt{Chini1997};
\citealt{Bachiller1998}; \citealt{Looney2000}; \citealt{Chen2009};
\citealt{Tobin2016}, and references therein).

The SVS13-A protostar has a bolometric luminosity of $L_{\rm bol}\simeq$
32.5 $L_{\rm sun}$, a low
$L_{\rm submm}$/$L_{\rm bol}$ ratio ($\sim$ 0.8\%) and a bolometric
temperature $T_{\rm bol}$ $\sim$ 188 K (\citealt{Tobin2016}).  
In addition, SVS13-A is associated with an extended outflow ($>$ 0.07
pc: \citealt{Lefloch1998a}; \citealt{Codella1999}) as well as with the
well-known chain of Herbig-Haro (HH) objects 7--11
\citep{Reipurth1993}. Thus, although still deeply embedded in a
large scale envelope \citep{Lefloch1998a}, SVS13-A is considered a
relatively evolved protostar, already entered in the Class I stage.

VLA observations have resolved SVS13-A as being a close binary system
(VLA4A and VLA4B), separated by 0$\farcs$3, corresponding to $\sim$ 70 au (\citealt{Rodriguez1999};
\citealt{Anglada2000}).  

A previous analysis of the HDO lines detected in the ASAI observations
revealed the presence of a hot ($\geq 150$ K), dense
($\geq 3\times10^7$ cm$^{-3}$) and compact ($\sim$ 25 au) region,
which could indicate the presence of a hot corino (Codella et
al. 2016).
The actual confirmation of the hot corino arrived soon after with the
detection of glycolaldehyde (HCOCH$_2$OH), which emits over a $\sim$
70 au diameter region centered on SVS13-A
\citep{Desimone2017}. Subsequent higher spatial resolution
observations by \citet{Lefevre2017} found that VLA4A is associated
with compact continuum emission, suggestive of a disk smaller than
about 50 au. In addition, VLA4A is associated with molecular emission,
indicating that it is the hot corino source. VLA4B, on the other hand,
seems deprived of molecular emission but it is driving a small
scale H$_2$/SiO microjet.

To conclude, based on all these observations, SVS13-A seems to be a
very good target where to obtain a census of the iCOMs present in a
Class I hot corino.

\section{Observations and line identification}\label{Sec:Observations}

\subsection{Observations}

Observations were performed with the IRAM-30m telescope at Pico Veleta
(Spain) in the framework of the ASAI large program (see
Introduction). Briefly, ASAI provided an unbiased spectral survey of
the 3 mm (80--116 GHz), 2 mm (129--173 GHz), and 1.3 mm (200--276 GHz)
bands, acquired during several runs between 2012 and 2014.  
The rms noise (in $T_{\rm MB}$ scale) is about 2
mK, 7 mK, 9 mK  in a channel of 0.6 km s$^{-1}$, 0.4 km s$^{-1}$ and 0.2 km s$^{-1}$ for the 3, 2 and 1 mm spectral windows, respectively.
The broad-band EMIR receivers were used, connected to the FTS200 backends, 
which provide a spectral resolution of 200 kHz. The observations were acquired in wobbler
switching mode (with a 180$\arcsec$ throw) and pointed towards
SVS13-A, namely at $\alpha_{\rm J2000}$ = 03$^{\rm h}$ 29$^{\rm m}$
03$\fs$76, $\delta_{\rm J2000}$ = +31$\degr$ 16$\arcmin$ 03$\farcs$0.
The pointing was found to be accurate to within 3$\arcsec$.  The
telescope HPBWs lies from $\simeq$ 9$\arcsec$ at 276 GHz to $\simeq$
30$\arcsec$ at 80 GHz.  For a more detailed description of the ASAI
observations we refer the reader to \citet{Lefloch2018}.

The data reduction was performed using the
GILDAS--CLASS\footnote{{\it http://www.iram.fr/IRAMFR/GILDAS}} package.  The
uncertainty of the calibration varies between $\sim$ 10\% at 3 mm and $\sim$ 20\%
at 1 mm.  Finally, the line intensities were converted from
antenna temperature to main beam temperature ($T_{\rm MB}$), using the
main beam efficiencies reported in the IRAM-30m
website\footnote{{\it http://www.iram.es/IRAMES/mainWiki/Iram30mEfficiencies}}.

\subsection{Line identification and spectroscopic properties}\label{Sub:lines_id}

Line identification was performed using the ULSA (Unbiased Line
Spectral Analysis) package developed at IPAG. ULSA allows to
automatically identify lines in the ASAI spectra using the Jet
Propulsor Laboratory (JPL\footnote{https://spec.jpl.nasa.gov/},
\citealt{Pickett1998}) and Cologne Database for Molecular Spectroscopy
(CDMS\footnote{http://www.astro.uni-koeln.de/cdms/};
\citealt{Muller2001}, \citealt{Muller2005}) molecular line databases.
We excluded from the analysis those lines peaking at velocities
displaced with respect to the systemic velocity by more than 0.6 km
s$^{-1}$, to minimise effects due to line blending.

We looked for the iCOMs commonly detected so far towards low-mass
hot corinos. More specifically, we searched for lines from
 acetaldehyde (CH$_{3}$CHO), methyl formate
(HCOOCH$_{3}$), dimethyl ether (CH$_{3}$OCH$_{3}$), ethanol
(CH$_{3}$CH$_{2}$OH), propynal (HC$_{\rm 2}$CHO),
glycoaldehyde (HCOCH$_{\rm 2}$OH), methylamine (CH$_{\rm 3}$NH$_{\rm 2}$)
 and acetone (CH$_{\rm 3}$COCH$_{\rm 3}$).
We also added some important iCOMs precursors
such as ketene (H$_{\rm 2}$CCO), formic acid (HCOOH) and methoxy (CH$_{\rm 3}$O).

We summarise here the spectroscopic properties of some species.
Because of the symmetries due to the presence of a methyl group (CH$_3$), CH$_{3}$CHO,
HCOOCH$_{3}$ and CH$_{3}$CH$_{2}$OH are associated with the {\it
  A}-type and {\it E}-type forms, while CH$_{3}$OCH$_{3}$, having two
methyl groups, can be found in four forms ({\it AA}, {\it AE}, {\it
  EA}, {\it EE}).  
The spin statistical weights of the {\it A} and {\it E} type levels is
the same (\citealt{Turner1991} and references therein).  The spin
statistical weight of the ({\it AA}, {\it AE}, {\it EA}, {\it EE})
type levels are 6 ({\it AA}), 16 ({\it EE}), 2 ({\it AE}), 4 ({\it
  EA}) and 10 ({\it AA}), 16 ({\it EE}), 6 ({\it AE}), 4 ({\it EA})
for ee-oo and eo-oe rotational transitions\footnote{e (even) and o
  (odd) refer to the $K_{\rm a}$, $K_{\rm c}$ labels of the energy
  states of an asymmetric rotor.}, respectively \citep{Myers1960}.

In the case of ethanol (CH$_{3}$CH$_{2}$OH), the internal rotation is
negligible in most instances.  However, ethanol exists under the form
of three rotamers: anti, gauche+ and gauche- (see Sect. \ref{Sec:ethanol}
and Fig. \ref{Fig:ethanol-chem}).  

Finally, H$_{\rm 2}$CCO exists in two isomers, ortho (transition with
{\it K} odd) and para (transitions with {\it K} even), that have
different nuclear spin states.  For kinetic temperature much larger
than 15 K, the ortho-to-para ratio is expected to be close to the
statistical value of 3 (e.g. \citealt{Ohishi1991}).

\section{Results}\label{Sec:Results}

\subsection{Overview}

Table \ref{Table:detected} summarises the list of the iCOMS detected in SVS13-A,
with the number of identified lines and their upper level energy
$E_{\rm u}$ ranges.  We detected more than 100 lines with $E_{\rm u}$
up to 254 K emitted by H$_{2}$CCO, CH$_{3}$CHO, HCOOCH$_{3}$,
CH$_{3}$OCH$_{3}$ and CH$_{3}$CH$_{2}$OH.  
The considered detection limit is 3$\sigma$ in the integrated intensity.

\begin{figure*}
\begin{center}
\includegraphics[width=12cm]{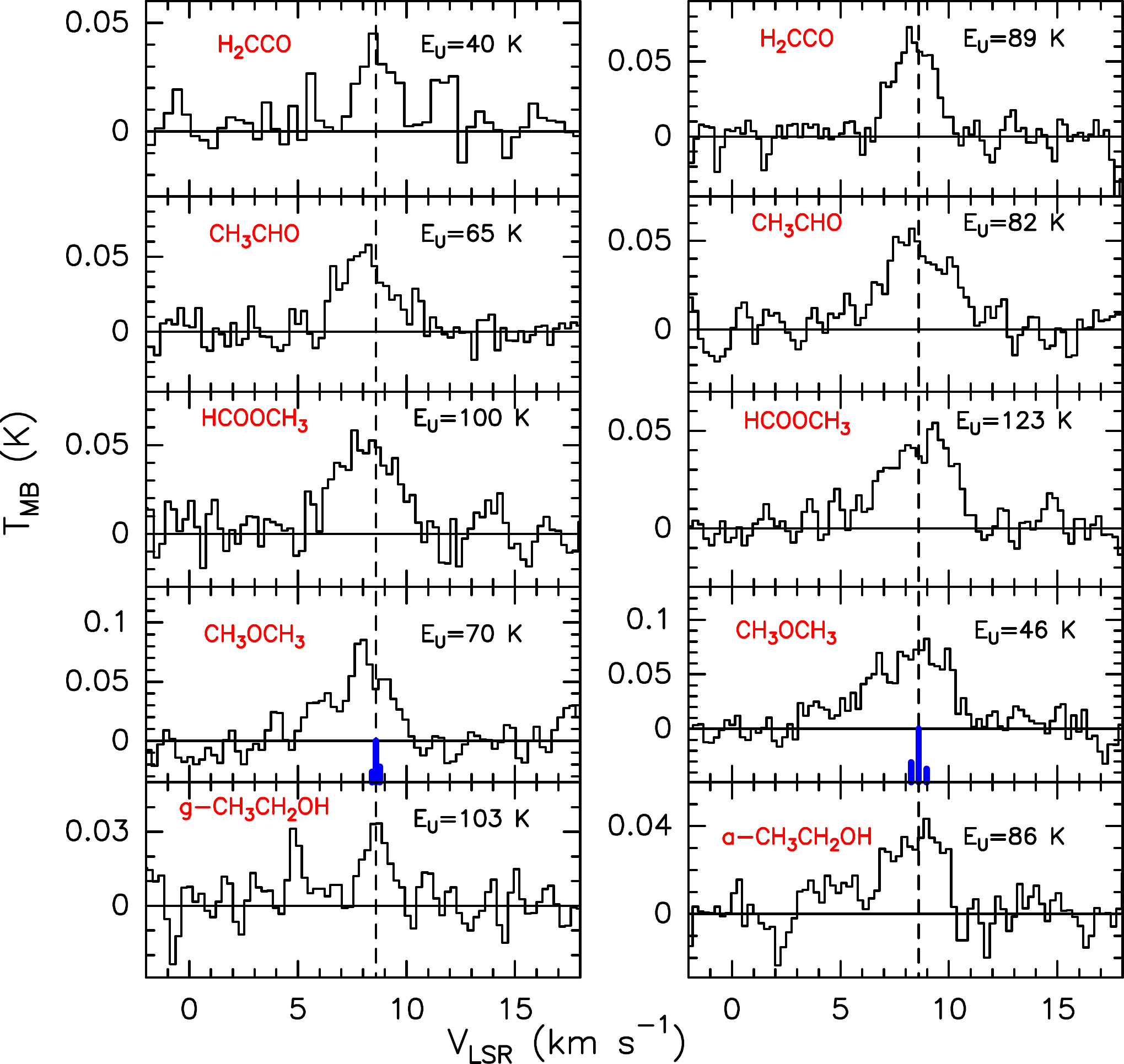}
\caption{Examples of the observed spectra, in $T_{\rm MB}$ scale (i.e. not
  corrected for the beam dilution): species and excitation energies
  are reported in the upper left and right corner of each panel,
  respectively. The vertical dashed line stands fo the ambient LSR
  velocity (+ 8.6 Km s$^{-1}$: \citealt{Chen2009}). In the
  CH$_3$OCH$_3$ panels, the blue lines indicate the presence of
  blended transitions with the same upper level energy. The complete
  observed spectra for each species are reported in
  Appendix \ref{appendix}.}
\label{Fig:spectra-example}
\end{center}
\end{figure*}
 
We fitted the detected lines with a Gaussian function using the GILDAS CLASS software.  
Their spectral
parameters as well as the results from the Gaussian fits are presented
in Tables \ref{Table:ketene}, \ref{Table:Ace}, \ref{Table:MF},
\ref{Table:DME}, \ref{Table:ethanol}.  Examples of the detected line
profiles (in $T_{\rm MB}$ scale) are shown in
Fig. \ref{Fig:spectra-example}.  The peak velocities of the detected
lines are between +8 and +9 km s$^{-1}$, consistent, within the
uncertainties, with the systemic velocity of both SVS13-A and SVS13-B
(+8.6 km s$^{-1}$, \citealt{Chen2009}; \citealt{Lopez2015}).
However, we note that the vast majority of the iCOMs lines are
detected in the 1.3 mm band, namely with a beam whose HPBW is $\sim$
10$\arcsec$.  Therefore, the detected iCOMs emission is mostly from
SVS13-A and the contribution from SVS13-B will be neglected in the
following analysis.
As regard the methyl formate lines, the noise on the spectra is too large hampering a more sophisticated analysis. Specifically, the sometime apparent double peaks are always within the noise of the spectrum, as well as possible wings. For this reason, we decided to not push further the analysis.

\begin{table}
\caption{List of the detected iCOMs (upper half table) and upper limits
  of the non-detected ones (lower half table) towards SVS13-A .}
\begin{center}

\begin{tabular}{lrcrrr}
  \hline
  \multicolumn{1}{c}{Species} &
                                \multicolumn{1}{c}{N$_{lines}$$^a$} &
                                                                  \multicolumn{1}{c}{$E_{\rm u}$} &
                                                                                                    \multicolumn{1}{c}{$T_{\rm rot}$$^b$} &
                                                                                                                                            \multicolumn{1}{c}{$N_{\rm tot}$$^b$} &
                                                                                                                                            			 \multicolumn{1}{c}{$X_{\rm H2}$$^c$}  \\
  \multicolumn{1}{c}{} &
                         \multicolumn{1}{c}{} &
                                                \multicolumn{1}{c}{(K)} &
                                                                          \multicolumn{1}{c}{(K)} &
                                                                                                    \multicolumn{1}{c}{(cm$^{-2}$)} &
                                                                                                    							    \multicolumn{1}{c}{} \\
  \hline
  $^{13}$CH$_3$OH$^d$ & 18 & 20--175 & 100(10) & 10(2) $\times$ 10$^{16}$  & 3 $\times$ 10$^{-8}$ \\
  
  H$_2$CCO           &   13 & 40--206 &  65(10) &  13(3) $\times$ 10$^{15}$  & 4 $\times$ 10$^{-9}$ \\
  
  CH$_3$CHO         &  13 & 61--108 &   35(10) &  12(7) $\times$ 10$^{15}$ &   4 $\times$ 10$^{-9}$\\
  
  HCOOCH$_3$       &  37 & 20--149 &   66(5) &  13(1) $\times$ 10$^{16}$ &   4 $\times$ 10$^{-8}$ \\
  
  CH$_3$OCH$_3$   & 11 & 33--254 & 110(10) &  14(4) $\times$ 10$^{16}$ &  5 $\times$ 10$^{-8}$ \\
  
  CH$_3$CH$_2$OH &  5 & 35--137 &  105(60) & 11(5) $\times$ 10$^{16}$  &  4 $\times$ 10$^{-8}$\\
  
  NH$_2$CHO$^e$  & 13 & 15--102 &   45(8)  &  26(9) $\times$ 10$^{14}$ &  9 $\times$ 10$^{-10}$\\
  \hline
  \multicolumn{6}{c}{Upper limits$^f$}\\
  \hline
  HCOOH   &--& --&  80 & $\leq$ 5 $\times$ $10^{15}$  & $\leq$ 2 $\times$ $10^{-9}$\\
  HC$_{2}$CHO  & --& --& 80 & $\leq$ 1 $\times$ $10^{16}$ & $\leq$ 3 $\times$ $10^{-9}$\\
  HCOCH$_{2}$OH  &-- &--&  80 & $\leq$ 8 $\times$ $10^{15}$ & $\leq$ 3 $\times$ $10^{-9}$\\
  CH$_{3}$NH$_{2}$ &--&-- &  80 &  $\leq$ 1 $\times$ $10^{16}$  & $\leq$ 3 $\times$ $10^{-9}$\\
  CH$_{3}$COCH$_{3}$  &--&--& 80 &  $\leq$ 1 $\times$ $10^{16}$ & $\leq$ 3 $\times$ $10^{-9}$\\
  CH$_{3}$O  &--&--&  80 & $\leq$ 3 $\times$ $10^{15}$ & $\leq$ 1 $\times$ $10^{-9}$\\
  \hline
  \noalign{\vskip 2mm}
\end{tabular}
  \end{center}
\small{
$^a$ Number of lines used in the analysis. Additional detected transitions, 
excluded from the analysis because contaminated, are reported in Appendix \ref{appendix}.\\
$^b$ Parameters are derived for a source size of 0$\farcs$3, as
  measured by \citet{Desimone2017}.\\   
  $^c$ We assume N(H$_2$) = 3 $\times$ 24 cm$^{-2}$ from \citet{Chen2009}
  (see Sec. \ref{sec:hot-corino-svs13} for details).\\
   $^d$ From \citet{Bianchi2017a}.\\
  $^e$ We repeated the analysis previously reported by
  \citet{Lopez2015}, which was obtained assuming a source size of 
  1$\arcsec$ instead of 0$\farcs$3. \\
  $^f$To derive the upper limit to the column density of each non-detected species we
  assumed $T_{\rm rot}$ equal to 80 K. The upper limits refer to 1$\sigma$.}

\label{Table:detected}
\end{table}

Having lines with a large $E_{\rm u}$ range can be used to identify
different regions in the field of view that have different excitation
conditions (density and temperature). For example, using a non-LTE
analysis of methanol lines towards SVS13-A we disentangled their
emission from the envelope and the hot corino, respectively
\citep{Bianchi2017a}.
Unfortunately, the collisional rates for the observed iCOMs transitions are not
available in literature, so that we used the standard Rotational
Diagram (RD) analysis to estimate the temperature and the column
density. Yet, even in this case, the occurence of multiple components
would be visible in the RD as straight lines with different slopes.

To construct the RD of each detected species, we included also lines
that consist of several transitions with the same upper level energy,
but different Einstein coefficients and statistical weights.  In this
case, we adopted the formalism described in Appendix \ref{sec:RD}. 

As a first step, we performed the RD analysis without beam filling 
factor corrections. The obtained rotational temperatures are:
(240 $\pm$ 95) K for H$_2$CCO, (50 $\pm$ 10) K for CH$_3$CHO, (400 $\pm$ 200) K for HCOOCH$_3$,
(110 $\pm$ 10) K for CH$_3$OCH$_3$, (140 $\pm$ 60) K for NH$_2$CHO, and (100 $\pm$ 30) K for CH$_3$CH$_2$OH.
The obtained values, always higher than 50 K,
 implies that the iCOMs
emission is very likely dominated by the hot corino of SVS13-A.
Therefore, in the subsequent analysis, we assume a source size of
0$\farcs$3 for all iCOMs, based on the interferometric observations
of the HCOCH$_2$OH emission by \citet{Desimone2017} and 
on the non-LTE analysis of the $^{13}$CH$_3$OH lines \citep{Bianchi2017a}.
Table
\ref{Table:detected} reports the list of the derived rotational
temperatures and column densities for each detected species.
The next subsection describes the
results for each detected iCOM as well as some general considerations
and comparison with previous observations.

\subsection{Ketene}
Ketene (H$_{\rm 2}$CCO), a near-prolate asymmetric top rotor, is one
of the numerous carbon-chain molecules that were observed in the
interstellar medium.  It was discovered for the first time by
\citep{Turner1977}.  Although it is not an iCOM in the strict sense
(it has only 5 atoms), it is tought to be involved in grain-surface
reactions to form iCOMs such as formic acid, ethanol and
acetaldehyde (\citealt{Charnley1997}; \citealt{Garrod2008};
\citealt{Hudson2013} and references therein), so we considered it in
this study.

We detected six lines of the para and eleven of the ortho form of
H$_{\rm 2}$CCO, respectively.  Five of the detected transitions are in
the 2 mm band, namely they are observed with a HPBW of $\sim$
15$\arcsec$, while the rest of the lines are detected in the 1.3 mm
band (HPBW $\sim$10$\arcsec$).  The line upper level energies
($E_{\rm u}$) are in the 40--206 K range.  The observed spectra are
shown in Fig. \ref{Fig:H2CCO-spectra}, while the fit to the spectral line parameters are reported in Tab. \ref{Table:ketene}.

The line profiles are close to a Gaussian with FWHM between 0.9
and 4.1 km s$^{-1}$.  The peak velocities are close to the systemic
source velocity (with values between +7.6 and +8.9
km s$^{-1}$.  The two transitions 8$_{\rm 3,5}$--7$_{\rm 3,4}$ and
8$_{\rm 3,6}$--7$_{\rm 3,5}$ with $E_{\rm u}$ = 152 K, as well as the
two transitions 13$_{\rm 3,10}$--12$_{\rm 3,9}$ and
13$_{\rm 3,11}$--12$_{\rm 3,10}$ with $E_{\rm u}$ = 206 K, are blended
(see Tab. \ref{Table:ketene}).  These lines are therefore excluded
from the subsequent analysis.

In Fig. \ref{Fig:RD}, we show the RD of both ortho and para
transitions. Their distribution does not show any significant scatter
from a common linear fit, within the errors. In other words, the RD is
consistent with the high-temperature LTE limit o/p statistical values
of 3.  We found a rotational temperature $T_{\rm rot}$ = (65$\pm$10) K
and a column density $N_{\rm tot}$ = (13$\pm$3) $\times$ 10$^{15}$
cm$^{-2}$  (Tab. \ref{Table:detected}).

\subsection{Acetaldehyde}
Acetaldehyde (CH$_{\rm 3}$CHO), an asymmetric top molecule, was
detected in many star-formation environments, from cold dark
cores (e.g. \citealt{Bacmann2012, Cernicharo2012}, \citealt{Vastel2014};
\citealt{Jimenez2016}) to hot corinos (e.g. \citealt{Cazaux2003};
\citealt{Codella2016}) and protostellar shocks
(e.g. \citealt{Codella2015}; \citealt{Lefloch2017}).
Despite the several detections in a large range of interstellar
conditions, it is not clear yet if acetaldehyde is synthesised
directly on the grain surfaces (e.g. \citealt{Garrod2006}) or via
gas-phase reactions (e.g. \citealt{Charnley2004}). In particular, the
first route was recently questioned by quantum chemistry
calculations \citep{Enrique-Romero2016} but the picture is still far
from being clear.

We detected six transitions of CH$_{\rm 3}$CHO in the E form and ten
transitions in the A form, respectively. All the lines are detected
in the 1 mm band\footnote{The 5$_{\rm 1,5}$--4$_{\rm 1,4}$ E
  transition is detected in the 3 mm band, but it has been excluded from
  the analysis because it is contaminated by an unidentified emission
  line.}. The line upper level energies range from 16 to 108 K while
the peak velocities are consistent with the systemic velocity with
values between +7.8 and +9.0 km s$^{-1}$. The line shape is nearly
gaussian with FWHMs of $\sim$2--4 km s$^{-1}$.  The spectral
parameters and the gaussian fit results are reported in
Tab. \ref{Table:Ace}, while the spectra of the detected lines are
shown in Fig. \ref{Fig:CH3CHO-spectra}.  The CH$_{\rm 3}$CHO
12$_{\rm 3,9}$--11$_{\rm 3,8}$ A transition was also excluded
from the analysis since it is contaminated by the CH$_{\rm 2}$DOH
9$_{\rm 2,7}$--9$_{\rm 1,8}$ e0 transition (see
\citealt{Bianchi2017a}).  The RD analysis, shown in Fig. \ref{Fig:RD},
indicates a rotational temperature $T_{\rm rot}$ = (35$\pm$10) K and a
column density $N_{\rm tot}$ = (12$\pm$7) $\times$ 10$^{15}$
cm$^{-2}$.

\subsection{Methyl formate}
Another important iCOM which appears to be common in star
forming regions is methyl formate (HCOOCH$_{3}$), an asymmetric
top species.  It was observed towards Sgr B2 by \citet{Brown1975} and
then detected in both cold environments (e.g. \citealt{Cernicharo2012, Bacmann2012, Jimenez2016};
\citealt{Taquet2017}), in hot
corinos (e.g. \citealt{Cazaux2003}; \citealt{Bottinelli2004}) and
protostellar shocks (\citealt{Arce2008}; \citealt{Lefloch2017}).
Methyl formate is the most abundant among its isomers acetic acid
(CH$_{3}$COOH) and glycolaldehyde (HCOCH$_2$OH). In the Sgr B2
complex, a glycolaldehyde : acetic acid : methyl formate relative
abundance of $\sim$ 0.5:1:26 was measured by \citet{Hollis2001}.
The reason of this abundance differentiation is still unclear but it
is tought to be related to the respective formation processes
(e.g. \citealt{Bennett2007}, but see also \citealt{Lattelais2009}).  It
is not clear how methyl formate is synthesized. It could be a
grain-surface (e.g. \citealt{Chuang2017} and references there) or a
gas-phase (\citealt{Balucani2015}; \citealt{Taquet2016}; Skouteris et al. 2018, MNRAS, submitted.) reactions
product.

Methyl formate can exist in two different geometrical configurations:
the {\it cis} configuration, if the methyl group (CH$_{3}$) is on the
same side of the carbon chain; the {\it trans} configuration if the
methyl group is on opposing sides of the carbon chain.  The {\it
  trans} form is less stable than the {\it cis} form and, at a
temperature of 100 K, the population ratio of {\it cis} over {\it
  trans} is $\sim$ 10$^{13}$:1 \citep{Neill2012}.  Furthermore, the
involved energy barriers make unlikely the interconversion between
{\it cis} and {\it trans} forms in interstellar environments, which
explains why only the {\it cis} was detected \citep{Laas2011}.

We detected in total thirty seven lines of HCOOCH$_{3}$. Among them,
only one line was detected in the 3 mm band (with a HPBW
$\sim$27$\arcsec$) while eleven lines are detected in the 2 mm band
and the others in the 1 mm band (namely with HPBWs of
$\sim$15$\arcsec$ and $\sim$10$\arcsec$, respectively).  The lines
cover a large range of upper level energies $E_{\rm u}$ , from 20 K to
158 K. The line profiles are close to gaussian shape with typical
FWHM$\sim$2--4 km s$^{-1}$ and the peak velocities are close to the
systemic source velocity with values between +7.5 and +9.2 km
s$^{-1}$.  The spectral line parameters and the results of the
gaussian fit are reported in Tab. \ref{Table:MF} while the emission
line spectra are shown in Fig. \ref{Fig:CH3OCHO-spectra}.

The RD analysis gives a rotation temperature $T_{\rm rot}$ =
(66$\pm$5) K and a column density $N_{\rm tot}$ = (13$\pm$1) $\times$
10$^{16}$ cm$^{-2}$, as illustrated in Fig. \ref{Fig:RD} and reported
in Tab. \ref{Table:detected}.  Note that in some cases, because of the
limited spectral resolution, we observe only one line but it consists
of different transitions with the same upper level energy $E_{\rm u}$.
These unresolved multiplets are treated in the rotation diagram
analysis using the method illustrated in Appendix
\ref{Sec:Multiplets}.  On the other hand, the lines containing several
transitions with different upper level energies and different quantum
numbers are excluded from the analysis.

\subsection{Dimethyl ether}\label{Sub:DME}
Dimethyl ether (CH$_{3}$OCH$_{3}$) is one of the largest iCOMs
detected in the interstellar medium and, similarly to others, it seems
to be present in all the stages of the star formation process.
Dimethyl ether was detected for the first time by
\citet{Snyder1974} in the Orion nebula and successively, in the
context of Sun-like star forming regions, in different hot corinos
(e.g. \citealt{Cazaux2003}, \citealt{Bottinelli2004}) as well as in
cold prestellar cores (e.g. \citealt{Cernicharo2012, Bacmann2012, Jimenez2016}). As for other
iCOMs, dimethyl ether is predicted to be formed either on the grain
surfaces (e.g. \citealt{Cuppen2017} and references there) or in the gas
phase \citep{Balucani2015}.

Dimethyl ether is an asymmetric top molecule with two equivalent
methyl groups. The torsional movements along the CO-bond of the two
CH$_{3}$ rotors, cause the splitting of each rotational level into
four substates AA, EE, EA, and AE.

We detected eleven lines of dimethyl ether, with upper level energies
ranging from 33 to 254 K.  All the detected lines consist of several
lines not spectrally resolved (see Tab. \ref{Table:DME}).  As a
consequence, the line profiles do not show a gaussian shape. For
this reason, we do not perform a gaussian fit but instead we considered
the velocity-integrated intensity over the interval where the lines
are expected to emit (Tab. \ref{Table:DME}).  We perform the RD
analysis treating the multiplets with the method described in
Appendix \ref{Sec:Multiplets}. To be consistent, we use only the lines
composed by any of the four forms (AA, AE, EE, EA) merged in one line.
The values derived for the rotation temperature and column density are
$T_{\rm rot}$ = (110$\pm$10) K and $N_{\rm tot}$ = (14$\pm$4) $\times$
10$^{16}$ cm$^{-2}$, respectively (see Fig. \ref{Fig:RD} and
Tab. \ref{Table:detected}).

\subsection{Ethanol}\label{Sec:ethanol}
Ethanol (CH$_{\rm 3}$CH$_{\rm 2}$OH) was detected for the first time
in space in 1975, towards Sgr B2 \citep{Zuckerman1975}, and
successively in several high- and low- mass star forming regions.
Recently a correlation was observed between ethanol and
glycolaldehyde in Sun-like protostars \citep{Lefloch2017}.
%

Since the information in the literature is old, we performed 
new quantum chemistry computations (the details are reported in
Appendix \ref{appendix:ethanol}) in order to characterise the energy diagram of the three
ethanol rotamers (anti, gauche+ and gauche-).
Fig. \ref{Fig:ethanol-chem} shows their relative energies as well as
the interconversion path between them. Our new computations give an
energy difference between the anti and gauche forms of 71 K, slightly
larger than the 56 K value by \citet{Pearson2008}.
 Therefore, for large
enough temperatures (i.e. $\geq 70$ K), the abundance ratio between
the anti and gauche forms is equal to the respective statistical
weight ratio, namely 1.

\begin{figure*}
\begin{center}
\begin{subfigure}{\columnwidth}
\includegraphics[width=9cm]{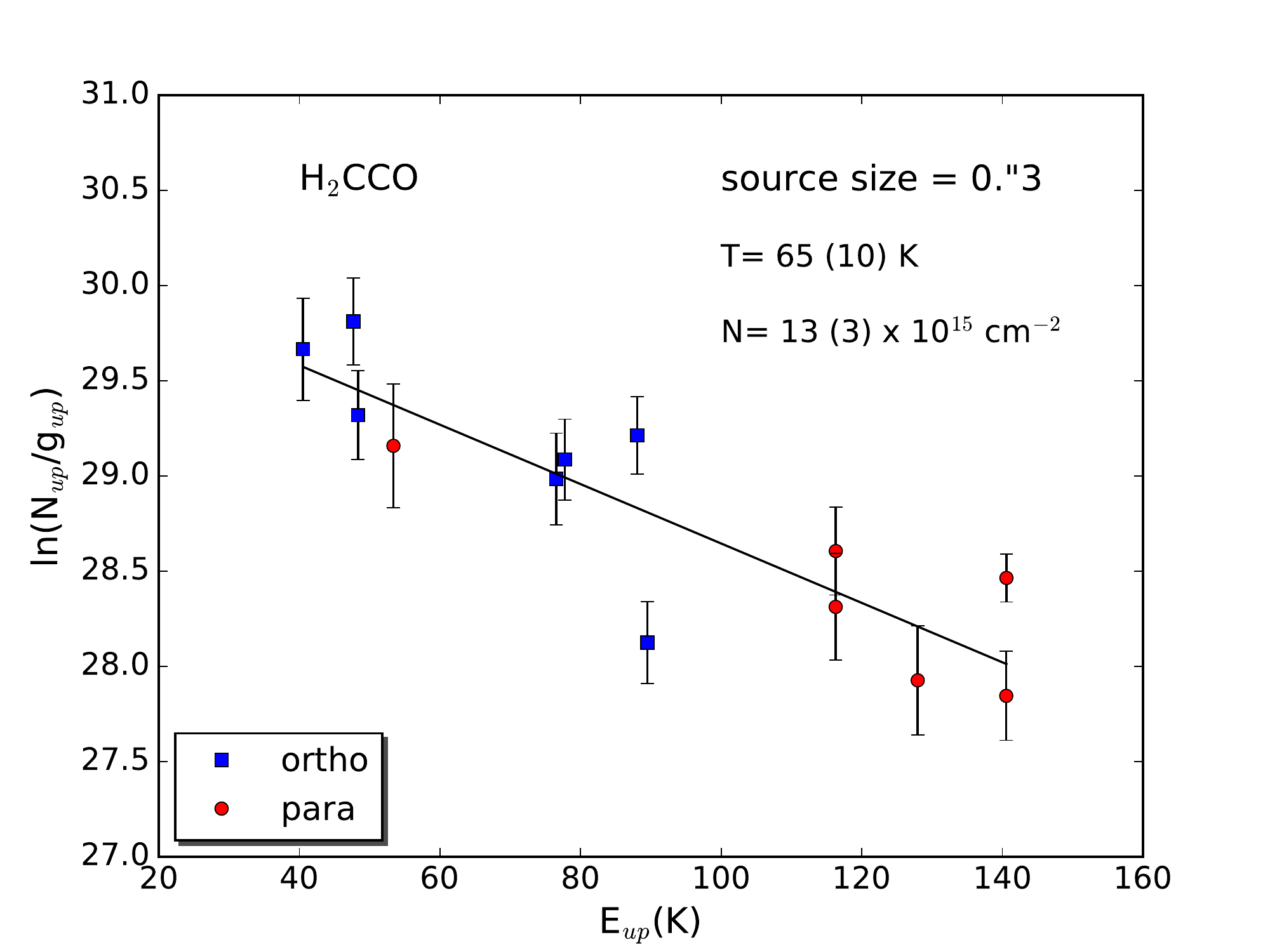}
\end{subfigure}\hfill
\begin{subfigure}{\columnwidth}
\includegraphics[width=9cm]{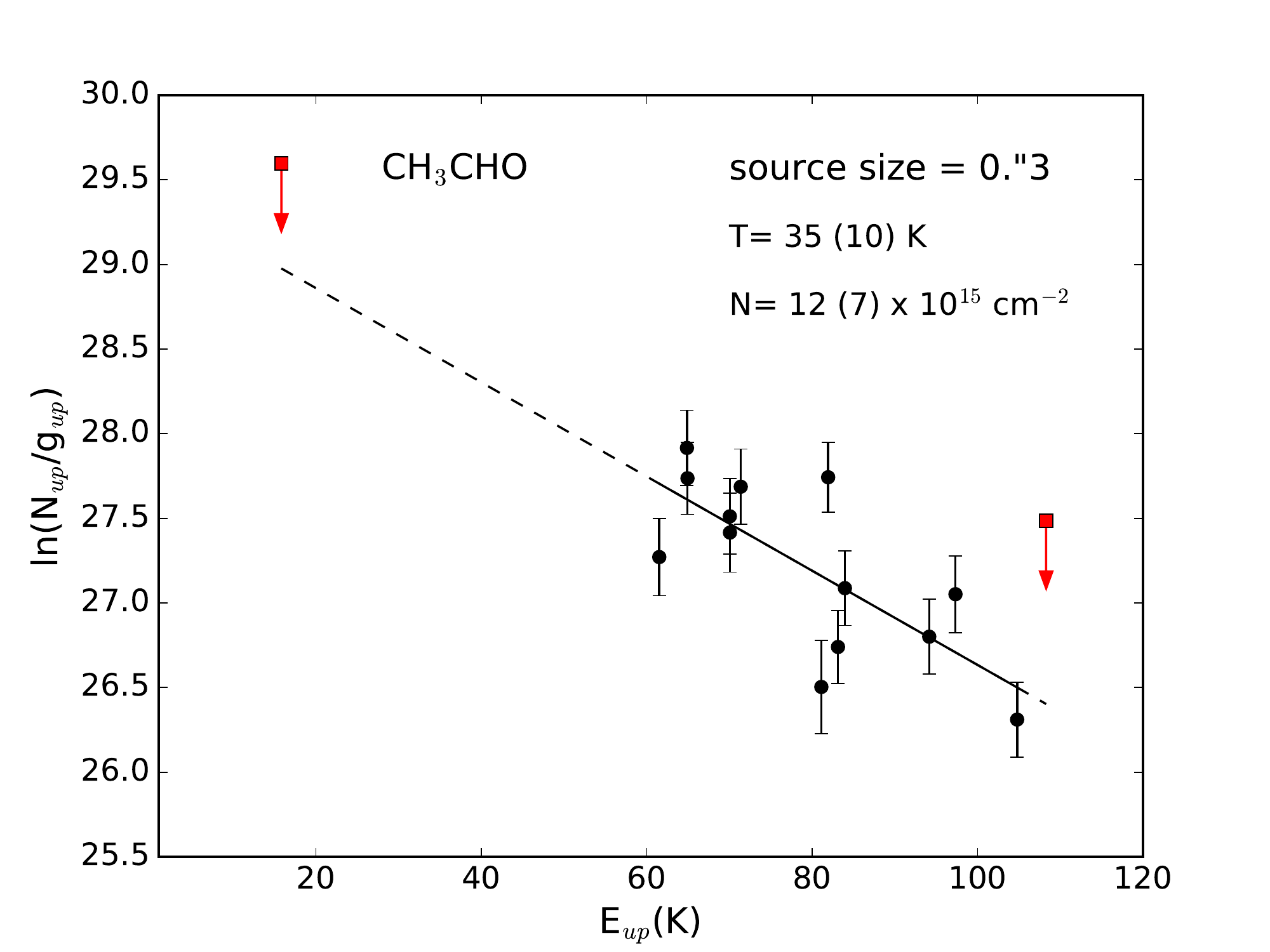}
\end{subfigure}\hfill
\begin{subfigure}{\columnwidth}
\includegraphics[width=9cm]{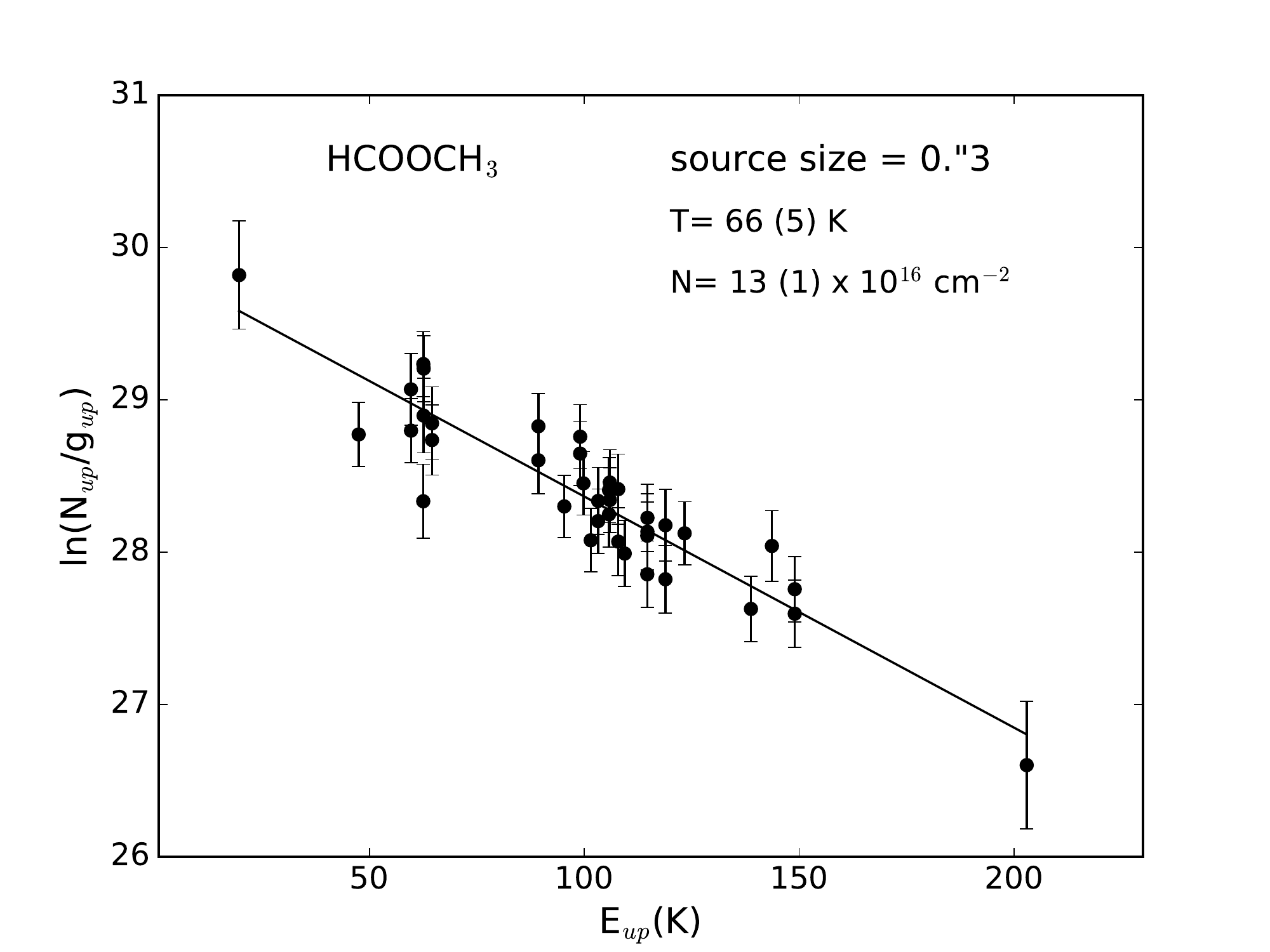}
\end{subfigure}\hfill
\begin{subfigure}{\columnwidth}
\includegraphics[width=9cm]{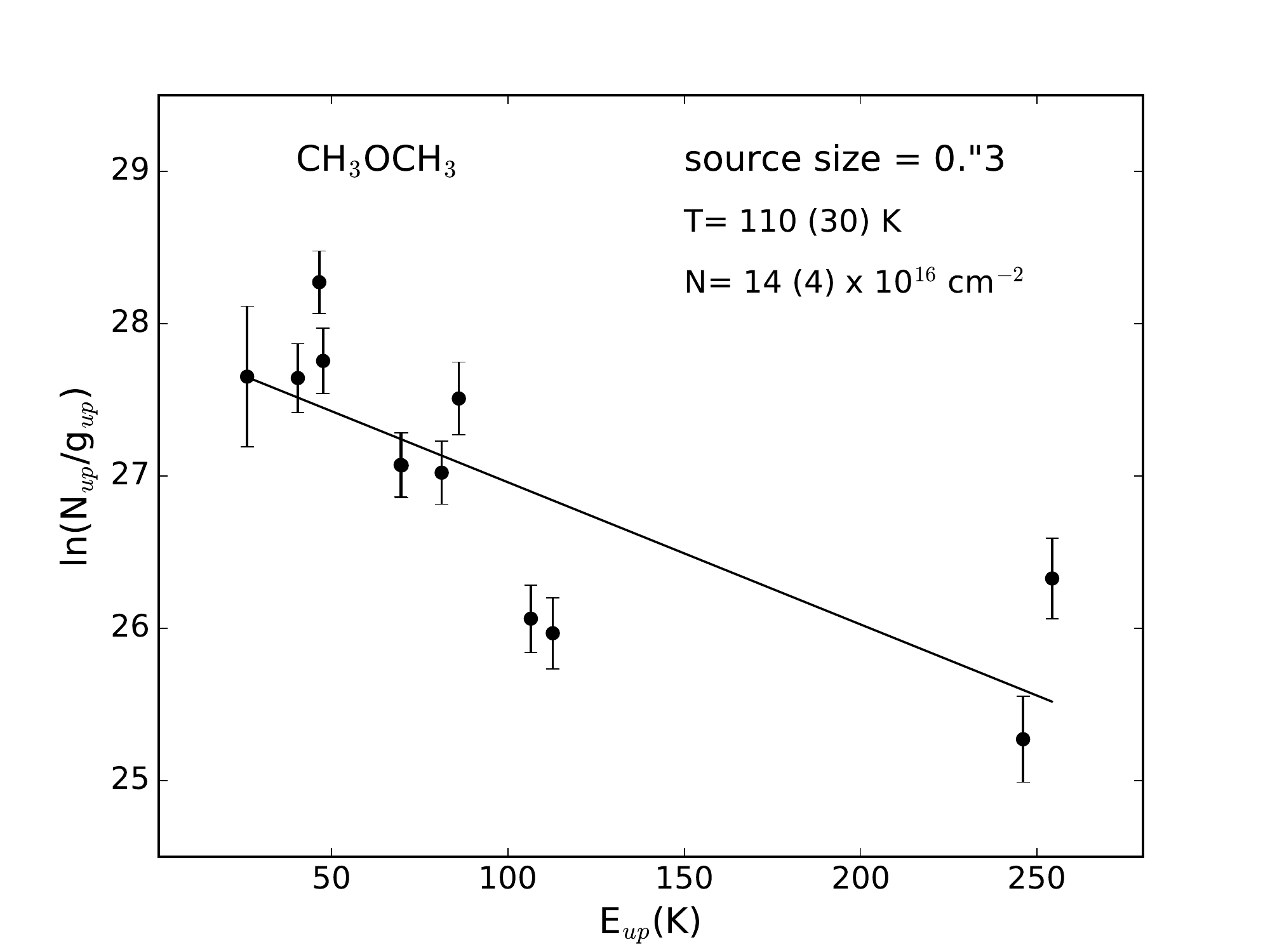}
\end{subfigure}\hfill
\begin{subfigure}{\columnwidth}
\includegraphics[width=9cm]{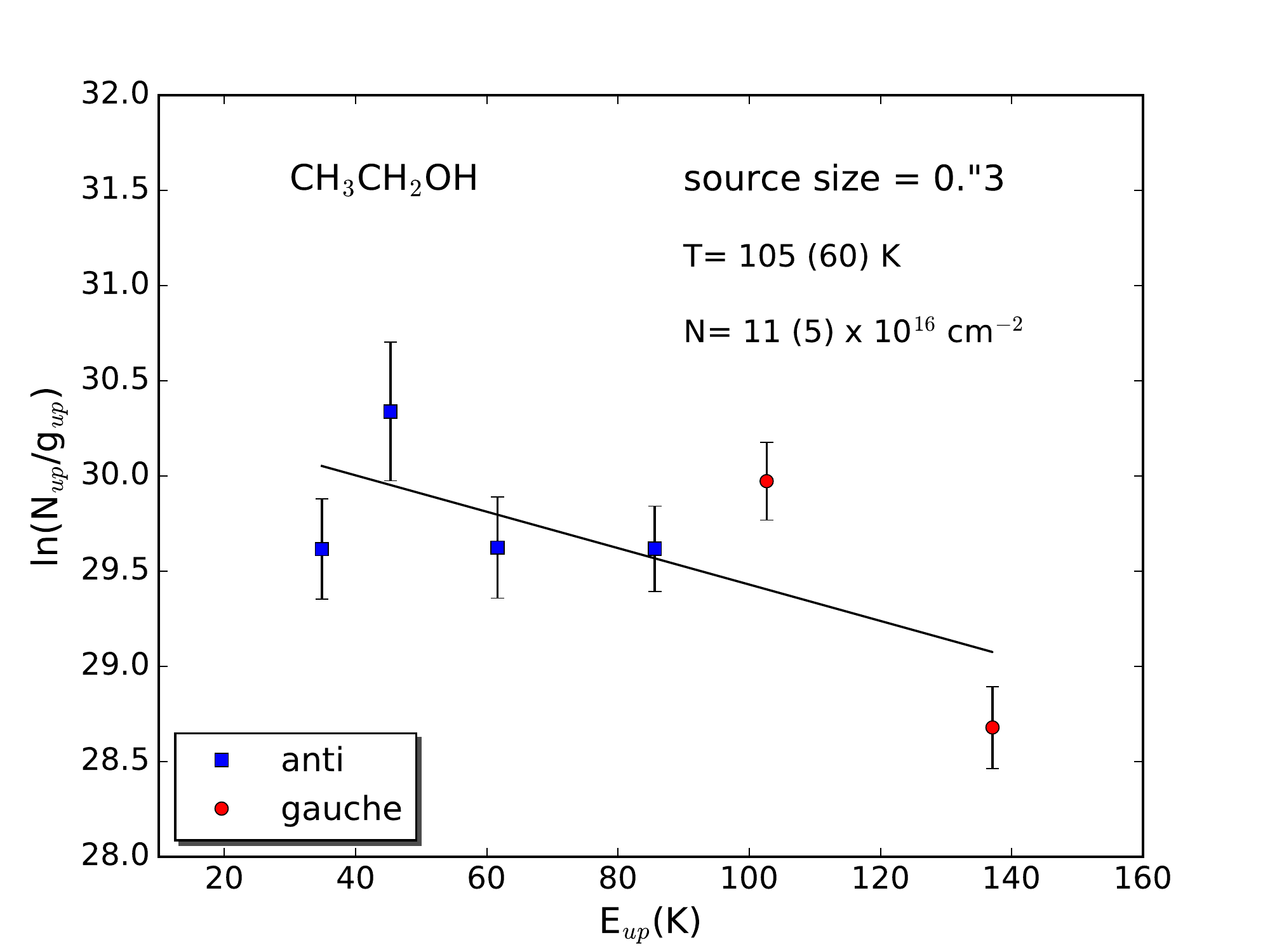}
\end{subfigure}\hfill
\begin{subfigure}{\columnwidth}
\includegraphics[width=9cm]{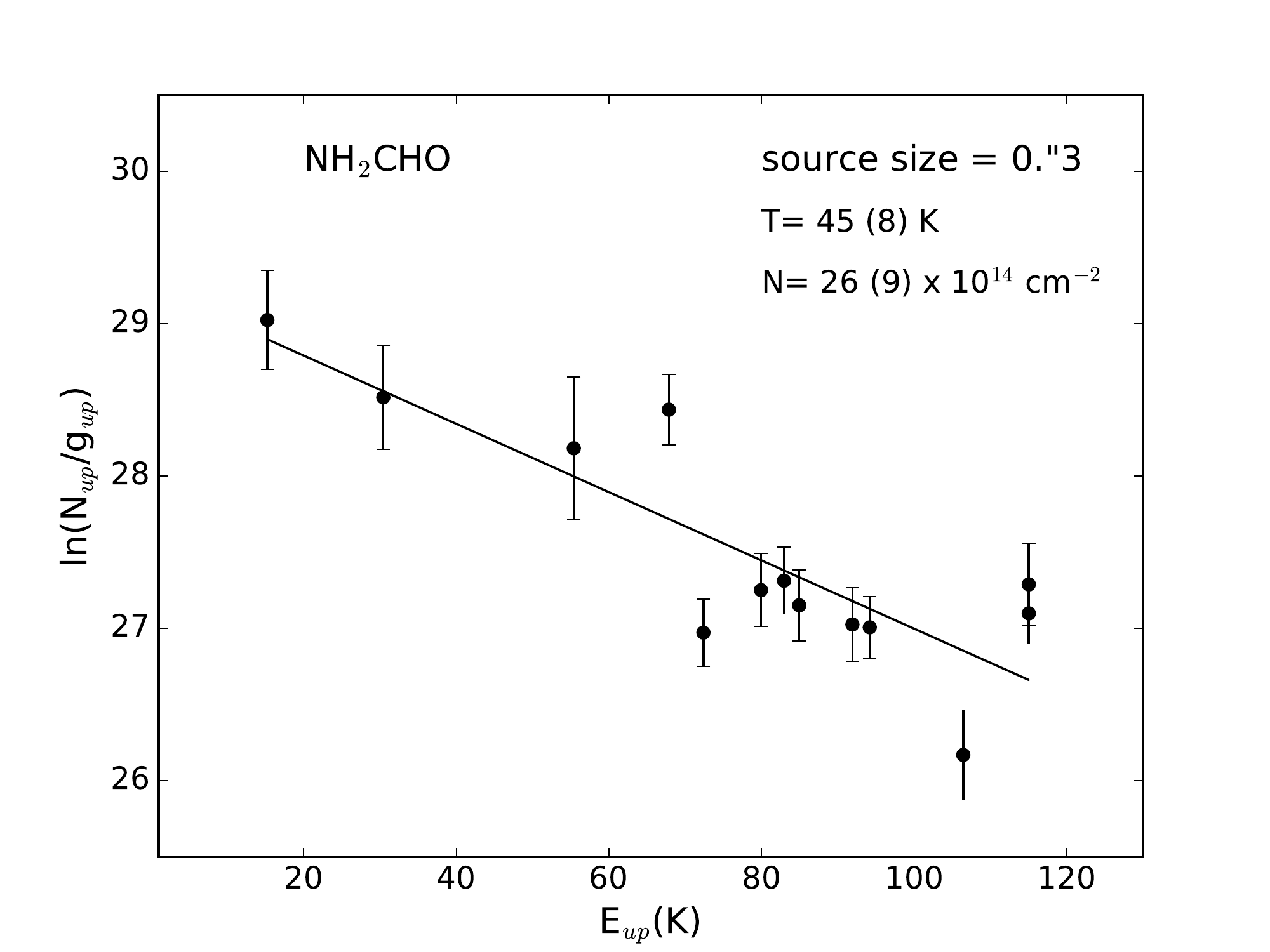}
\end{subfigure}\hfill
\end{center}
\caption{Rotation diagrams of ketene (upper left panel), acetaldehyde
  (upper right panel), methyl formate (left middle panel), dimethyl
  ether (right middle panel), ethanol (lower left panel) and formamide (lower right panel).
   An emitting region size of 0$\farcs$3 (corresponding to $\sim$ 70 au) is assumed
  (see text). The parameters N$_{u}$, g$_{u}$, and $E_{\rm u}$ are the
  column density, the degeneracy and the energy (with respect to the
  ground state of each symmetry) of the upper level, respectively. The
  derived values of the rotational temperature and the column density
  are reported in each panel for each species. For H$_2$CCO, blue
  squares and red dots indicate ortho- and para- transitions,
  respectively. Red arrows in the CH$_3$CHO panel indicate the
  5$_{\rm 1,5}$--4$_{\rm 1,4}$ E and 12$_{\rm 4,8}$--11$_{\rm 4,7}$ E
  transitions, which were not used in the fit because
  contaminated. For CH$_3$CH$_2$OH, blue squares and red dots indicate
  anti- and gauche- transitions, respectively. NH$_2$CHO data are from 
  \citet{Lopez2015} but they
were reanalysed assuming the size of 0$\farcs$3.}
\label{Fig:RD}
\end{figure*}

We detected two lines of the gauche and three lines of the anti forms.
All the transitions are detected in the
1.3 mm band (HPBW $\simeq$ $\sim$10--11$\arcsec$).  The upper level
energies of the detected lines are in the 35--137 K range.  The line
shape is nearly gaussian with FWHMs ranging between 1 and 3 km
s$^{-1}$.  The peak velocities are in the +8.2 -- +8.6 km s$^{-1}$
range, consistently with the systemic source velocity.  The observed
spectra are shown in Fig. \ref{Fig:CH3CH2OH-spectra}, while the
spectral line parameters are reported in Tab. \ref{Table:ethanol}.
The RD analysis gives a rotation temperature $T_{\rm rot}$ =
(105$\pm$60) K and a column density $N_{\rm tot}$ = (11$\pm$5)
$\times$ 10$^{16}$ cm$^{-2}$ (Fig. \ref{Fig:RD} and
Tab. \ref{Table:detected}).  The low number of detected lines and
their respective upper level energies do not allow us to put strong
constraints on the anti and gauche abundance ratio, so that we assumed the
theoretical high-temperature limit.

\subsection{Formamide}\label{formamide}

Finally, for comparison, we analysed also the lines from formamide,
previously reported by \citet{Lopez2015}. Formamide
is an interstellar molecule of great interest because it was
proposed as a pre-biotic precursor of genetic material
(e.g. \citealt{Saladino2012}, and references therein). It was
detected for the first time in space by \citet{Rubin1971} towards Sgr
B2, and only very recently it was reported in the context of
low-mass star formation: namely protostellar shocks
\citep{Yamaguchi2012, Mendoza2014}, and hot corinos
\citep{Kahane2013, Lopez2015, Coutens2016}.  

We detected thirteen lines and analysed them using the RD approach,
assuming here a size of 0$\farcs$3, as for the other species, instead of the 1$\arcsec$ assumed by \citet{Lopez2015}. The derived
values of rotation temperature and column density are $T_{\rm rot}$ =
(45$\pm$8) K and $N_{\rm tot}$ = (26 $\pm$9) $\times$ 10$^{14}$
cm$^{-2}$, respectively (see Tab. \ref{Table:detected}).


\subsection{Summary of the results} \label{summary} 

The summary of the derived iCOMs column densities and rotational
temperatures is given in Tab. \ref{Table:detected}.  In addition,
Tab. \ref{Table:detected} reports the upper limits on the column
density of non-detected iCOMs detected in some Class 0 hot corinos
(e.g. \citealt{Jaber2014}): HCOOH, HCCCOH, HCOCH$_{2}$OH,
CH$_{3}$NH$_{2}$, CH$_{3}$COCH$_{3}$ and CH$_{3}$O.  Note that, to
derive the upper limits, we assumed a rotational temperature of 80 K,
 i.e the average temperature measured using the detected iCOMs
 and a typical FWHM of 2.5 km s$^{-1}$.
We also notice that HCOCH$_{2}$OH was previously imaged towards
SVS13-A by \citet{Desimone2017} using the IRAM-NOEMA interferometer.
These authors derived a column density $N_{\rm tot}$ = (21$\pm$8)
$\times$ 10$^{14}$ cm$^{-2}$, consistent with the upper limit derived
here.

With the exception of acetaldehyde and formamide, all detected iCOMs
have a rotational temperature larger than about 45 K, indicating that
the gas of the hot corino region dominates the line emission of these
species. The lower temperature derived for acetaldehyde and formamide
could indicate that there is a not negligible emission from colder
and more extended gas, but always within the 10$\arcsec$ region where the
lines were detected. It is possible that this is gas associated with
the outflow. Indeed, these species were detected towards the L1157 outflow 
by \citet{Mendoza2014}, \citet{Codella2014,Codella2016, Codella2017}, \citet{Lefloch2017}.      

Regarding the column densities, methyl formate, dimethyl ether and ethanol are
the detected iCOMs with the largest (and similar) values, followed by
ketene and acetaldehyde (similar values) about a factor ten lower, and
then formamide, which is the least abundant detected iCOM in SVS13-A,
about a factor three lower.


\section{Discussion}\label{Sec:Discussion}

\subsection{The hot corino of SVS13-A}\label{sec:hot-corino-svs13}

The hot corino phenomenon was discovered about fifteen years ago towards
the Class 0 protostar IRAS16293--2422 \citep{Cazaux2003}. Since
then, less than fifteen hot corinos were found, some of which are
rather candidates than confirmed hot corinos (see later). Most of
them are in Class 0 protostars: NGC1333 IRAS4A
\citep{Bottinelli2004}; IRAS4B and IRAS2A \citep{Bottinelli2007};
Serpens SMM1 and SMM4 \citep{Oberg2011}; SVS 4-5 \citep{Oberg2014};
IRAS23238+7401, L1455 SMM1 and B1-c \citep{Bergner2017}; HH212-mm
\citep{Codella2016}; IRAS19347+0727 in B335 \citep{Imai2016} L483
\citep{Oya2017}; CepE-mm \citep{Ospina2018}.  
Even fewer hot corino candidates have so far been found in Class I
sources: B1-a \citep{Oberg2014} and IRAS03245+3002 (or L1455 IRS 1;
\citealt{Bergner2017}) in Perseus.  We notice, though, that all Class
I hot corinos were observed with the single-dish IRAM-30m
telescope and that the derived rotational temperatures of the few
detected iCOMs are very low, less than 20 K. These low temperatures
cast some doubts on whether the detected iCOMs emission originates in
the hot corinos of these sources or rather in a more extended
component, such as outflows or PDRs (Photo-Dissociation Regions).  

The case of SVS13-A, here studied, is a clear-cut Class I hot corino
for the following reasons: (1) eight iCOMs were detected (the
seven detected by this study, see Table \ref{Table:detected}, plus
glycolaldehyde detected by \citealt{Desimone2017}); (2) the emission
from one iCOM was imaged and showed to originate in a compact
region \citep{Desimone2017}; (3) the rotational temperatures of the
iCOMs here detected are all larger than 45 K (with the exception of
acetaldehyde, which however is also the species with the smallest 
interval of the transition upper level energy, from 61 to 108 K), which
ensures that the emission arises from hot gas. In other words, 
SVS13-A is the only confirmed Class I hot corino and it is also the
one with the largest number of detected iCOMs.

In Sect. \ref{Sec:Results}, we derived the iCOMs column densities from
the ASAI observations. In order to convert them into abundances, we
estimated the H$_2$ column density from the IRAM-PdBI\footnote{IRAM
  Plateau de Bure Interferometer: {\it
    http://www.iram-institute.org/}.} continuum observations obtained
by \citet{Chen2009}. These authors quote a gas mass of (0.75$\pm$0.12)
M$_{\odot}$ for SVS13-A over a 1$\farcs$1 diameter. Correcting for the
different distance assumed by \citet{Chen2009} (350 pc instead of 235
pc; see Sec. \ref{Sec:Source}) and the gas temperature (they assumed
20 K) gives an average H$_2$ density of $\sim$ 2.9 $\times$ 10$^9$
cm$^{-3}$. Therefore, assuming that the mass is mostly concentrated
towards the hot corino, we obtain a H$_2$ column density of $\sim$ 3
$\times$ 10$^{24}$ cm$^{-2}$. The measured abundance of the iCOMs
detected in SVS13-A ranges from $\sim$ 9 $\times$ 10$^{-10}$
(formamide) to $\sim$ 4 $\times$ 10$^{-8}$ (methyl formate and
ethanol). The comparison of these values with those previously
derived in Class 0 hot corinos is discussed in the next Section.

\subsection{Comparison with other hot corinos}\label{Sec:comparison}

The question that we want to address in this Section is whether the
SVS13-A hot corino shows any sign of diversity with respect to Class 0
hot corinos, which may indicate an evolution of the chemical
composition. 
To this end, we compared the iCOMs abundances in the known hot
corinos. We included in the comparison all candidate Class 0 and I hot
corinos identified in the literature, including those observed only with single-dish telescopes.
Note that, given the well-known relatively large uncertainty in the
derivation of the H$_{2}$ column density and, hence, absolute
abundances, we compared iCOMs abundance ratios of different species.

\begin{figure*}
\begin{center}
\includegraphics[width=20cm]{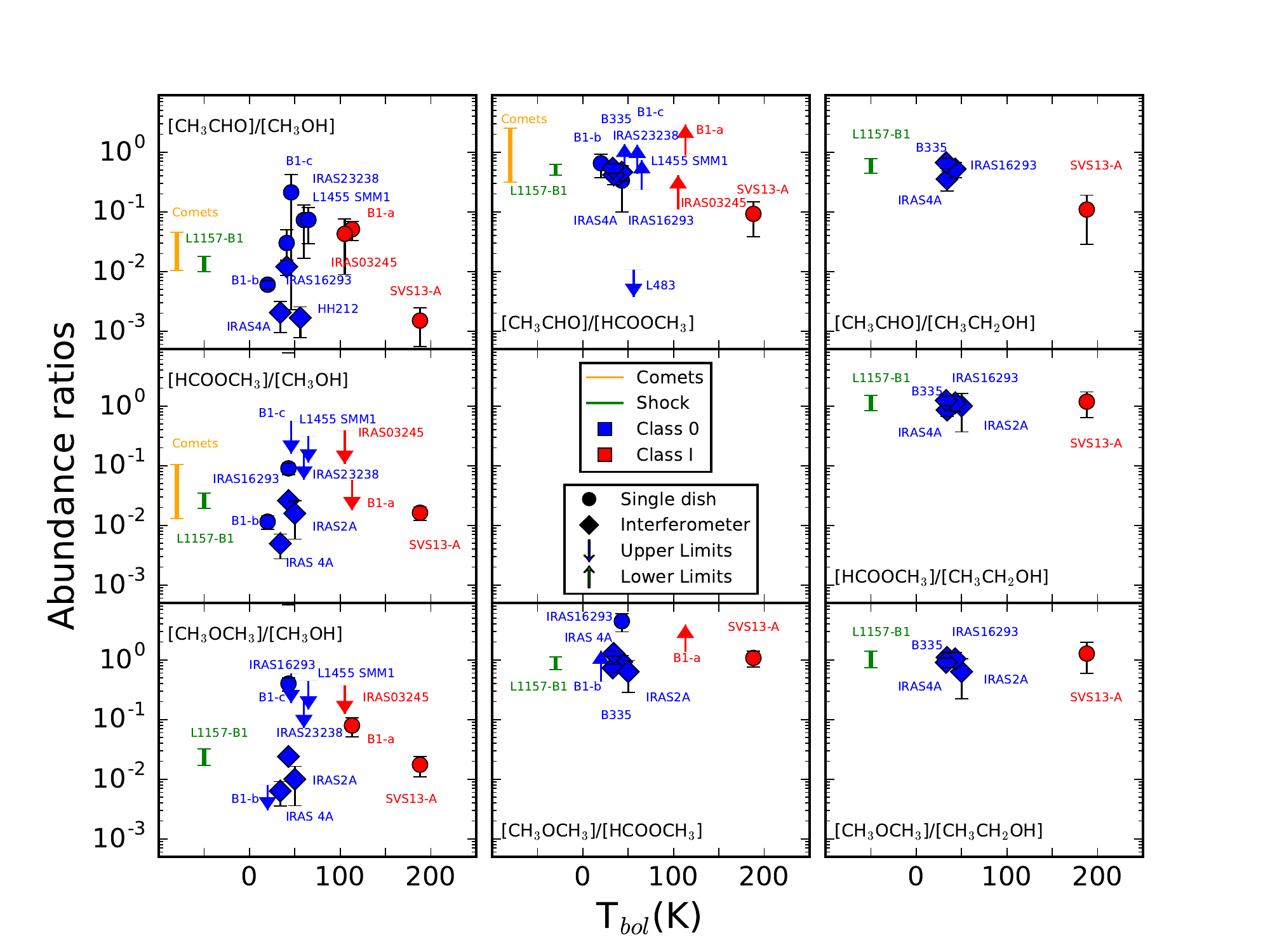}
\end{center}
\caption{Abundance ratios of the iCOMs detecetd in SVS13-A compared
  to different sources, as indicated in the upper panels. The sources
  are: the pre-protostar
  hydrostatic core B1-b \citep{Pezzuto2012, Gerin2015}; the Class 0
  sources IRAS4A, IRAS2A \citep{Taquet2015, Lopez2017}, IRAS16293--2422
  (e.g. \citealt{Jaber2014}, \citealt{Jorgensen2016, Jorgensen2018}), B1-c
  \citep{Oberg2014, Bergner2017}, IRAS23238+7401 \citep{Graninger2016,
    Bergner2017}, L1455 SMM1 \citep{Graninger2016, Bergner2017}, HH212
  \citep{Bianchi2017b, Codella2018}, IRAS19347+0727
  in B335 \citep{Imai2016} and L483 \citep{Oya2017}; the protostellar
  shock L1157-B1 \citep{Codella2010, Lefloch2017}, the Class I sources
  B1-a \citep{Oberg2014, Bergner2017}, IRAS03245+3002
  \citep{Graninger2016, Bergner2017} and SVS13-A (this work); comets
  \citep{LeRoy2015}. Blue symbols indicate Class 0 protostars while red symbols are for 
  Class I protostars. Circles indicate single-dish measurements while diamonds are for
  interferometric measurements. Arrows indicate upper limit measurements. 
  The abundance ratios of comets and the protostellar shock L1157-B1 
  are reported for comparison using an orange line and a green line, respectively. Note that 
  in these cases the x-value has no meaning.}
\label{Fig:COMs}
\end{figure*}

Figure \ref{Fig:COMs} shows the ratio of acetaldehyde, methyl formate
and dimethyl ether with respect to methanol, methyl formate and
ethanol, respectively. These species were selected because
they are the ones detected in the largest number of candidate hot
corinos.
Note that the abundance ratios are showed as a function of the source
bolometric temperature T$_{bol}$, which is considered an indicator of
the different evolutionary status of the source \citep{Myers1998}.

\begin{figure}
\begin{center}
\includegraphics[width=9cm]{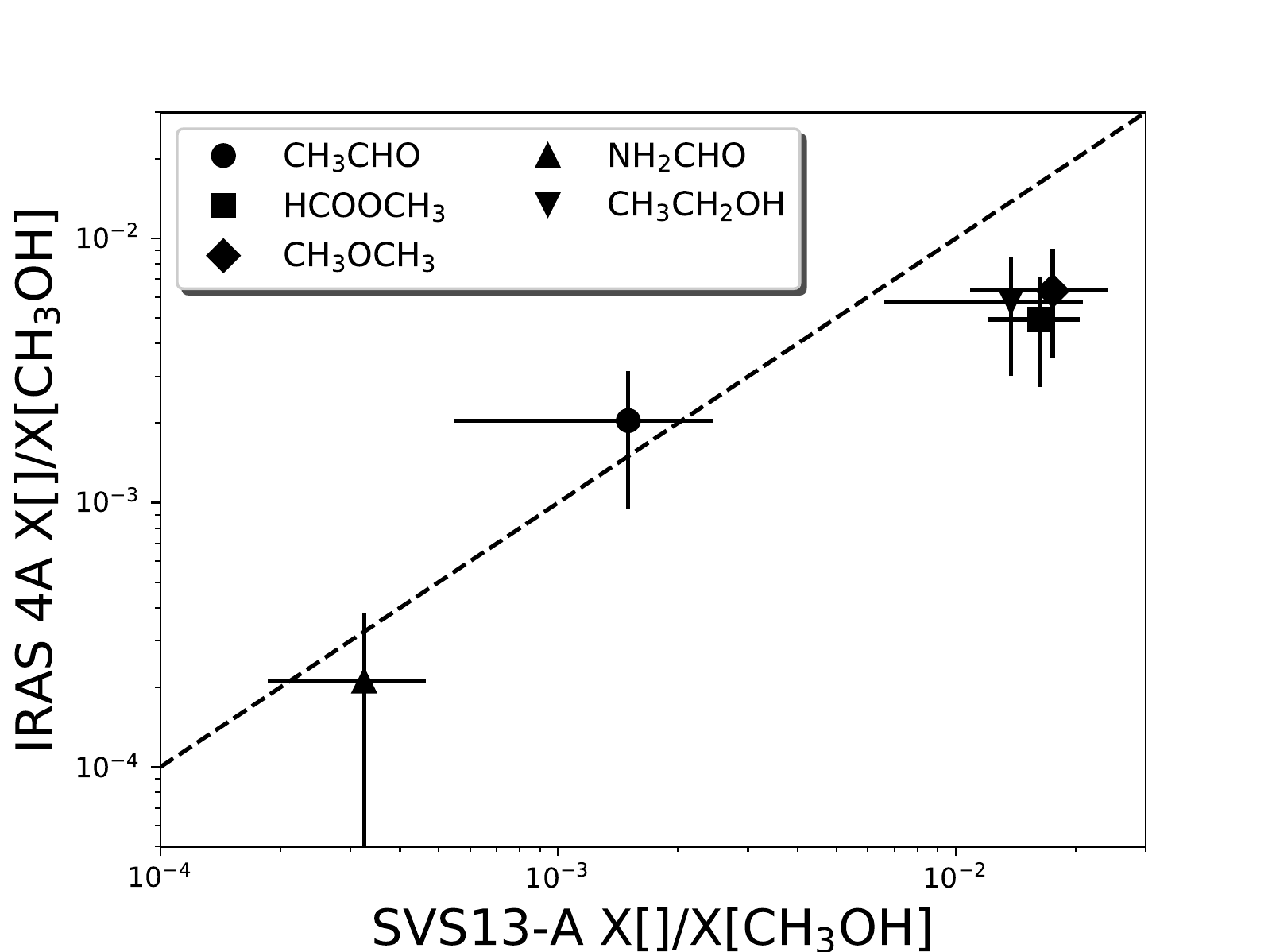}
\includegraphics[width=9cm]{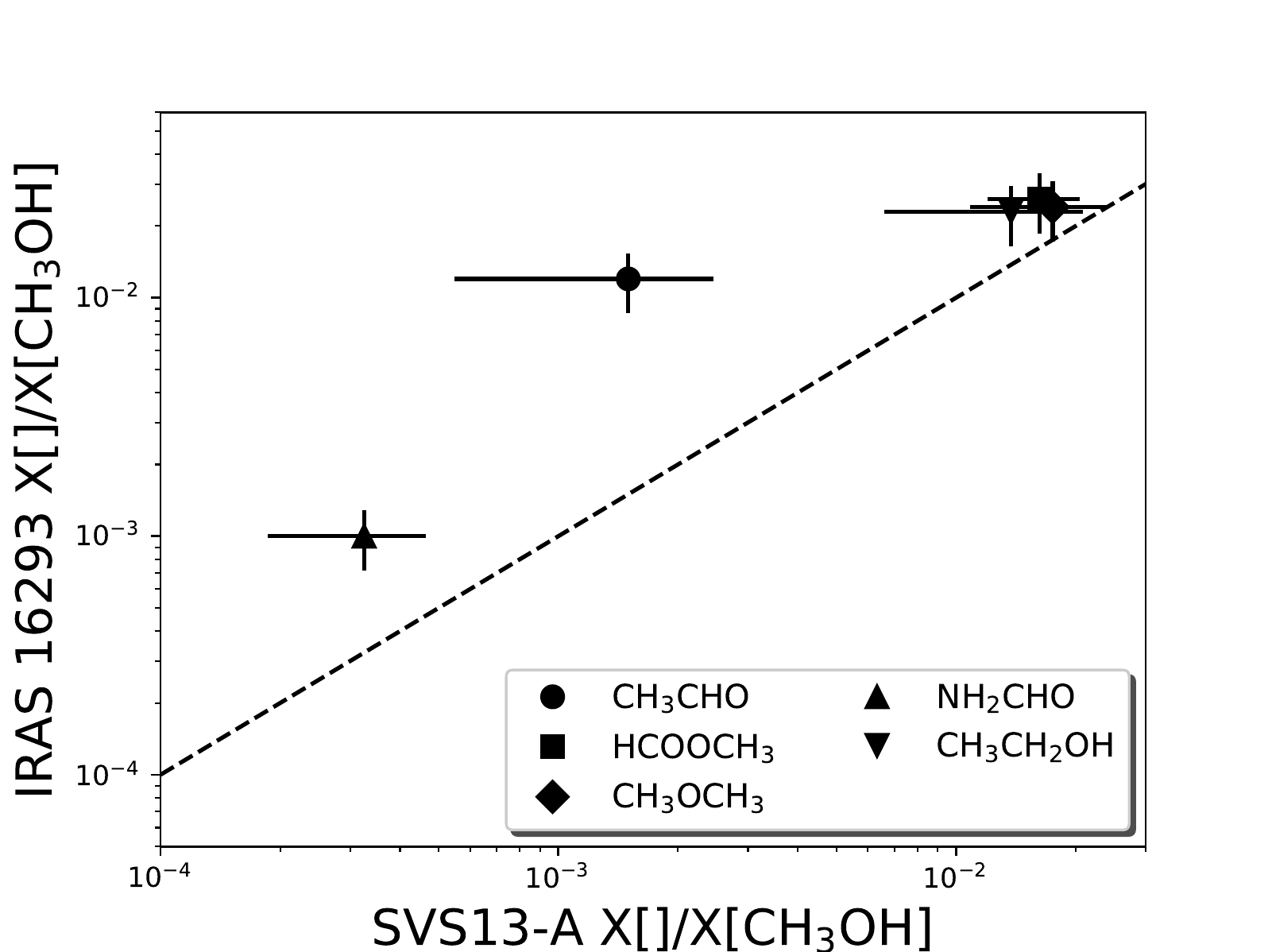}
\end{center}
\caption{Abundances, normalised to methanol, of the iCOMs detected in
  the hot corinos of IRAS4A (upper panel, y-axis), IRAS16293--2422 (lower panel, y-axis) 
and SVS13-A (x-axis). The
  different symbols represent different molecules, as marked in the
  figure inset: acetaldehyde (circles), methyl formate (squares),
  dimethyl ether (diamonds), formamide (up-triangles) and
  ethanol (low-triangles). The iCOMs abundances in IRAS4A are
  from \citet{Lopez2017} and \citet{Taquet2015}, those of IRAS16293--2422 are from 
  \citet{Jorgensen2018} and \citet{Coutens2016} while the SVS13-A abundances are from the present work.}
\label{Fig:SVS-IR4A}
\end{figure}

We will conservatively limit our analysis to the 
objects for which the sizes of the hot corino
has been determined using an interferometer:
namely the Class 0 IRAS16293--2422, IRAS4A, IRAS2A, and HH212,
and the Class I SVS13-A.
The observations of SVS13-A, even if single-dish, are supported by the previous image of the hot corino by \citet{Desimone2017}. This allow us to
safely correct for the appropriate filling factor, obtaining reliable measurements of the iCOMs abundances. 
Fig. \ref{Fig:COMs} does
not show a significant difference in the iCOMs relative abundances of the four
Class 0 hot corinos with respect to the (only) Class I one, SVS13-A.
This is also shown in Fig. \ref{Fig:SVS-IR4A}, where the abundances,
normalised with respect to methanol, of the six iCOMs detected in
SVS13-A are compared with those detected in IRAS4A and IRAS16293--2422\footnote{Note that
  these are the only Class 0 hot corinos in the literature where all the six
  iCOMs detected in SVS13-A are available and observed with an interferometer.} (\citealt{Taquet2015, Lopez2017, Coutens2016, Jorgensen2018}). In
practice, the six iCOMs have similar abundance ratios, within one order of magnitude,
in both Class 0 and Class I hot corinos.

We conclude that, so far based on only less than a handful sources and
molecules, Class 0 and Class I hot corinos do not differ in their
iCOMs relative abundances. Needless to say, more observations of both hot
corinos and iCOMs are necessary to confirm or invalidate this
conclusion.
In this respect, it is worth noticing that the molecular deuteration
seems, on the other hand, to decrease from Class 0 to Class I sources
\citet{Bianchi2017a}. Also in this case, however, more measurements,
especially with interferometers able to disentangle the various
components in a single-dish telescope beam, are needed before drawing
firm conclusions.

\subsection{Comparison with comets}\label{Sec:comets}

An important open question of astro- and cosmo-chemistry is whether
there is a link between the molecules formed in the protostellar phase
and, particularly, in the hot corino and those found in cometary
material (see e.g. \citealt{Caselli2012, Ceccarelli2014, Drozdovskaya2018}.
Particularly important is understanding
the possible heritage of iCOMs, for the hypothetical link with the
life emergence on the Earth.

\begin{figure*}
\begin{center}
\includegraphics[width=18cm]{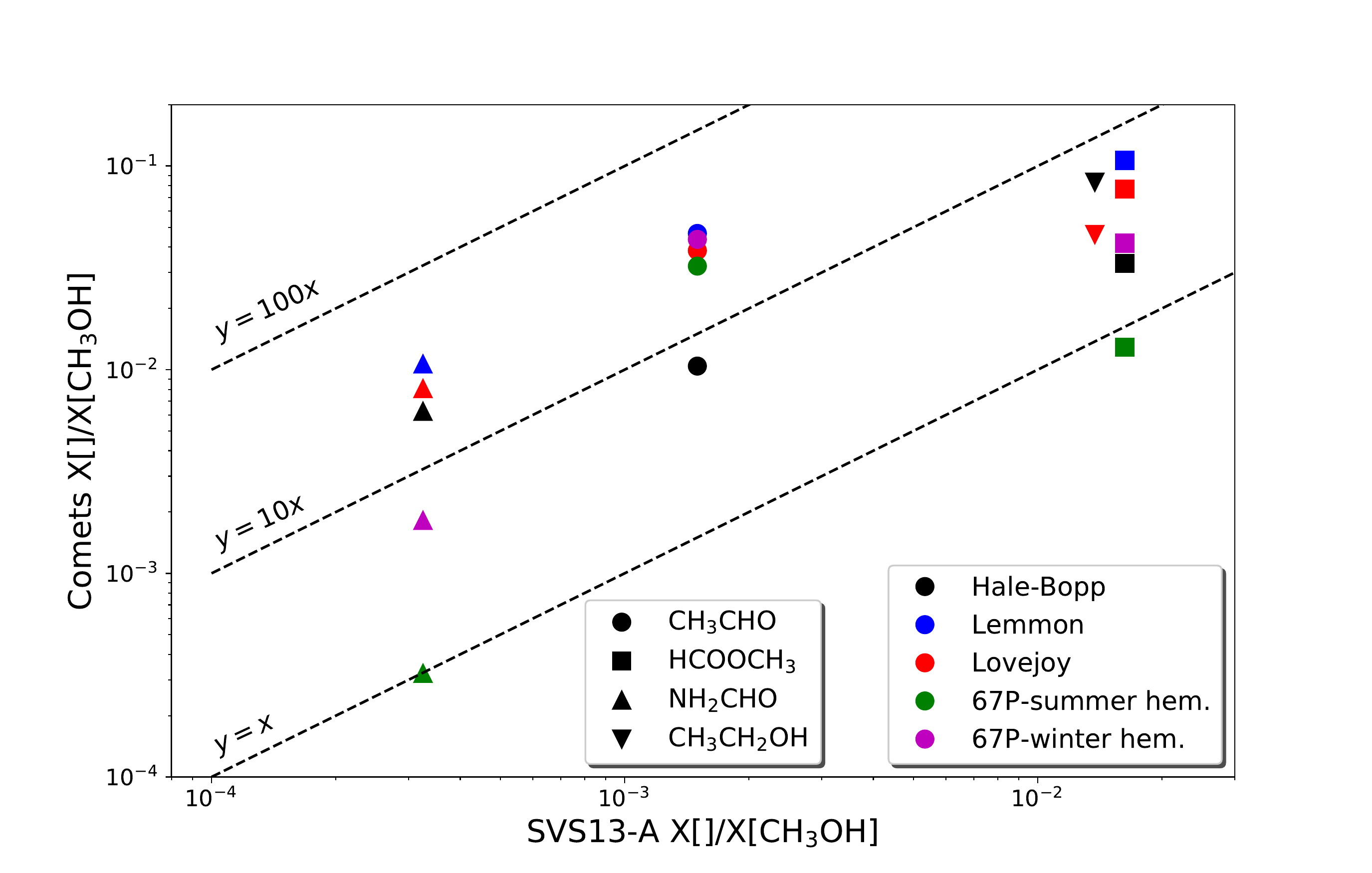}
\end{center}
\caption{Abundances, normalised to methanol, of the iCOMs detected in various comets (y-axis) and SVS-13A (x-axis). The
  different symbols represent different molecules, as marked in the
  figure inset: acetaldehyde (circles), methyl formate (squares) and
 formamide (up-triangles) and
  ethanol (low-triangles). Different color correspond to different
  comets: Hale-Bopp (black), Lemmon (blue), Lovejoy (red) and 67P in
  the summer (green) and winter (magenta) hemisphere, respectively.
  The iCOMs abundances in comets are taken from the compilation of \citep{LeRoy2015},
   except for ethanol which is from \citep{Biver2015}, while the SVS13-A abundances are from the present
  work.}
\label{Fig:comets}
\end{figure*}

Figure \ref{Fig:comets} shows the abundance ratio of the iCOMs
detected in SVS13-A and the four comets where some of the same
molecules were detected: Hale-Bopp, Lemmon, Lovejoy and 67P. In
the latter, measurements exist of the winter and summer hemispheres
\citep{LeRoy2015}. 
The first remark is that in the 67P summer hemisphere formamide and
methyl formate have the same abundances, normalised to methanol, than
in SVS13-A, whereas acetaldehyde is more than a factor ten more
abundant in 67P. A similar trend is present also in the 67P winter
hemisphere, with acetaldehyde more abundant than formamide and methyl
formate, although the latter abundances are also larger with respect
to those in SVS13-A. In the Lemmon and Lovejoy comets, acetaldehyde 
and formamide are about a constant factor twenty larger than in SVS13-A,
while methyl formate is less than a factor ten larger.
In Hale-Bopp the difference with respect to SVS13-A decreases
from formamide to acetaldehyde and to methyl formate.
For ethanol the abundance ratios obtained in the two comets
Lemmon and Lovejoy are higher with respect to what measured 
in SVS13-A by less than a factor ten.

Therefore, we can conclude that the relative abundances of
 methyl formate and ethanol do not seem to be
substantially different between SVS13-A and comets
within a factor ten.
Acetaldehyde shows instead higher relative abundances in comets, while
formamide shows the highest spread.

\subsection{The case of Methoxy}\label{sec:case-methoxy}

The unprecedented results provided by ASAI, i.e. a large number of
iCOMs detected through a large number of emission lines put severe
constraints on the iCOMs relative abundances.  An interesting case is
represented by methoxy (CH$_{3}$O), which is not detected in SVS13-A,
with an upper limit in the column density of $\leq$ 3 $\times$
$10^{15}$ cm$^{-2}$.  Similarly, methoxy is not detected in the "warm
sources'' observed with the IRAM-NOEMA interferometer\footnote{{\it
    http://www.iram-institute.org/}} within the SOLIS project ({\it
  https://solis.osug.fr/}: \citealt{Ceccarelli2017}), while it is, so
far, only detected towards few cold sources (e.g. \citealt{Cernicharo2012}; 
\citealt{Jimenez2016}; \citealt{Bacmann2016}).

Methoxy was recently proposed as a possible precursor of
methyl formate and dimethyl ether in gas-phase processes
\citep{Balucani2015}.  Given the high abundances of HCOOCH$_{3}$
($N_{\rm tot}$ $\sim$ 1.3 $\times$ 10$^{17}$ cm$^{-2}$) and
CH$_{3}$OCH$_{3}$ ($N_{\rm tot}$ $\sim$ 1.4 $\times$ 10$^{17}$
cm$^{-2}$), the present CH$_{3}$O non detection seems at odds with the
proposed gas formation route. However, the low methoxy abundance
with respect to that of methyl formate and dimethyl ether may just
indicate that all methoxy is consumed to form these two species.

So far, only models adapted to cold objects were published
(\citealt{Balucani2015}; \citealt{Vasyunin2017}), so that in order to
reach firm conclusions new models adapted to describe hot corino
should be considered.  These models will have to account for the upper
limits here found: [CH$_{3}$O]/[HCOOCH$_{3}$] and
[CH$_{3}$O]/[CH$_{3}$OCH$_{3}$]$<$0.02.

\section{Conclusions}\label{Sec:Conclusions}

The analysis of the Class I object SVS13-A provides us the opportunity
to characterize the chemical content of a more evolved source, trying
to determine a possible evolutionary trend in the comparison with
Class 0 protostars. Even if the angular resolution of single-dish
data is not enough to disentangle the protostellar components, the
large number of lines provided by unbiased spectral surveys (such as
ASAI) allows us to analyse the global chemical complexity in a source
previously unexplored.  Note that we previously \citep{Bianchi2017a}
measured towards SVS13-A a methanol and formaldehyde deuteration up to
two orders of magnitude lower than the values measured in Class 0
sources located in the same star-forming region and observed using the
same telescope.  This was indeed a first indication of a modified
chemical content possibly due to the different evolutionary stages.
Thanks to the wide observed bandwidth, we 
detected several iCOMs
such as CH$_3$CHO, HCOOCH$_3$, CH$_3$OCH$_3$ and CH$_3$CH$_2$OH, typical of 
a hot corino.
SVS13-A appears to be as chemically rich as previously studied Class 0
protostars. In addition, it seems that the iCOMs abundances do not significantly vary 
during the protostellar phase.
On the other hand, the comparison of the relative iCOMs abundances measured towards SVS13-A and in comets
shows different behaviours. More specifically, methyl formate and ethanol do not seem to be substantially different, within a factor ten, in SVS13-A and comets, while for the other iCOMs the difference is larger up to a factor 30.

\section*{Acknowledgments}

The authors are grateful to the IRAM staff for its help in the
calibration of the 30-m data. 
The research leading to these results has received funding from the
European Commission Seventh Framework Programme (FP/2007-2013) under
grant agreement N¡ 283393 (RadioNet3).  
This work was supported by by
the MIUR (Italian Ministero dell'Istruzione, Universit\`a e Ricerca)
through the grants: Progetti Premiali 2012 - iALMA (CUP
C52I13000140001), the program PRIN-MIUR 2015 STARS in the CAOS
(Simulation Tools for Astrochemical Reactivity and Spectroscopy in the
Cyberinfrastructure for Astrochemical Organic Species) (2015F59J3R)e).
This project has also been supported by the PRIN-INAF 2016 "The Cradle
of Life - GENESIS-SKA (General Conditions in Early Planetary Systems
for the rise of life with SKA)".  
RB acknowledges the financial support from from Spanish MINECO (through
project FIS2012-32096).  
Finally, we acknowledge the funding from the European Research Council (ERC) under the European Union's Horizon 2020 research and innovation programme, for the Project "The Dawn of Organic Chemistry" (DOC), grant agreement No 741002.

\newpage
\bibliographystyle{mnras} 
\bibliography{Mybib.bib}

\newpage

\appendix
\section{Rotational diagram analysis}\label{sec:RD}

Given that for the observed iCOMs, the collisional rates
are not available in literature, we used the standard Rotational Diagram (RD) analysis
to estimate the temperature and the column density. 
We assumed optically thin emission and Local Thermodynamic Equilibrium
(LTE) conditions.
The relative population distribution of
all the energy levels is then described by the rotational temperature $T_{\rm rot}$.
The upper level column density can be written as:

\begin{equation} 
N_{u} = \frac{8 \pi k \nu^{2}}{h c^{3} A_{ul}} \frac{1}{ff} \int T_{mb} dV
\label{eq:col_dens}
\end{equation}

where $k$ and $h$ are, respectively, the Boltzmann and Planck constants, $\nu$ is the frequency of the transition, 
$c$ is the light speed, A$_{\rm ul}$ is the Einstein coefficient of the $u$ $\to$ $\l$ transition, 
$ff$\footnote{$ff = \theta_{s}^{2} \times (\theta_{s}^{2}+\theta_{b}^{2})^{-1}$; $\theta_{s}$ and $\theta_{b}$ 
are the source (assumed to be a circular gaussian) and the beam sizes.} is the beam-filling factor 
and the integral is the integrated line intensities.
$N_{\rm u}$ is related to the rotational temperature $T_{\rm rot}$, as follows:

\begin{equation} 
ln \frac{N_{u}}{g_{u}} = ln N_{tot} - ln Q(T_{rot}) - \frac{E_{up}}{k T_{rot}} 
\end{equation}

where $g_{\rm u}$ and $E_{\rm u}$ are, respectively, the generacy and the energy of the upper level, 
$N_{\rm tot}$ is the total column 
density of the molecule, and $Q$($T_{\rm rot}$) is the partition function, depending on the rotational 
temperature. 
We assumed an angular size of 0$\farcs$3, in agreement with \citet{Desimone2017} IRAM-PdBI HCOCH$_2$OH images.

\subsection{Analysis of multiplets}\label{Sec:Multiplets}

In the case of some species, as for example CH$_3$OCH$_3$, the spectral resolution of the observations does not allow us 
to resolve the single transitions. 
We observe only one emission line which consists of several transitions $i$ with 
the same upper level energy $E_u$, but different Einstein coefficients 
$A_{ul} (i)$ and degeneracies $g_u (i)$.
These unresolved multiplets require to be treated using the following method in the rotation diagram analysis.
The observed intensity is the sum of the integrated intensity of the single $i$ transitions, so that

\begin{equation}
I=\sum_{i} I(i).
\end{equation}

Given that the transitions have the same $E_u$, Eq. \ref{eq:col_dens} become

\begin{equation}
\dfrac{n_u(i)}{g_u(i)}=\dfrac{8 \pi k \nu^2}{h c^3}\dfrac{\int T_b dV}{\sum_i (A_{ul}(i) g_u(i))}.
\end{equation}

\section{Ethanol: quantum chemistry computation}\label{appendix:ethanol}

Ethanol exists under the form of three rotamers: anti, gauche+ and
gauche- ethanol.  Since relatively only old computations exist in the
literature, we performed ourselves ab initio quantum chemistry
calculations in order to characterise the ethanol conformes geometry
and energetic. The computations were performed using a better level of theory in 
order to improve the accuracy of the values.

To this end, we carried out calculations based on the Density
Functional Theory (DFT) using the Gaussian16 suite of programs
\citep{g16}. Full geometry optimizations followed by vibrational
computations were carried out for all three minima and both transition
states to make sure that the first ones were true minima on the
Potential Energy Surface (PES) and that the latter exhibited a single
imaginary frequency. Those calculations were achieved with the B2PLYP
double hybrid functional \citep{Grimme2006} in conjunction with the
m-aug-cc-pVTZ triple-$\xi$ basis set \citep{Papajak2009,
  Dunning1989}. Semiempirical dispersion contributions were also
included by means of the D3BJ model of Grimme, leading to the
so-called B2PLYP-D3 computational model \citep{Goerigk2011,
  Grimme2011}.

\begin{figure}
\begin{center}
\includegraphics[width=8cm]{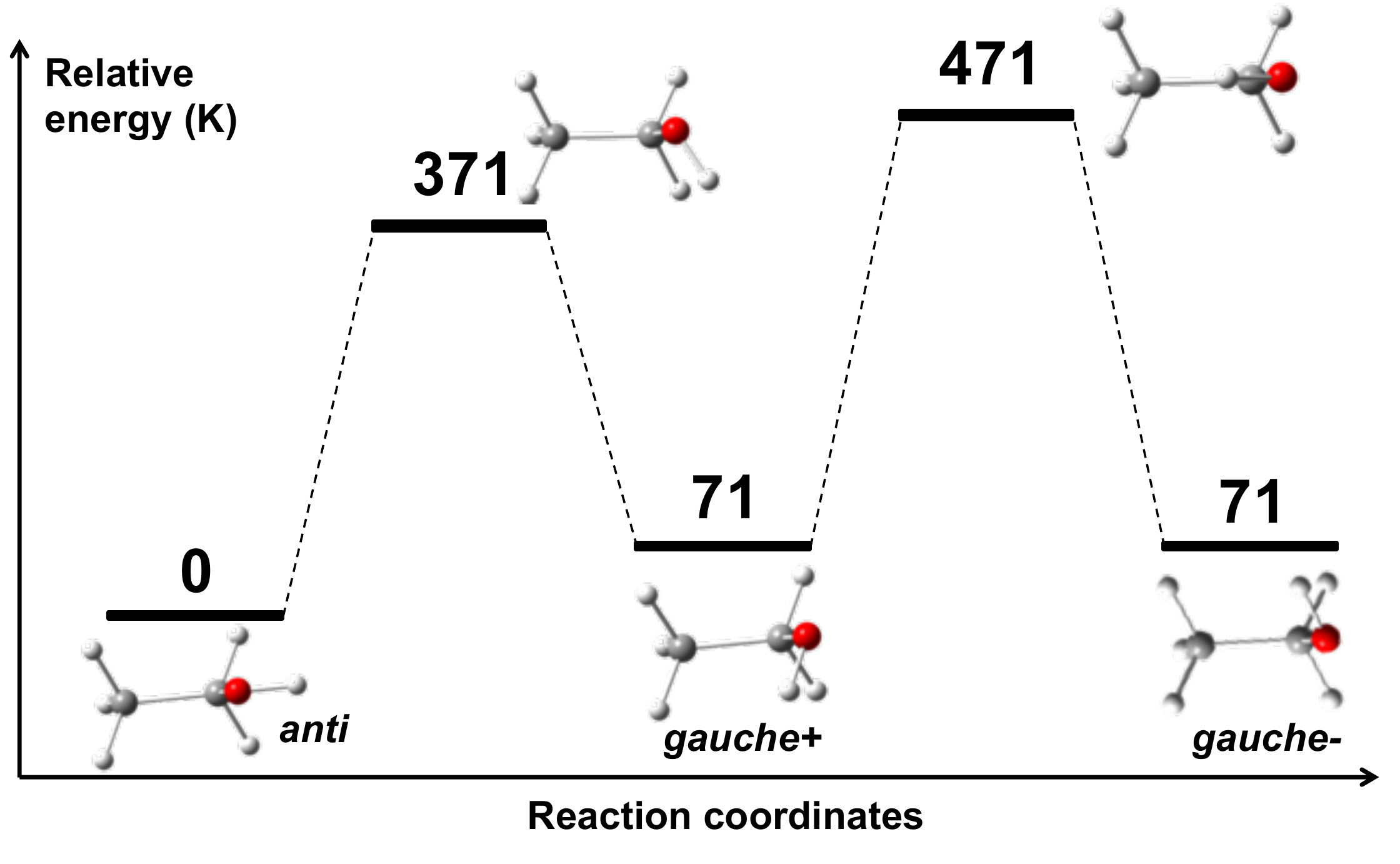}
\end{center}
\caption{Interconversion path between conformers of ethanol. }
\label{Fig:ethanol-chem}
\end{figure}

The results of the calculations are shown in
Fig. \ref{Fig:ethanol-chem}.  Both gauche+ and gauche- rotamers, that
are energetically equal for symmetry reasons, are 71 K less stable
than their anti counterpart. The isomerisation (namely to reach the
gauche rotamers from the anti one) requires an energy of 371 K, while
471 K are necessary to go from one gauche rotamer to the other. 

\section{iCOMs spectra and tables}\label{appendix}

\begin{figure*}
\begin{center}
\includegraphics[width=14cm]{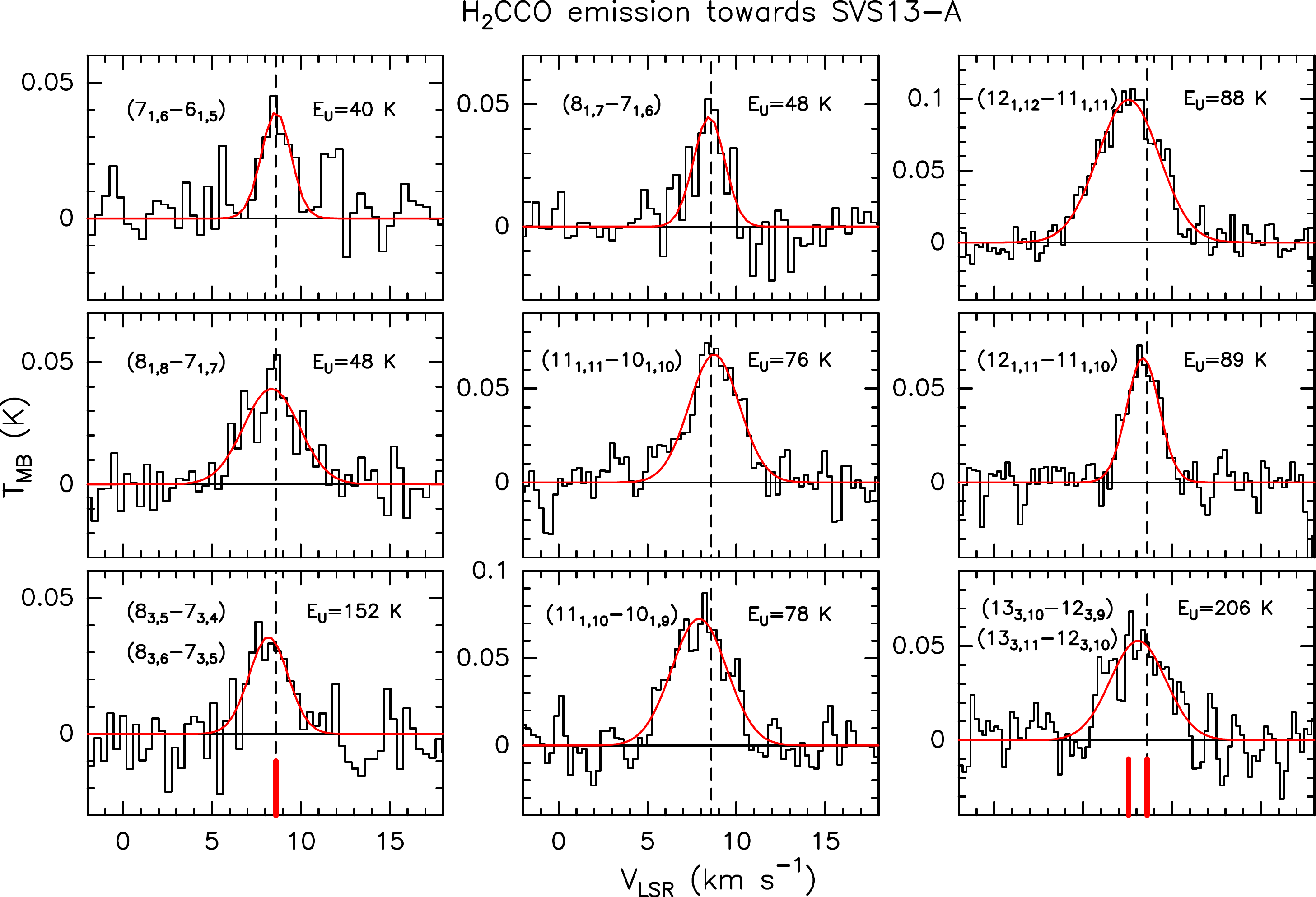}
\includegraphics[trim=0 0 0 -1cm, width=10cm]{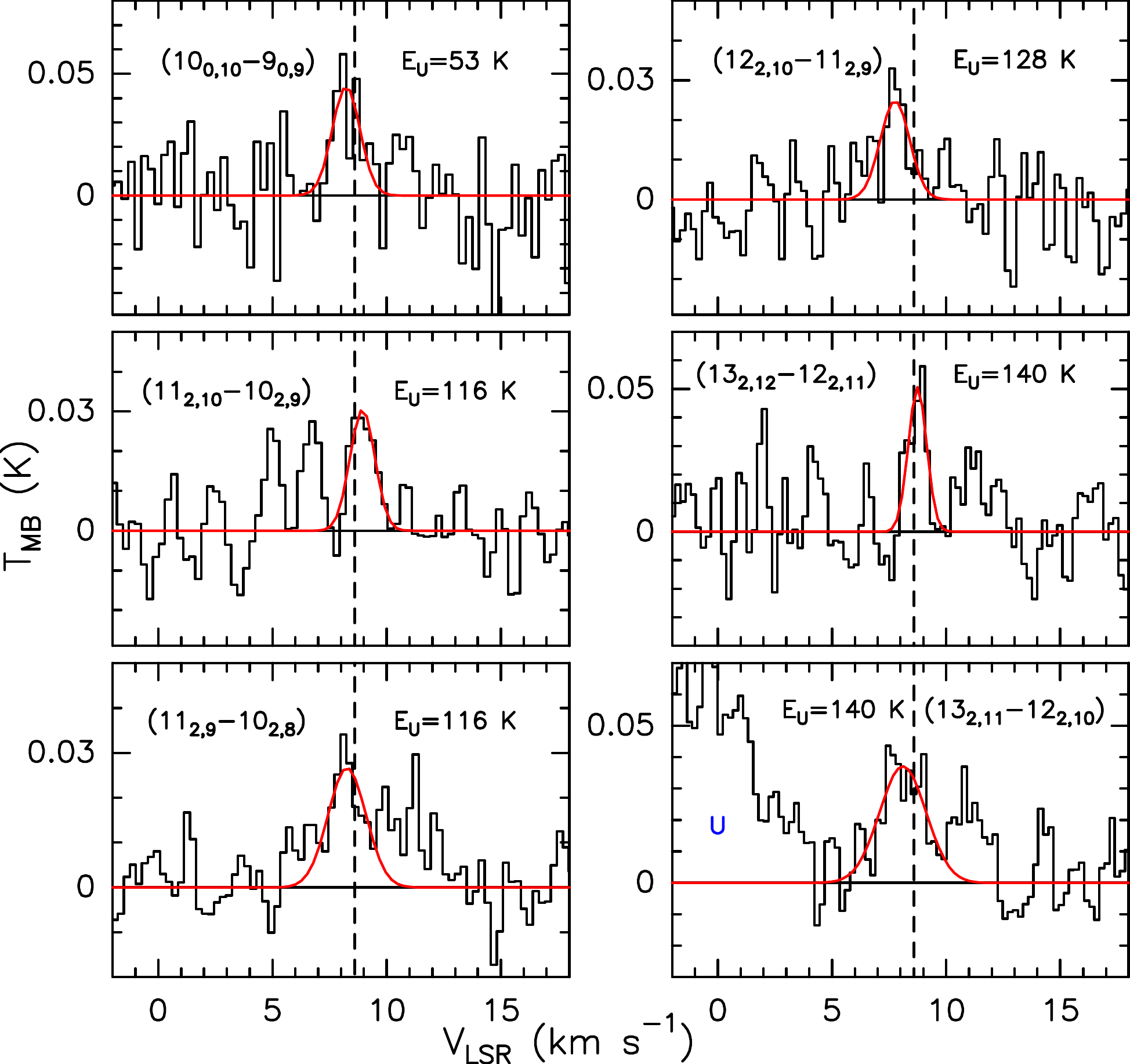}
\end{center}
\caption{H$_{2}$CCO line profiles  in $T_{\rm MB}$ scale (not corrected for the beam dilution); transitions are reported. The vertical dashed line stands for the ambient LSR velocity (+ 8.6 Km s$^{-1}$, \citealt{Chen2009}). For blended lines, the vertical red solid lines indicate the different transitions. In the upper and lower panel are reported the ortho and it para transitions respectively. Note that the intensity of the para transitions is lower than the ortho ones, consistently with the ratio 1:3 of the statistical weights in the high temperature limit (T $\gg$15 K).}
\label{Fig:H2CCO-spectra}
\end{figure*}

\begin{figure*}
\begin{center}
\includegraphics[width=14cm]{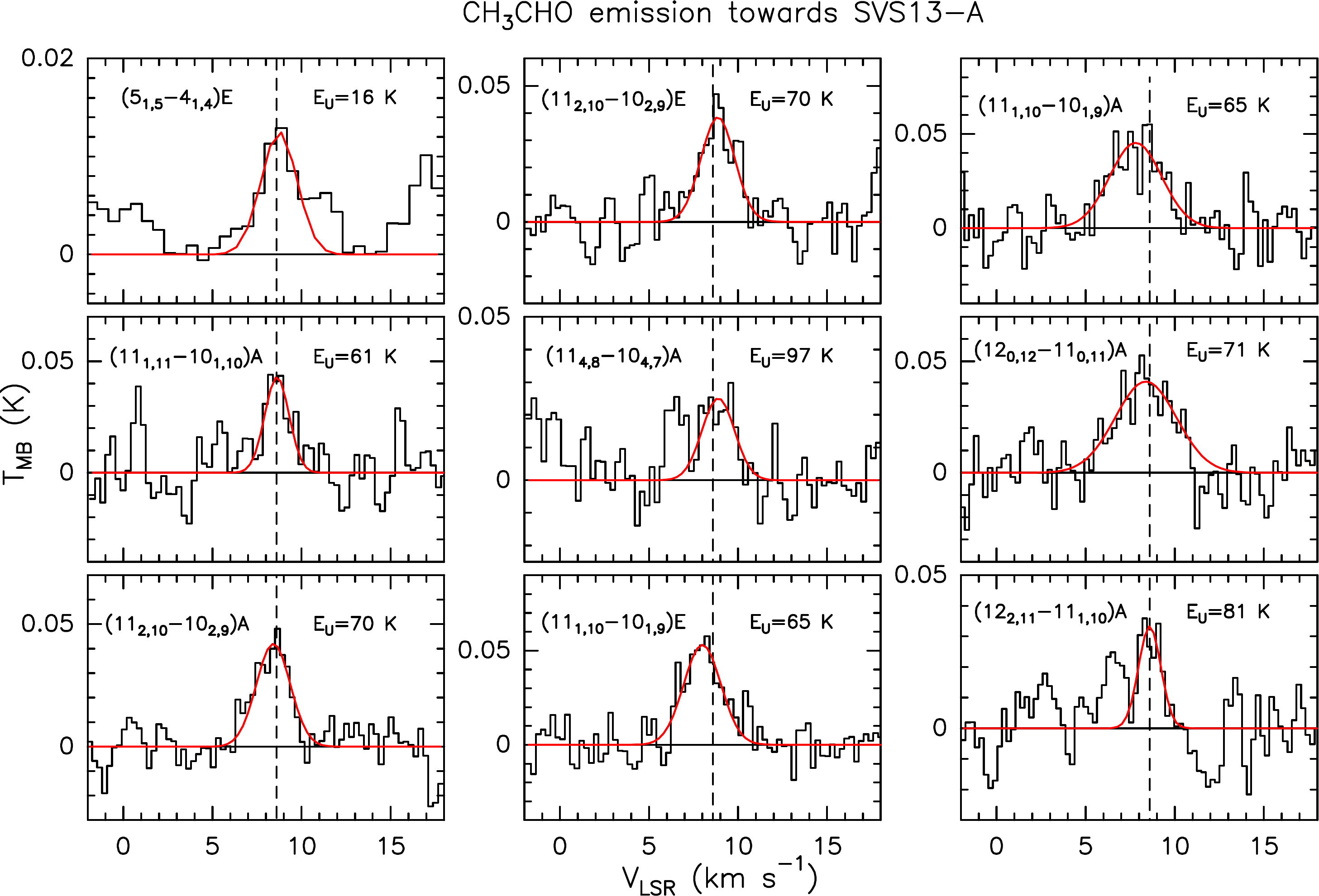}
\includegraphics[trim=0 0 0 -1cm, width=10cm]{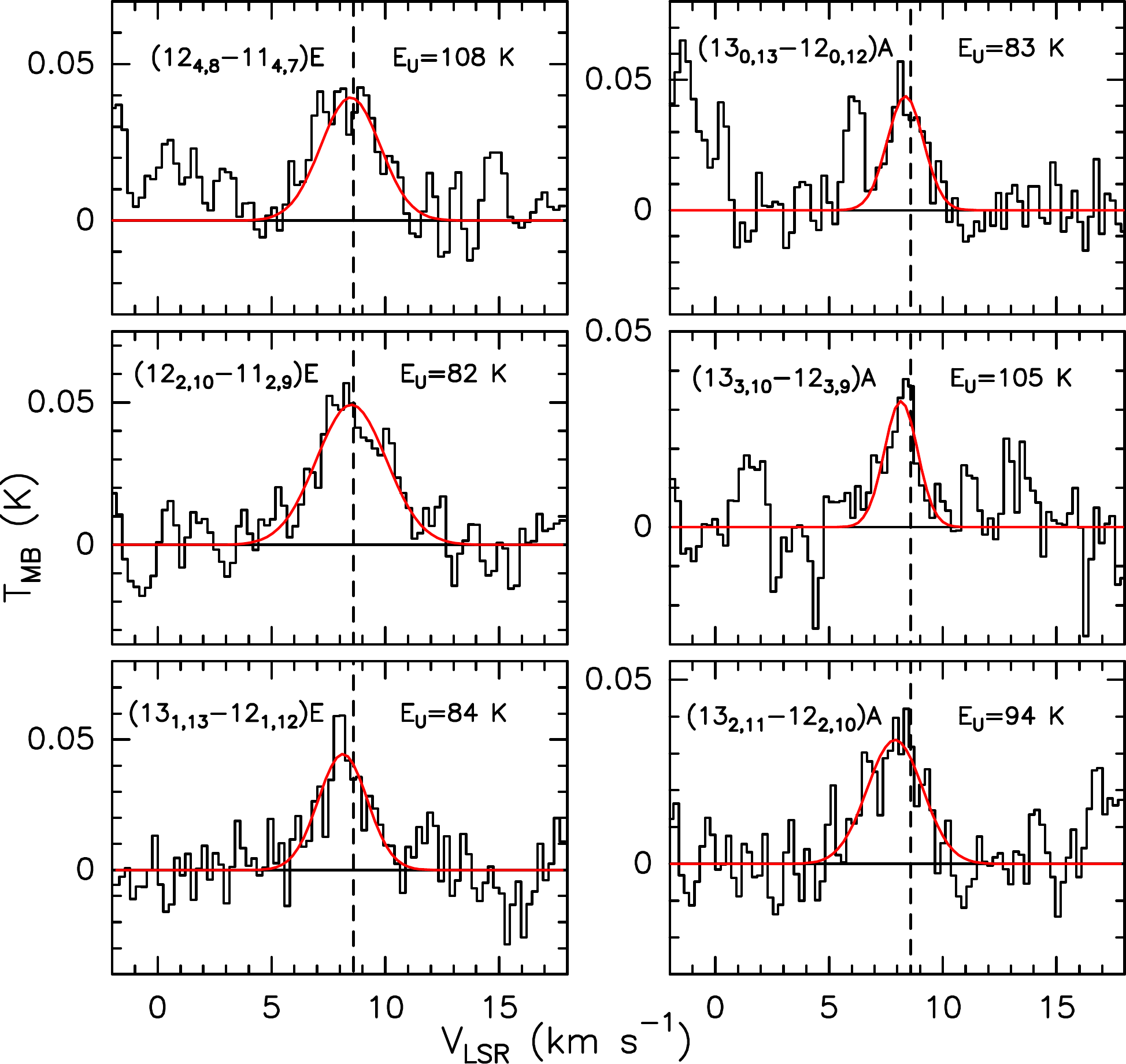}
\end{center}
\caption{CH$_{3}$CHO line profiles  in $T_{\rm MB}$ scale (not corrected for the beam dilution); transitions are reported. The vertical dashed line stands for the ambient LSR velocity (+ 8.6 Km s$^{-1}$, \citealt{Chen2009}). The 5$_{\rm 1,5}$--4$_{\rm 1,4}$ E transition at frequency 93.5953 GHz and the 12$_{\rm 4,8}$--11$_{\rm 4,7}$ E transition at 231.4844 GHz are contaminated by unidentified lines and thus excluded from the further analysis.} 
\label{Fig:CH3CHO-spectra}
\end{figure*}

We report hereafter the iCOMs emission line spectra in Fig. \ref{Fig:H2CCO-spectra},  \ref{Fig:CH3CHO-spectra}, \ref{Fig:CH3OCHO-spectra},
\ref{Fig:CH3OCH3-spectra}, \ref{Fig:CH3CH2OH-spectra} and the Tables \ref{Table:ketene},  \ref{Table:Ace},  \ref{Table:MF},  \ref{Table:DME},  \ref{Table:ethanol}, reporting all the spectral line parameters and all the results of the gaussian fit for each species.

\begin{figure*}
\begin{center}
\includegraphics[width=14cm]{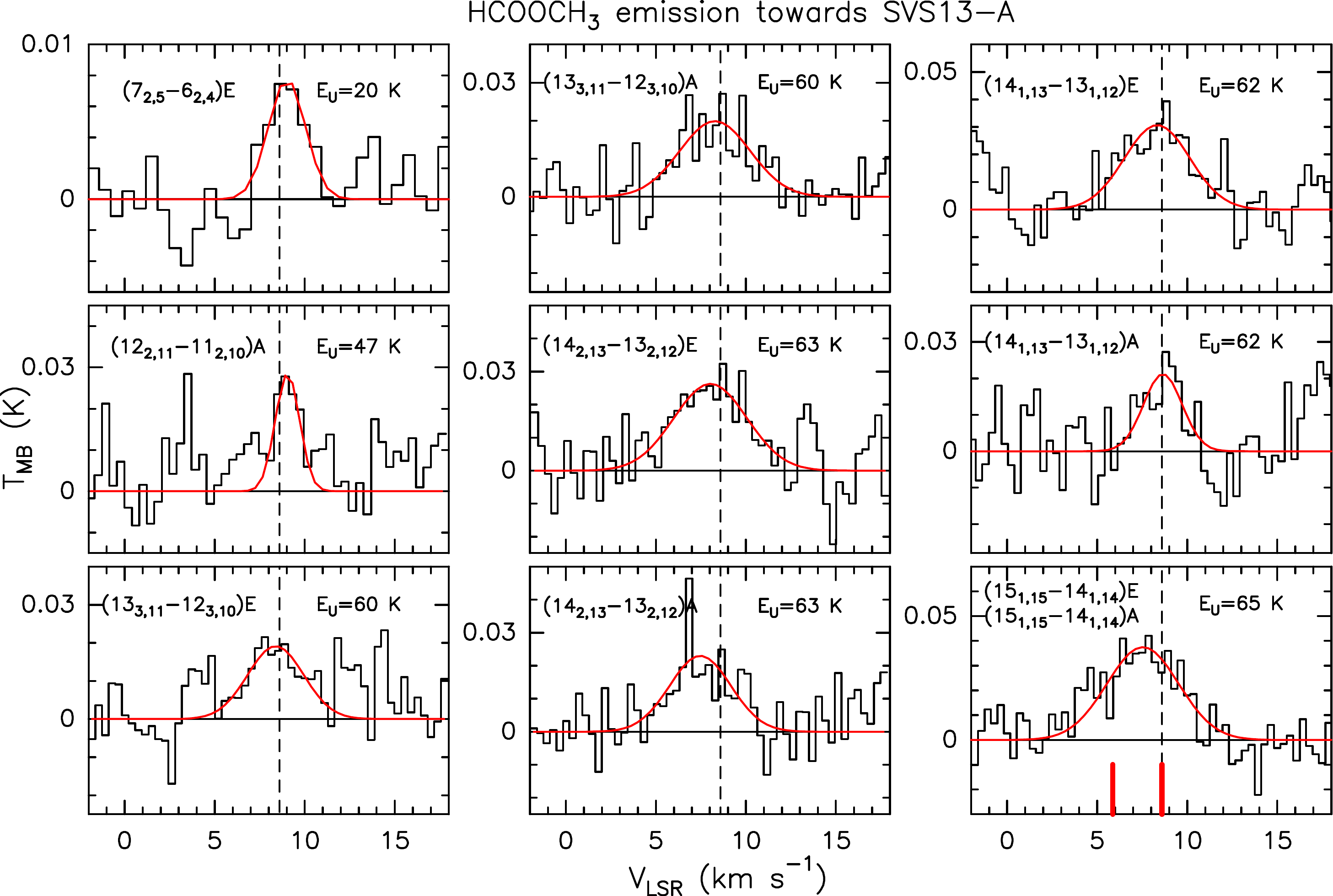}
\includegraphics[trim=0 0 0 -1cm, width=14cm]{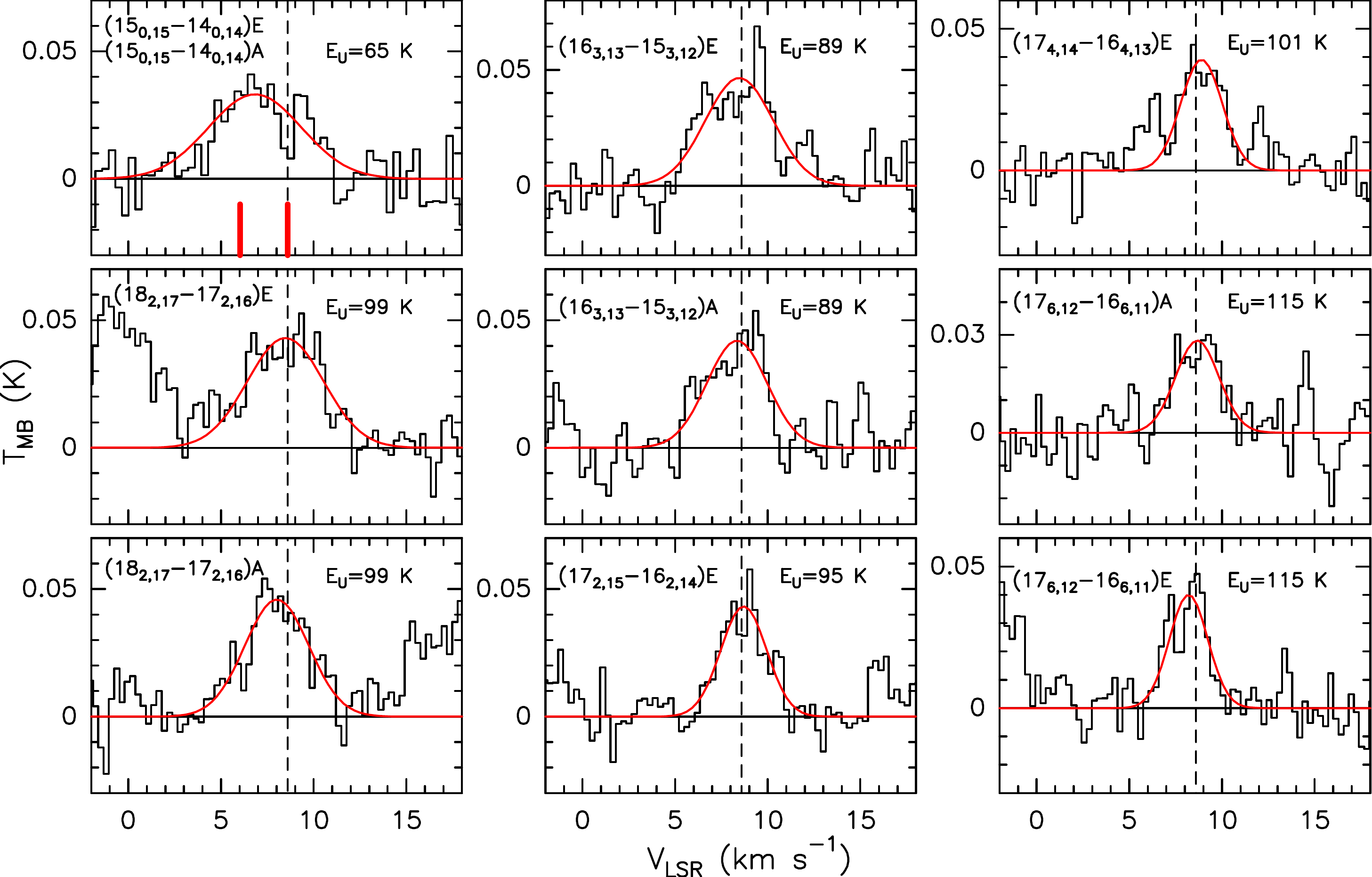}
\end{center}
\caption{HCOOCH$_{3}$ line profiles  in $T_{\rm MB}$ scale (not corrected for the beam dilution); transitions are reported. The vertical dashed line stands for the ambient LSR velocity (+ 8.6 Km s$^{-1}$, \citealt{Chen2009}). For blended lines, the vertical red solid lines indicate the different transitions. All the line profiles
due to several transitions with different upper level energies are excluded from the analysis (see text).}
\label{Fig:CH3OCHO-spectra}
\end{figure*}

\addtocounter{figure}{-1}
\begin{figure*}
\begin{center}
\includegraphics[width=14cm]{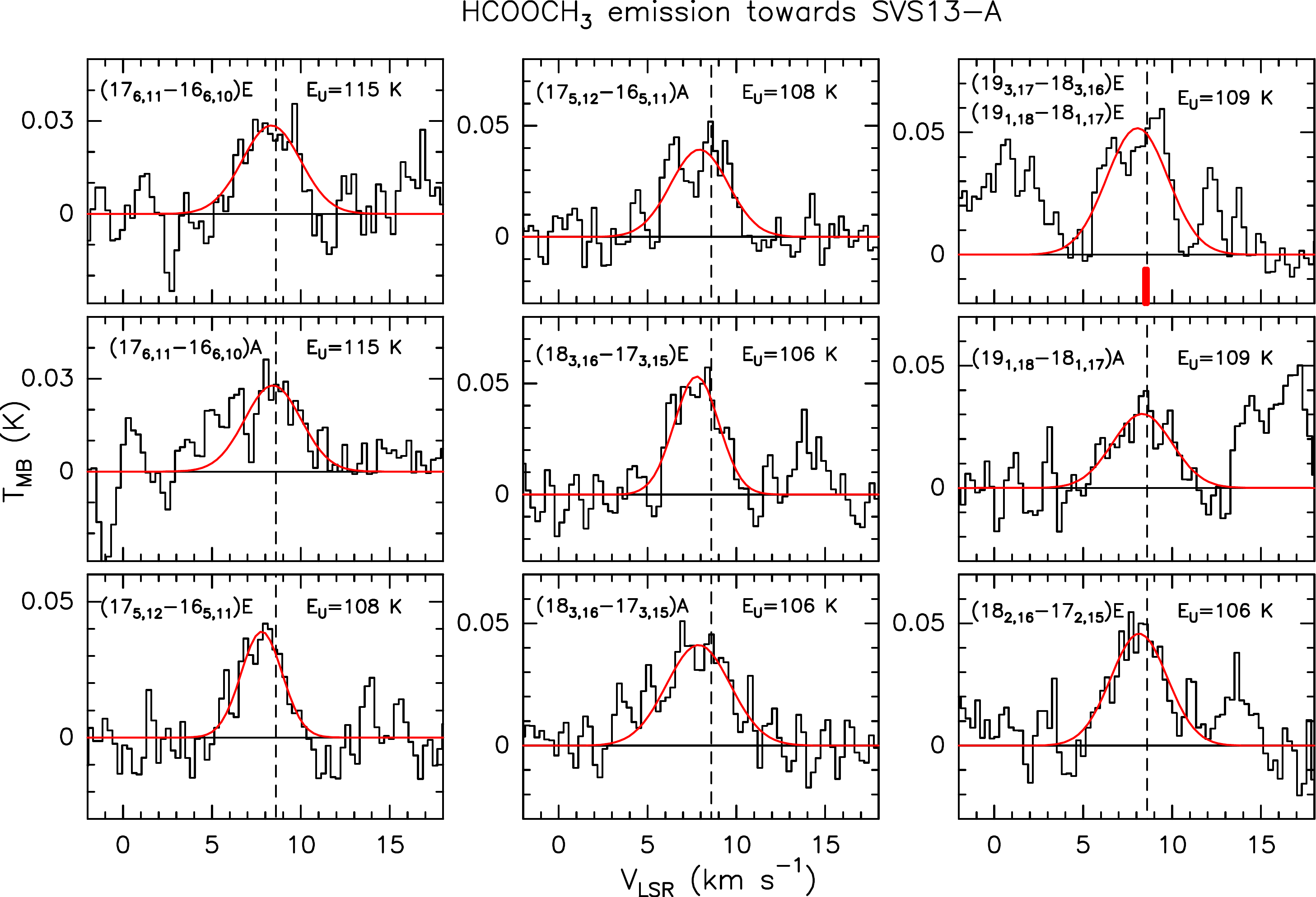}
\includegraphics[trim=0 0 0 -1cm, width=14cm]{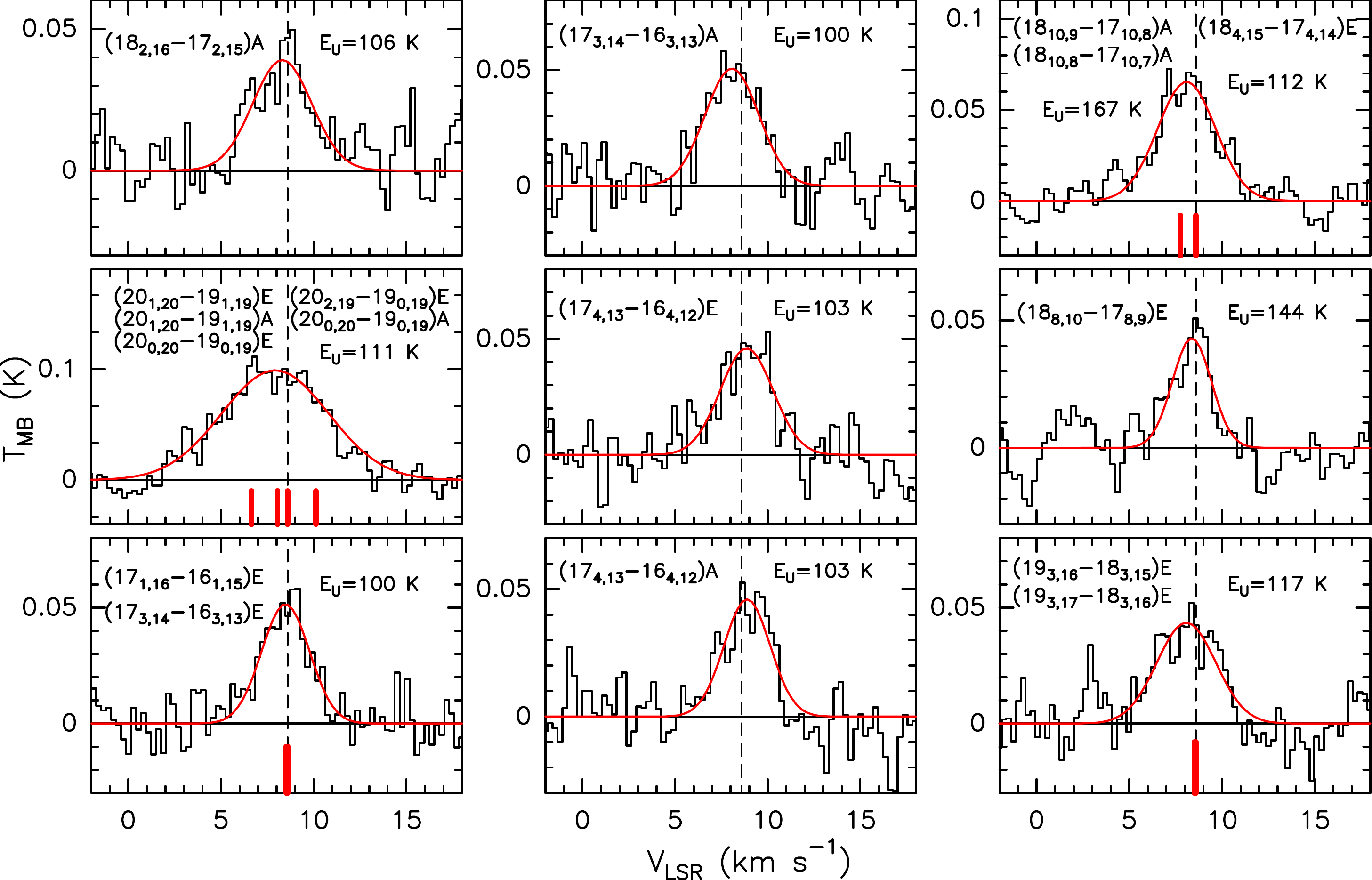}
\end{center}
\caption{{\it Continued.}}
\end{figure*}
\addtocounter{figure}{-1}
\begin{figure*}
\begin{center}
\includegraphics[width=14cm]{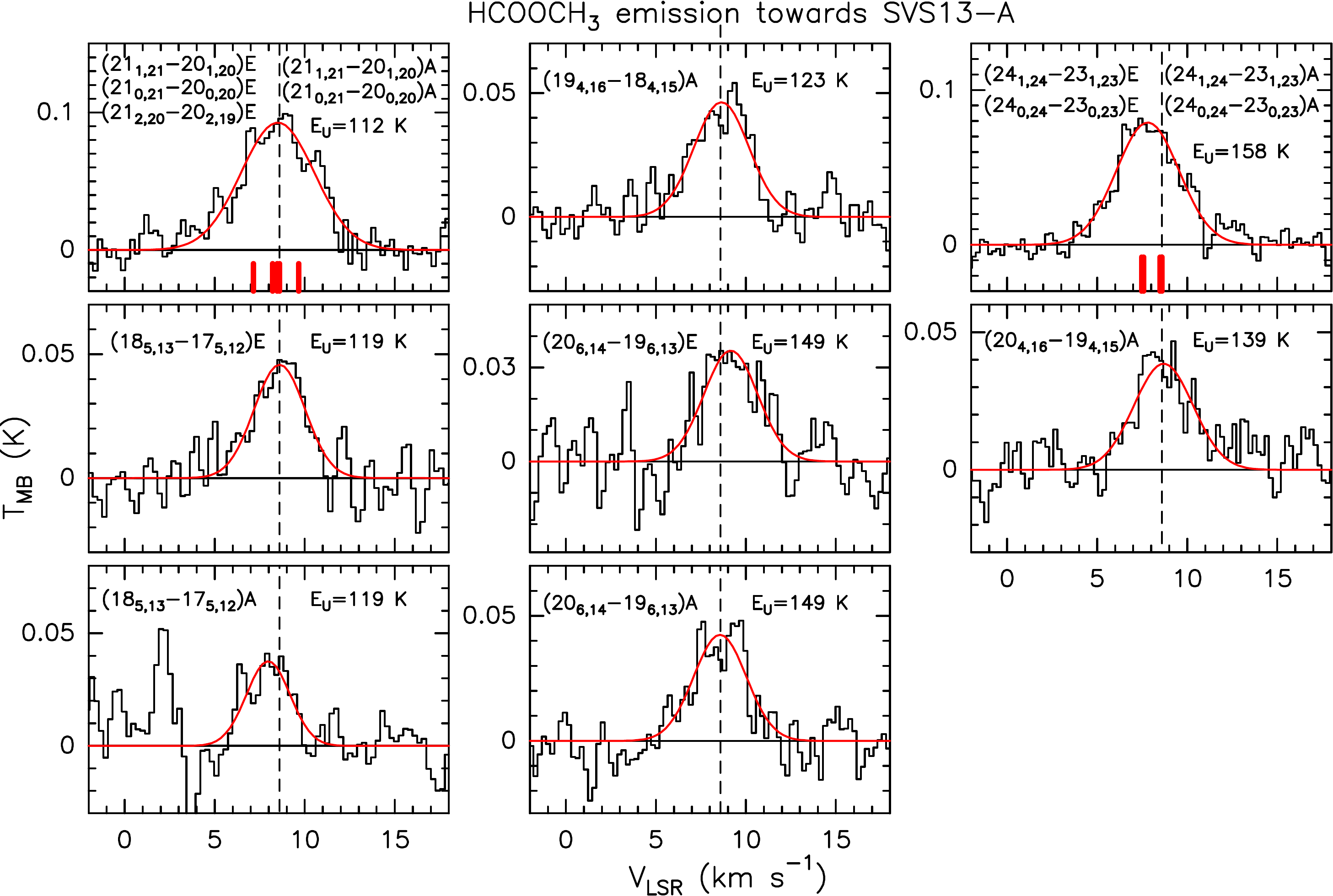}
\end{center}
\caption{{\it Continued.}}
\end{figure*}

\begin{figure*}
\begin{center}
\includegraphics[width=14cm]{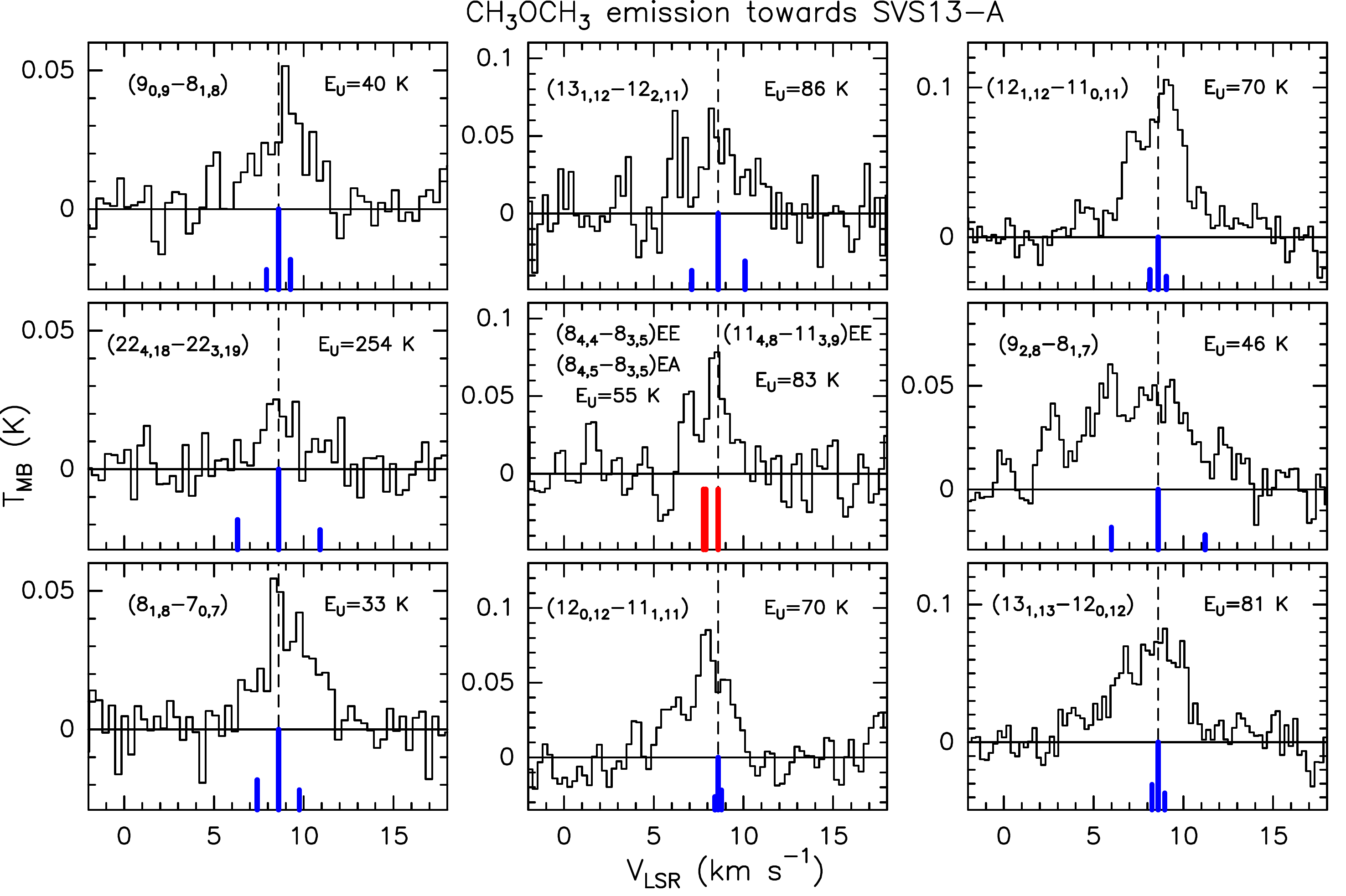}
\includegraphics[trim=0 0 0 -1cm, width=10cm]{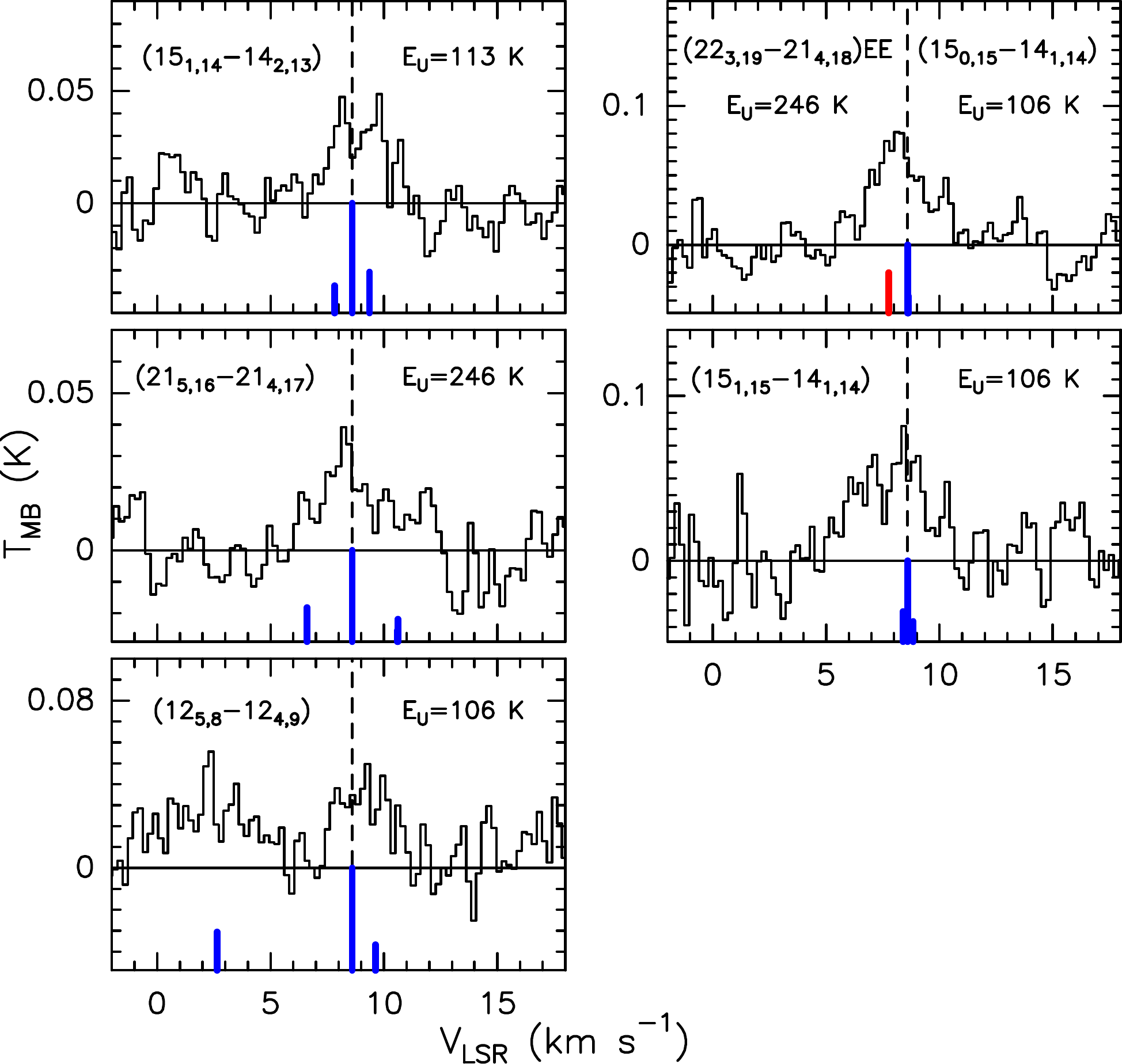}
\end{center}
\caption{CH$_{3}$OCH$_{3}$ line profiles  in $T_{\rm MB}$ scale (not corrected for the beam dilution); transitions are reported. The vertical dashed line stands for the ambient LSR velocity (+ 8.6 Km s$^{-1}$, \citealt{Chen2009}). The blue lines indicate transitions with the same upper level energies and quantum numbers. The different lenght 
is related to the different spin statistical weight of each transitions (see also Subsection \ref{Sub:lines_id}).
Red lines indicate transitions with different upper level energies and quantum numbers. Gaussian fit is not performed given the asymmetric line profiles.}
\label{Fig:CH3OCH3-spectra}
\end{figure*}

\begin{figure*}
\begin{center}
\includegraphics[width=10cm]{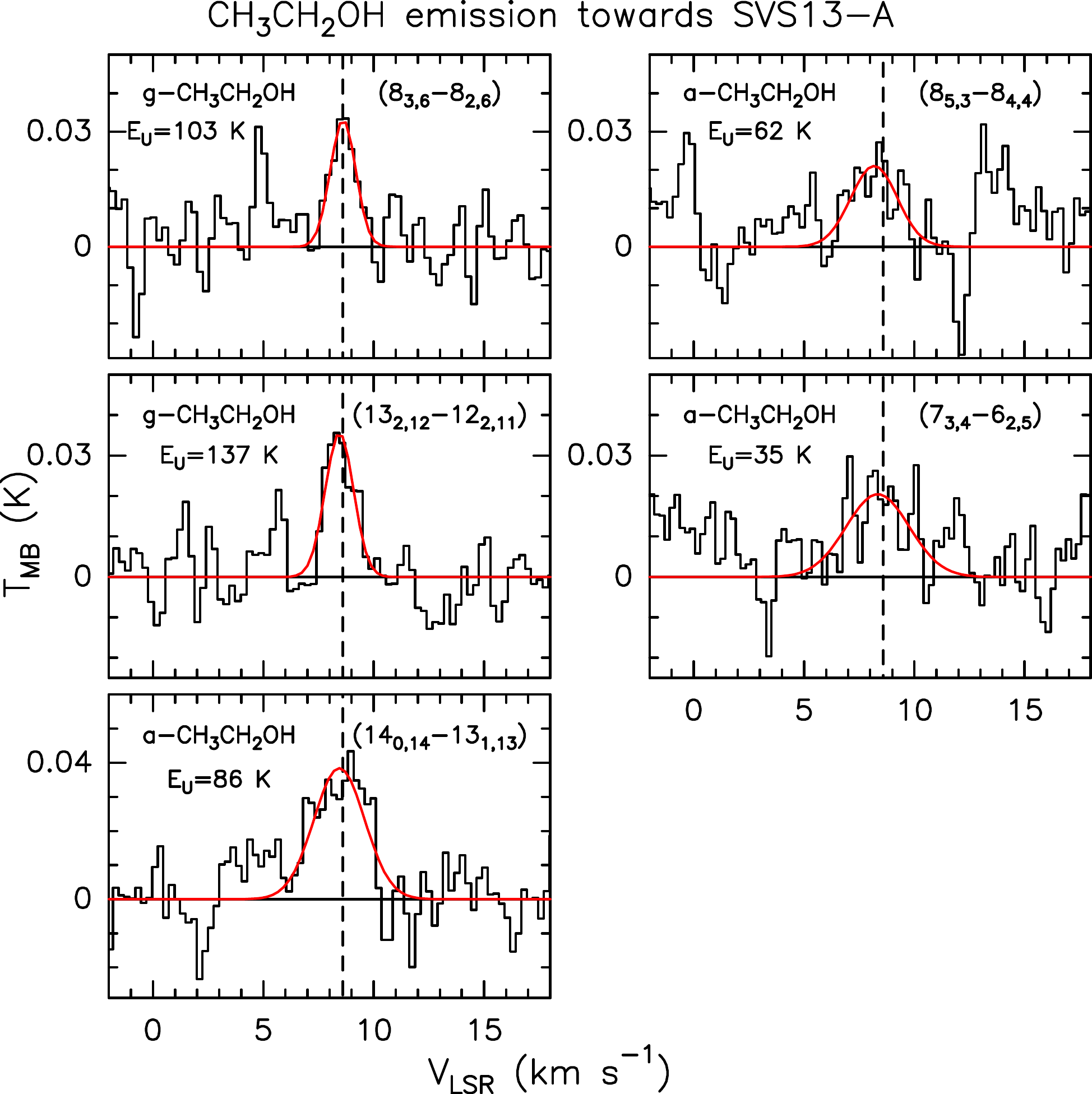}
\end{center}
\caption{CH$_{\rm 3}$CH$_{\rm 2}$OH line profiles  in $T_{\rm MB}$ scale (not corrected for the beam dilution); transitions are reported. The vertical dashed line stands for the ambient LSR velocity (+ 8.6 Km s$^{-1}$, \citealt{Chen2009}).}
\label{Fig:CH3CH2OH-spectra}
\end{figure*}

\begin{table*}
\resizebox{\textwidth}{!}{
\begin{tabular}{lccccccccc}
\hline
\multicolumn{1}{c}{Transition} &
\multicolumn{1}{c}{$\nu$$^{\rm a}$} &
\multicolumn{1}{c}{$HPBW$} &
\multicolumn{1}{c}{$E_{\rm u}$$^a$} &
\multicolumn{1}{c}{$S\mu^2$$^a$} &
\multicolumn{1}{c}{rms} &
\multicolumn{1}{c}{$T_{\rm peak}$$^b$} &
\multicolumn{1}{c}{$V_{\rm peak}$$^b$} &
\multicolumn{1}{c}{$FWHM$$^b$} &
\multicolumn{1}{c}{$I_{\rm int}$$^b$} \\
\multicolumn{1}{c}{ } &
\multicolumn{1}{c}{(GHz)} &
\multicolumn{1}{c}{($\arcsec$)} &
\multicolumn{1}{c}{(K)} &
\multicolumn{1}{c}{(D$^2$)} &
\multicolumn{1}{c}{(mK)} &
\multicolumn{1}{c}{(mK)} &
\multicolumn{1}{c}{(km s$^{-1}$)} &
\multicolumn{1}{c}{(km s$^{-1}$)} &
\multicolumn{1}{c}{(mK km s$^{-1}$)} \\
\hline

o-H$_{\rm 2}$CCO 7$_{\rm 1,6}$--6$_{\rm 1,5}$ & 142.7690 & 17 & 40 & 41   &  10 & 39 (04)   & +8.6 (0.2) & 2.0 (0.5) & 80 (20)\\
\vspace{0.25cm} 
o-H$_{\rm 2}$CCO 8$_{\rm 1,8}$--7$_{\rm 1,7}$ & 160.1422 & 15 & 48 & 48   &  9 & 40 (10)   & +8.4 (0.2) & 3.7 (0.5) & 150 (20)\\
o-H$_{\rm 2}$CCO 8$_{\rm 3,5}$--7$_{\rm 3,4}$ $^c$ & 161.6022 &\multirow{2}{*}{15}  &\multirow{2}{*}{152} &\multirow{2}{*}{42}&\multirow{2}{*}{10}&\multirow{2}{*}{36 (7)}&\multirow{2}{*}{+8.2 (0.2)}&\multirow{2}{*}{2.6 (0.4)}&\multirow{2}{*}{100 (20)}\\
\vspace{0.25cm} 
o-H$_{\rm 2}$CCO 8$_{\rm 3,6}$--7$_{\rm 3,5}$ $^c$  & 161.6022\\

o-H$_{\rm 2}$CCO 8$_{\rm 1,7}$--7$_{\rm 1,6}$ & 163.1609 & 15 & 48 & 48   &  8 & 50 (10)   & +8.5 (0.1) & 2.1 (0.3) & 100 (10)\\

p-H$_{\rm 2}$CCO 10$_{\rm 0,10}$--9$_{\rm 0,9}$ & 202.0143 & 12 & 53 & 20 & 18 & 50 (10) & +8.2 (0.2) & 1.4 (0.4) & 70 (20)\\


o-H$_{\rm 2}$CCO 11$_{\rm 1,11}$--10$_{\rm 1,10}$ & 220.1776 & 11 & 76 & 66   &  21 & 70 (10)   & +8.8 (0.2) & 3.3 (0.6) & 240 (30)\\


p-H$_{\rm 2}$CCO 11$_{\rm 2,10}$--10$_{\rm 2,9}$ & 222.2286 &  11 & 116 & 21 & 9 &  31 (04) &  +8.9 (0.1) & 1.3 (0.3) & 41 (08)\\

p-H$_{\rm 2}$CCO 11$_{\rm 2,9}$--10$_{\rm 2,8}$ & 222.3144 & 11 & 116 & 21 & 5 & 27 (05) & +8.3 (0.1) & 2.0 (0.3) & 56 (06)\\

o-H$_{\rm 2}$CCO 11$_{\rm 1,10}$--10$_{\rm 1,9}$ & 224.3273 & 11 & 78 & 66   &  14 & 70 (10)   & +7.9 (0.1) & 3.7 (0.3) & 280 (20)\\

o-H$_{\rm 2}$CCO 12$_{\rm 1,12}$--11$_{\rm 1,11}$ & 240.1858 & 10 & 88 & 72   &  11 & 99 (09)   & +7.6 (0.1) & 4.1 (0.2) & 430 (20)\\

p-H$_{\rm 2}$CCO 12$_{\rm 2,10}$--11$_{\rm 2,9}$ & 242.5362 & 10 &  128 & 24 & 7 & 25 (07) & +7.8 (0.1) & 1.5 (0.5) & 40 (08)\\

\vspace{0.25cm} 

o-H$_{\rm 2}$CCO 12$_{\rm 1,11}$--11$_{\rm 1,10}$ & 244.7123 & 10 & 89 & 72   &  11 & 66 (06)   & +8.4 (0.1) & 2.2 (0.2) & 150 (10)\\

o-H$_{\rm 2}$CCO 13$_{\rm 3,10}$--12$_{\rm 3,9}$ $^c$ & 262.5974 &\multirow{2}{*}{9}  &\multirow{2}{*}{206} &\multirow{2}{*}{74}&\multirow{2}{*}{10}&\multirow{2}{*}{50 (10)}&\multirow{2}{*}{+8.1 (0.2)}&\multirow{2}{*}{3.7 (0.4)}&\multirow{2}{*}{210 (20)}\\
\vspace{0.25cm} 
o-H$_{\rm 2}$CCO 13$_{\rm 3,11}$--12$_{\rm 3,10}$ $^c$  & 262.5966\\

p-H$_{\rm 2}$CCO 13$_{\rm 2,12}$--12$_{\rm 2,11}$ & 262.6190 & 9 & 140 & 26 & 9 & 51 (09) & +8.8 (0.1) & 0.9 (0.1) & 51 (06)\\

p-H$_{\rm 2}$CCO 13$_{\rm 2,11}$--12$_{\rm 2,10}$ & 262.7609 & 9 & 140 & 26 & 8 & 37 (08) & +8.1 (0.1) & 2.4 (0.3) & 95 (09)\\


\hline

\noalign{\vskip 2mm}

\end{tabular}}
\caption{List of transitions and line properties (in $T_{\rm MB}$ scale) of the H$_{\rm 2}$CCO emission detected towards SVS13-A. We report the frequency of each transition (GHz),
the telescope HPBW ($\arcsec$),
the excitation energies of the upper level $E_{\rm up}$ (K),
the S$\mu^{2}$ product (D$^{2}$), the line rms (mK),
the peak temperature (mK), the peak velocities (km s$^{-1}$), the line full width at half maximum (FWHM) (km s$^{-1}$)
and the velocity integrated line intensity $I_{\rm int}$ (mK km s$^{-1}$).}
\small {$^a$ Frequencies and spectroscopic parameters are extracted from the Cologne Database
 for Molecular Spectroscopy (CDMS\footnote{http://www.astro.uni-koeln.de/cdms/};\citealt{Muller2001},\citealt{Muller2005}).
$^b$ The errors in brackets are the gaussian fit uncertainties.
$^c$ The lines cannot be distinguished with the present spectral resolution.} \\
\label{Table:ketene}
\end{table*}

\begin{table*}
\resizebox{\textwidth}{!}{
\begin{tabular}{lccccccccc}

\hline
\multicolumn{1}{c}{Transition} &
\multicolumn{1}{c}{$\nu$$^{\rm a}$} &
\multicolumn{1}{c}{$HPBW$} &
\multicolumn{1}{c}{$E_{\rm u}$$^a$} &
\multicolumn{1}{c}{$S\mu^2$$^a$} &
\multicolumn{1}{c}{rms} &
\multicolumn{1}{c}{$T_{\rm peak}$$^b$} &
\multicolumn{1}{c}{$V_{\rm peak}$$^b$} &
\multicolumn{1}{c}{$FWHM$$^b$} &
\multicolumn{1}{c}{$I_{\rm int}$$^b$} \\
\multicolumn{1}{c}{ } &
\multicolumn{1}{c}{(GHz)} &
\multicolumn{1}{c}{($\arcsec$)} &
\multicolumn{1}{c}{(K)} &
\multicolumn{1}{c}{(D$^2$)} &
\multicolumn{1}{c}{(mK)} &
\multicolumn{1}{c}{(mK)} &
\multicolumn{1}{c}{(km s$^{-1}$)} &
\multicolumn{1}{c}{(km s$^{-1}$)} &
\multicolumn{1}{c}{(mK km s$^{-1}$)} \\

\hline

CH$_{\rm 3}$CHO 5$_{\rm 1,5}$--4$_{\rm 1,4}$ E $^c$ & 93.5953 & 26 & 16 & 61 & 1 & 13 (01) & +8.7 (0.4) &
2.4 (0.1) & 32 (01)\\



CH$_{\rm 3}$CHO 11$_{\rm 1,11}$--10$_{\rm 1,10}$ A & 205.1707 & 12 & 61 & 138 & 7 & 43 (07) & +8.6 (0.1) & 1.6 (0.2) & 74 (08)\\

CH$_{\rm 3}$CHO 11$_{\rm 2,10}$--10$_{\rm 2,9}$ A & 211.2430 & 12 & 70 & 134 & 5 & 42 (05) & +8.4 (0.1) & 2.3 (0.3) & 100 (10)\\

CH$_{\rm 3}$CHO 11$_{\rm 2,10}$--10$_{\rm 2,9}$ E & 211.2738 & 12 &  70 & 134 & 6 & 38 (06) & +9.0 (1.0) & 2.2 (0.4) & 90 (10)\\



CH$_{\rm 3}$CHO 11$_{\rm 4,8}$--10$_{\rm 4,7}$ A & 212.1284 &  12 &  97 & 121 & 5 & 25 (05) & +8.9 (0.1) & 2.2 (0.3) & 57 (06)\\

CH$_{\rm 3}$CHO 11$_{\rm 1,10}$--10$_{\rm 1,9}$ E & 216.5819 & 11 & 65 & 138 & 11 & 53 (08) & +8.0 (0.1) & 2.5 (0.2) & 140 (10)\\
CH$_{\rm 3}$CHO 11$_{\rm 1,10}$--10$_{\rm 1,9}$ A & 216.6302 & 11 & 65 & 138 & 8 &  50 (10) & +7.8 (0.2) & 3.4 (0.4) & 170 (20)\\

CH$_{\rm 3}$CHO 12$_{\rm 0,12}$--11$_{\rm 0,11}$ A & 226.5927 & 11 & 71 & 151 & 8 & 41 (08) & +8.4 (0.2) & 4.1 (0.5) & 180 (20)\\

CH$_{\rm 3}$CHO 12$_{\rm 2,11}$--11$_{\rm 2,10}$ A & 230.3019 & 11 & 81 & 147 & 11 & 33 (06) & +8.6 (0.2) & 1.5 (0.3) & 50 (10)\\

CH$_{\rm 3}$CHO 12$_{\rm 4,8}$--11$_{\rm 4,7}$ E   $^c$ &       231.4844 & 11 & 108 &  135 & 6 & 39 (06) & + 8.5 (0.2) & 3.1 (0.4) & 130 (20)\\

CH$_{\rm 3}$CHO 12$_{\rm 3,9}$--11$_{\rm 3,8}$ A $^d$ & 231.9684 & 11 & 93 & 142 & 7 & 98 (07) & +8.1 (0.1) & 2.8 (0.3) & 300 (30)\\

CH$_{\rm 3}$CHO 12$_{\rm 2,10}$--11$_{\rm 2,9}$ E & 234.7955 & 10 & 82 & 147 & 7 & 49 (07) & +8.5 (0.1) & 3.6 (0.2) & 190 (10)\\

CH$_{\rm 3}$CHO 13$_{\rm 1,13}$--12$_{\rm 1,12}$ E & 242.1060 & 10 & 84 & 163 & 9 & 44 (09) & +8.1 (0.1) & 2.5 (0.3) & 120 (10)\\

CH$_{\rm 3}$CHO 13$_{\rm 0,13}$--12$_{\rm 0,12}$ A & 244.8322 & 10 & 83 & 164 & 7 & 44 (07) & +8.4 (0.1) & 1.9 (0.2) & 88 (07)\\

CH$_{\rm 3}$CHO 13$_{\rm 3,10}$--12$_{\rm 3,9}$ A & 251.4893 & 10 & 105 & 156 & 5 & 32 (05) & +8.2 (0.1) & 1.7 (0.2) & 59 (06)\\

CH$_{\rm 3}$CHO 13$_{\rm 2,11}$--12$_{\rm 2,10}$ A & 254.8505 & 10 & 94 & 161 & 7 & 34 (07) & +7.9 (0.1) & 2.9 (0.3) & 100 (10)\\
\hline
\noalign{\vskip 2mm}
\end{tabular}}
\caption{List of transitions and line properties (in $T_{\rm MB}$ scale) of the CH$_{\rm 3}$CHO emission detected towards SVS13-A. We report the frequency of each transition (GHz),
the telescope HPBW ($\arcsec$),
the excitation energies of the upper level $E_{\rm up}$ (K),
the S$\mu^{2}$ product (D$^{2}$), the line rms (mK),
the peak temperature (mK), the peak velocities (km s$^{-1}$), the line full width at half maximum (FWHM) (km s$^{-1}$)
and the velocity integrated line intensity $I_{\rm int}$ (mK km s$^{-1}$).}
\small {$^a$ Frequencies and spectroscopic parameters are extracted from the Jet Propulsion Laboratory database \citep{Pickett1998}.
$^b$ The errors in brackets are the gaussian fit uncertainties.
$^c$ Irregular profile probably due to contamination by unidentified lines: excluded from the further analysis.
$^d$The transition CH$_{\rm 3}$CHO 12$_{\rm 3,9}$--11$_{\rm 3,8}$ A at frequency 231.9684 GHz is observed but excluded from the analysis because contaminated by the CH$_{\rm 2}$DOH 9$_{\rm 2,7}$--9$_{\rm 1,8}$ e0 transition (see \citealt{Bianchi2017a})
} \\
\label{Table:Ace}
\end{table*}

\begin{table*}
\resizebox{\textwidth}{!}{
\begin{tabular}{lccccccccc}
\hline
\multicolumn{1}{c}{Transition} &
\multicolumn{1}{c}{$\nu$$^{\rm a}$} &
\multicolumn{1}{c}{$HPBW$} &
\multicolumn{1}{c}{$E_{\rm u}$$^a$} &
\multicolumn{1}{c}{$S\mu^2$$^a$} &
\multicolumn{1}{c}{rms} &
\multicolumn{1}{c}{$T_{\rm peak}$$^b$} &
\multicolumn{1}{c}{$V_{\rm peak}$$^b$} &
\multicolumn{1}{c}{$FWHM$$^b$} &
\multicolumn{1}{c}{$I_{\rm int}$$^b$} \\
\multicolumn{1}{c}{ } &
\multicolumn{1}{c}{(GHz)} &
\multicolumn{1}{c}{($\arcsec$)} &
\multicolumn{1}{c}{(K)} &
\multicolumn{1}{c}{(D$^2$)} &
\multicolumn{1}{c}{(mK)} &
\multicolumn{1}{c}{(mK)} &
\multicolumn{1}{c}{(km s$^{-1}$)} &
\multicolumn{1}{c}{(km s$^{-1}$)} &
\multicolumn{1}{c}{(mK km s$^{-1}$)} \\

\hline

HCOOCH$_{\rm 3}$ 7$_{\rm 2,5}$--6$_{\rm 2,4}$ E & 90.1457 & 27 & 20 & 17 & 3 & 8 (01) & +9.0 (0.4) & 2.4 (0.7) & 20 (06)\\


HCOOCH$_{\rm 3}$ 12$_{\rm 2,11}$--11$_{\rm 2,10}$ A & 141.0443 & 17 & 47 &  31 &  2 & 29 (02) & +9.0 (0.1) & 1.6 (0.1) & 48 (03)\\

HCOOCH$_{\rm 3}$ 13$_{\rm 3,11}$--12$_{\rm 3,10}$ E & 158.6937 & 16 & 60 &  33 & 3 & 19 (03) & +8.4 (0.1) & 3.7 (0.3) & 74 (05)\\

HCOOCH$_{\rm 3}$ 13$_{\rm 3,11}$--12$_{\rm 3,10}$ A & 158.7044 & 16 & 60 &  33 & 6 & 20 (06) & +8.3 (0.3) & 4.6 (0.7) & 100 (10)\\

HCOOCH$_{\rm 3}$ 14$_{\rm 2,13}$--13$_{\rm 2,12}$ E & 162.7689 & 15 & 63 & 36 & 6 & 26 (06) & +8.0 (0.2) & 4.8 (0.5) & 130 (10)\\

HCOOCH$_{\rm 3}$ 14$_{\rm 2,13}$--13$_{\rm 2,12}$ A & 162.7753 & 15 & 63 & 36 & 8 & 23 (08) & +7.5 (0.3) & 4.0 (0.6) & 100 (10)\\

HCOOCH$_{\rm 3}$ 14$_{\rm 1,13}$--13$_{\rm 1,12}$ E & 163.8297 & 15 & 62 & 36 & 5 & 31 (05) & +8.3 (0.2) & 4.3 (0.4) & 140 (10)\\

\vspace{0.25cm} 
HCOOCH$_{\rm 3}$ 14$_{\rm 1,13}$--13$_{\rm 1,12}$ A & 163.8355 & 15 & 62 & 36 & 6 & 21 (06) & +8.7 (0.2) & 2.5 (0.4) & 57 (08)\\

HCOOCH$_{\rm 3}$ 15$_{\rm 1,15}$--14$_{\rm 1,14}$ E  $^c$ & 163.9604 & \multirow{2}{*}{15} & \multirow{2}{*}{65} & \multirow{2}{*}{39} & \multirow{2}{*}{6} & \multirow{2}{*}{38 (06)} & \multirow{2}{*}{+7.5 (0.3)} & \multirow{2}{*}{4.7 (0.6)} & \multirow{2}{*}{190 (20)}\\

\vspace{0.25cm} 
HCOOCH$_{\rm 3}$ 15$_{\rm 1,15}$--14$_{\rm 1,14}$ A  $^c$ & 163.9619\\

HCOOCH$_{\rm 3}$ 15$_{\rm 0,15}$--14$_{\rm 0,14}$ E  $^c$ & 163.9875 & \multirow{2}{*}{15} & \multirow{2}{*}{65} & \multirow{2}{*}{39} & \multirow{2}{*}{8} & \multirow{2}{*}{33 (08)} & \multirow{2}{*}{+6.8 (0.4)} & \multirow{2}{*}{6.0 (1.0)} & \multirow{2}{*}{210 (30)}\\

HCOOCH$_{\rm 3}$ 15$_{\rm 0,15}$--14$_{\rm 0,14}$ A  $^c$ & 163.9889\\

HCOOCH$_{\rm 3}$ 18$_{\rm 2,17}$--17$_{\rm 2,16}$ E  & 205.4957 & 12 & 99 &  47 & 7 & 43 (07) & +8.5 (0.2) & 4.8 (0.4) & 220 (20)\\
HCOOCH$_{\rm 3}$ 18$_{\rm 2,17}$--17$_{\rm 2,16}$ A  & 205.5016 & 12 & 99 &  47 & 7 & 46 (07) & +8.0 (0.1) & 4.1 (0.3) & 200 (10)\\

HCOOCH$_{\rm 3}$ 16$_{\rm 3,13}$--15$_{\rm 3,12}$ E  & 206.6012 & 12 & 89 & 41 & 11 & 50 (10) & +8.5 (0.2) & 4.3 (0.4) & 210 (20)\\

HCOOCH$_{\rm 3}$ 16$_{\rm 3,13}$--15$_{\rm 3,12}$ A  & 206.6195 & 12 & 89 & 41 & 10 & 40 (10) & +8.3 (0.2) & 3.8 (0.4) & 170 (20)\\

HCOOCH$_{\rm 3}$ 17$_{\rm 2,15}$--16$_{\rm 2,14}$ E  & 206.7109 & 12 & 95 & 43 & 4 & 43 (07) & +8.7 (0.1) & 2.9 (0.1) & 130 (60) \\

HCOOCH$_{\rm 3}$ 17$_{\rm 4,14}$--16$_{\rm 4,13}$ E  & 209.9185 & 12 & 101 & 43 & 5 & 39 (05) & +8.9 (0.1) & 2.6 (0.2) & 110 (06)\\

HCOOCH$_{\rm 3}$ 17$_{\rm 6,12}$--16$_{\rm 6,11}$ A  & 211.2550 & 12 & 115 & 40 & 6 & 28 (06) & +8.7 (0.1) & 2.8 (0.3) & 83 (07)\\

HCOOCH$_{\rm 3}$ 17$_{\rm 6,12}$--16$_{\rm 6,11}$ E  & 211.2661 & 12 & 115 & 39 & 8 & 40 (08) & +8.2 (0.1) & 2.5 (0.3) & 100 (10)\\

HCOOCH$_{\rm 3}$ 17$_{\rm 6,11}$--16$_{\rm 6,10}$ E  & 211.5372 & 12 & 115 & 39 & 6 & 29 (06) & +8.4 (0.2) & 3.9 (0.5) & 120 (10)\\

HCOOCH$_{\rm 3}$ 17$_{\rm 6,11}$--16$_{\rm 6,10}$ A  & 211.5751 & 12 & 115 & 39 & 5 & 28 (05) & +8.4 (0.3) & 3.7 (0.6) & 110 (20)\\

HCOOCH$_{\rm 3}$ 17$_{\rm 5,12}$--16$_{\rm 5,11}$ E  & 214.6317 & 11 & 108 & 41 & 9 & 39 (06) & +7.8 (0.1) & 2.7 (0.3) & 110 (10)\\

HCOOCH$_{\rm 3}$ 17$_{\rm 5,12}$--16$_{\rm 5,11}$ A  & 214.6526 &  11 & 108 & 41 & 12 & 40 (10) & +7.9 (0.2) & 3.8 (0.5) & 160 (20)\\
HCOOCH$_{\rm 3}$ 18$_{\rm 3,16}$--17$_{\rm 3,15}$ E  & 214.7824 & 11 & 106 & 46 & 9 & 53 (09) & +7.8 (0.1) & 3.0 (0.2) & 170 (10)\\

\vspace{0.25cm} 
HCOOCH$_{\rm 3}$ 18$_{\rm 3,16}$--17$_{\rm 3,15}$ A  & 214.7925 & 11 & 106 & 46 & 8 & 41 (08) & +7.9 (0.2) & 4.2 (0.4) & 190 (20)\\

HCOOCH$_{\rm 3}$ 19$_{\rm 3,17}$--18$_{\rm 3,16}$ E  $^{c,}$$^d$ & 216.2109 &  \multirow{2}{*}{11} & \multirow{2}{*}{109} & \multirow{2}{*}{49} & \multirow{2}{*}{11} & \multirow{2}{*}{ 50 (10)} & \multirow{2}{*}{+8.0 (0.2)} & \multirow{2}{*}{4.1 (0.4)} & \multirow{2}{*}{220 (20)}\\
\vspace{0.25cm} 
HCOOCH$_{\rm 3}$ 19$_{\rm 1,18}$--18$_{\rm 1,17}$ E $^{c,}$$^d$ & 216.2110 \\

HCOOCH$_{\rm 3}$ 19$_{\rm 1,18}$--18$_{\rm 1,17}$ A & 216.2165 & 11 & 109 &  49 &  6 & 30 (06) & +8.3 (0.2) & 3.9 (0.4) & 130 (10)\\

HCOOCH$_{\rm 3}$ 18$_{\rm 2,16}$--17$_{\rm 2,15}$ E & 216.8302 & 11 & 106 & 46 & 8 & 46 (08) & +8.2 (0.1) & 3.8 (0.4) & 180 (10)\\
\vspace{0.25cm} 
HCOOCH$_{\rm 3}$ 18$_{\rm 2,16}$--17$_{\rm 2,15}$ A & 216.8389 & 11 & 106 & 46 & 8 & 39 (08) & +8.3 (0.1) & 3.8 (0.4) & 160 (10)\\

HCOOCH$_{\rm 3}$ 20$_{\rm 1,20}$--19$_{\rm 1,19}$ E $^{c,}$$^d$ & 216.9648 &  \multirow{5}{*}{11} & \multirow{5}{*}{111} & \multirow{5}{*}{53} & \multirow{5}{*}{9} & \multirow{5}{*}{99 (09)} & \multirow{5}{*}{+7.9 (0.1)} & \multirow{5}{*}{6.8 (0.5)} & \multirow{5}{*}{720 (40)}\\

HCOOCH$_{\rm 3}$ 20$_{\rm 1,20}$--19$_{\rm 1,19}$ A  $^{c,}$$^d$ & 216.9659 \\

HCOOCH$_{\rm 3}$ 20$_{\rm 0,20}$--19$_{\rm 0,19}$ E $^{c,}$$^d$ & 216.9663 \\

HCOOCH$_{\rm 3}$ 20$_{\rm 2,19}$--19$_{\rm 0,19}$ E $^{c,}$$^d$ & 216.9663 \\
\vspace{0.25cm} 
HCOOCH$_{\rm 3}$ 20$_{\rm 0,20}$--19$_{\rm 0,19}$ A $^{c,}$$^d$ & 216.9673\\

HCOOCH$_{\rm 3}$ 17$_{\rm 1,16}$--16$_{\rm 1,15}$ E  $^{c,}$$^d$ & 218.2808 &  \multirow{2}{*}{11} & \multirow{2}{*}{100} & \multirow{2}{*}{44} & \multirow{2}{*}{7} & \multirow{2}{*}{51 (07)} & \multirow{2}{*}{+8.5 (0.1)} & \multirow{2}{*}{3.0 (0.2)} & \multirow{2}{*}{167 (09)}\\
\vspace{0.25cm} 
HCOOCH$_{\rm 3}$ 17$_{\rm 3,14}$--16$_{\rm 3,13}$ E  $^{c,}$$^d$ & 218.2809 \\

HCOOCH$_{\rm 3}$ 17$_{\rm 3,14}$--16$_{\rm 3,13}$ A & 218.2979 & 11 & 100 & 44 & 7 & 51 (07) & +8.1 (0.1) & 3.4 (0.2) & 180 (10)\\

HCOOCH$_{\rm 3}$ 17$_{\rm 4,13}$--16$_{\rm 4,12}$ E & 220.1669 & 11 & 103 & 43 & 10 & 50 (10) & +8.9 (0.1) & 3.4 (0.4) & 160 (20)\\
\vspace{0.25cm} 
HCOOCH$_{\rm 3}$ 17$_{\rm 4,13}$--16$_{\rm 4,12}$ A & 220.1903 & 11 & 103 & 43 & 8 & 46 (08) & +8.9 (0.1) & 2.9 (0.2) & 140 (10)\\

HCOOCH$_{\rm 3}$ 18$_{\rm 4,15}$--17$_{\rm 4,14}$ E  $^{c,}$$^d$ & 221.6605  & 11 & 112 & 45 & \multirow{3}{*}{8} & \multirow{3}{*}{65 (08)} & \multirow{3}{*}{+8.1 (0.1)} & \multirow{3}{*}{3.7 (0.2)} & \multirow{3}{*}{260 (10)}\\

HCOOCH$_{\rm 3}$ 18$_{\rm 10,9}$--17$_{\rm 10,8}$ A  $^{c,}$$^d$ & 221.6611 & 11 & 167 & 33\\
\vspace{0.25cm} 
HCOOCH$_{\rm 3}$ 18$_{\rm 10,8}$--17$_{\rm 10,7}$ A  $^{c,}$$^d$ & 221.6611& 11 & 167 & 33\\
\vspace{0.25cm} 
HCOOCH$_{\rm 3}$ 18$_{\rm 8,10}$--17$_{\rm 8,9}$ E   & 222.4214 & 11 & 144 & 38 & 11 & 43 (07) & +8.4 (0.2) & 2.5 (0.4) & 110 (10)\\
HCOOCH$_{\rm 3}$ 19$_{\rm 3,16}$--18$_{\rm 3,15}$ E  $^{c,}$$^d$ & 225.6088  & \multirow{2}{*}{11} & \multirow{2}{*}{117} & \multirow{2}{*}{49} & \multirow{2}{*}{8} & \multirow{2}{*}{44 (08)} & \multirow{2}{*}{+8.1 (0.1)} & \multirow{2}{*}{3.7 (0.3)} & \multirow{2}{*}{173 (10)}\\
HCOOCH$_{\rm 3}$ 19$_{\rm 3,17}$--18$_{\rm 3,16}$ E  $^{c,}$$^d$ & 225.6088\\
\hline
\noalign{\vskip 2mm}
\end{tabular}}
\caption{List of transitions and line properties (in $T_{\rm MB}$ scale) of the HCOOCH$_{\rm 3}$ emission detected towards SVS13-A. We report the frequency of each transition (GHz),
the telescope HPBW ($\arcsec$),
the excitation energies of the upper level $E_{\rm up}$ (K),
the S$\mu^{2}$ product (D$^{2}$), the line rms (mK),
the peak temperature (mK), the peak velocities (km s$^{-1}$), the line full width at half maximum (FWHM) (km s$^{-1}$)
and the velocity integrated line intensity $I_{\rm int}$ (mK km s$^{-1}$).}
\small {$^a$ Frequencies and spectroscopic parameters are extracted from the Jet Propulsion Laboratory database \citep{Pickett1998}.
$^b$ The errors in brackets are the gaussian fit uncertainties.
$^c$ The lines cannot be distinguished with the present spectral resolution.
$^d$ The line is excluded from the further analysis because blended by other transitions with different upper level energies (see text).
} \\
\label{Table:MF}
\end{table*}

\addtocounter{table}{-1}
\begin{table*}
\resizebox{\textwidth}{!}{
\begin{tabular}{lccccccccc}
\hline
\multicolumn{1}{c}{Transition} &
\multicolumn{1}{c}{$\nu$$^{\rm a}$} &
\multicolumn{1}{c}{$HPBW$} &
\multicolumn{1}{c}{$E_{\rm u}$$^a$} &
\multicolumn{1}{c}{$S\mu^2$$^a$} &
\multicolumn{1}{c}{rms} &
\multicolumn{1}{c}{$T_{\rm peak}$$^b$} &
\multicolumn{1}{c}{$V_{\rm peak}$$^b$} &
\multicolumn{1}{c}{$FWHM$$^b$} &
\multicolumn{1}{c}{$I_{\rm int}$$^b$} \\
\multicolumn{1}{c}{ } &
\multicolumn{1}{c}{(GHz)} &
\multicolumn{1}{c}{($\arcsec$)} &
\multicolumn{1}{c}{(K)} &
\multicolumn{1}{c}{(D$^2$)} &
\multicolumn{1}{c}{(mK)} &
\multicolumn{1}{c}{(mK)} &
\multicolumn{1}{c}{(km s$^{-1}$)} &
\multicolumn{1}{c}{(km s$^{-1}$)} &
\multicolumn{1}{c}{(mK km s$^{-1}$)} \\

\hline

HCOOCH$_{\rm 3}$ 21$_{\rm 1,21}$--20$_{\rm 1,20}$ E  $^{c,}$$^d$ & 227.5609 &  \multirow{5}{*}{11} & \multirow{5}{*}{112} & \multirow{5}{*}{55} & \multirow{5}{*}{12} & \multirow{5}{*}{90 (10)} & \multirow{5}{*}{+8.5 (0.1)} & \multirow{5}{*}{4.8 (0.3)} & \multirow{5}{*}{470 (20)}\\

HCOOCH$_{\rm 3}$ 21$_{\rm 0,21}$--20$_{\rm 0,20}$ E  $^{c,}$$^d$ & 227.5617 \\

HCOOCH$_{\rm 3}$ 21$_{\rm 2,20}$--20$_{\rm 2,19}$ E  $^{c,}$$^d$ & 227.5618\\

HCOOCH$_{\rm 3}$ 21$_{\rm 1,21}$--20$_{\rm 1,20}$ A  $^{c,}$$^d$ & 227.5620\\
\vspace{0.25cm} 
HCOOCH$_{\rm 3}$ 21$_{\rm 0,21}$--20$_{\rm 0,20}$ A  $^{c,}$$^d$ & 227.5628\\

HCOOCH$_{\rm 3}$ 18$_{\rm 5,13}$--17$_{\rm 5,12}$ E  & 228.6289 & 11 & 119 & 44 & 14 & 46 (06) & +8.6 (0.2) & 3.3 (0.5) & 160 (20)\\

HCOOCH$_{\rm 3}$ 18$_{\rm 5,13}$--17$_{\rm 5,12}$ A   & 228.6514 &  11 & 119 & 44 & 9 & 38 (08) & +8.0 (0.1) & 2.8 (0.3) & 110 (10)\\

HCOOCH$_{\rm 3}$ 19$_{\rm 4,16}$--18$_{\rm 4,15}$ A   & 233.2268 & 11 & 123 & 48 & 7 & 46 (07) & +8.7 (0.1) & 3.6 (0.2) & 180 (10)\\






HCOOCH$_{\rm 3}$ 20$_{\rm 6,14}$--19$_{\rm 6,13}$ E   & 251.2645 & 10 & 149 & 48 & 10 & 35 (07) & +9.2 (0.2) & 3.5 (0.3) & 130 (10)\\

\vspace{0.25cm} 
HCOOCH$_{\rm 3}$ 20$_{\rm 6,14}$--19$_{\rm 6,13}$ A   & 251.2857 & 10 & 149 & 48 & 9 & 42 (09) & +8.6 (0.1) & 3.4 (0.3) & 160 (10)\\


HCOOCH$_{\rm 3}$ 24$_{\rm 1,24}$--23$_{\rm 1,23}$ E  $^{c,}$$^d$ & 259.3420 &  \multirow{4}{*}{9} & \multirow{4}{*}{158} & \multirow{4}{*}{64} & \multirow{4}{*}{8} & \multirow{4}{*}{79 (08)} & \multirow{4}{*}{+7.8 (0.1)} & \multirow{4}{*}{4.1 (0.2)} & \multirow{4}{*}{350 (10)}\\

HCOOCH$_{\rm 3}$ 24$_{\rm 0,24}$--23$_{\rm 0,23}$ E  $^{c,}$$^d$ & 259.3421\\
HCOOCH$_{\rm 3}$ 24$_{\rm 1,24}$--23$_{\rm 1,23}$ A  $^{c,}$$^d$ & 259.3429\\
\vspace{0.25cm} 
HCOOCH$_{\rm 3}$ 24$_{\rm 0,24}$--23$_{\rm 0,23}$ A  $^{c,}$$^d$ & 259.3430\\

HCOOCH$_{\rm 3}$ 20$_{\rm 4,16}$--19$_{\rm 4,15}$ A   & 259.5217& 9 & 139 &  51 & 8 & 39 (07) & +8.7 (0.2) & 3.9 (0.4) & 160 (10)\\
\hline
\noalign{\vskip 2mm}
\end{tabular}}
\caption{{\it Continued.}}
\small {$^a$ Frequencies and spectroscopic parameters are extracted from the Jet Propulsion Laboratory database \citep{Pickett1998}.
$^b$ The errors in brackets are the gaussian fit uncertainties.
$^c$ The lines cannot be distinguished with the present spectral resolution.
$^d$ The line is excluded from the further analysis because blended by other transitions with different upper level energies (see text).
} \\
\end{table*}

\begin{table*}
\resizebox{\textwidth}{!}{
\begin{tabular}{lccccccccc}

\hline
\multicolumn{1}{c}{Transition} &
\multicolumn{1}{c}{$\nu$$^a$} &
\multicolumn{1}{c}{$HPBW$} &
\multicolumn{1}{c}{$E_{\rm u}$$^a$} &
\multicolumn{1}{c}{$S\mu^2$$^a$} &
\multicolumn{1}{c}{rms} &
\multicolumn{1}{c}{$T_{\rm peak}$$^b$} &
\multicolumn{1}{c}{$V_{\rm peak}$$^b$} &
\multicolumn{1}{c}{$FWHM$$^b$} &
\multicolumn{1}{c}{$I_{\rm int}$$^b$} \\
\multicolumn{1}{c}{ } &
\multicolumn{1}{c}{(GHz)} &
\multicolumn{1}{c}{($\arcsec$)} &
\multicolumn{1}{c}{(K)} &
\multicolumn{1}{c}{(D$^2$)} &
\multicolumn{1}{c}{(mK)} &
\multicolumn{1}{c}{(mK)} &
\multicolumn{1}{c}{(km s$^{-1}$)} &
\multicolumn{1}{c}{(km s$^{-1}$)} &
\multicolumn{1}{c}{(mK km s$^{-1}$)} \\

\hline


CH$_{\rm 3}$OCH$_{\rm 3}$ 9$_{\rm 0,9}$--8$_{\rm 1,8}$ AA  & 153.0545 &  \multirow{4}{*}{16} & \multirow{4}{*}{40} & 100 & \multirow{4}{*}{8} & \multirow{4}{*}{--} & \multirow{4}{*}{--} & \multirow{4}{*}{--} & \multirow{4}{*}{120 (10)}\\
CH$_{\rm 3}$OCH$_{\rm 3}$ 9$_{\rm 0,9}$--8$_{\rm 1,8}$ EE  & 153.0548& & & 160\\
CH$_{\rm 3}$OCH$_{\rm 3}$ 9$_{\rm 0,9}$--8$_{\rm 1,8}$ AE  & 153.0552& & & 60\\
\vspace{0.25cm} 
CH$_{\rm 3}$OCH$_{\rm 3}$ 9$_{\rm 0,9}$--8$_{\rm 1,8}$ EA  & 153.0552& & & 40\\

CH$_{\rm 3}$OCH$_{\rm 3}$ 22$_{\rm 4,18}$--22$_{\rm 3,19}$ AE  & 162.4095 &  \multirow{4}{*}{15} & \multirow{4}{*}{254} & 51 & \multirow{4}{*}{7} & \multirow{4}{*}{--} & \multirow{4}{*}{--} & \multirow{4}{*}{--} & \multirow{4}{*}{70 (10)}\\
CH$_{\rm 3}$OCH$_{\rm 3}$ 22$_{\rm 4,18}$--22$_{\rm 3,19}$ EA  & 162.4095& & & 102\\
CH$_{\rm 3}$OCH$_{\rm 3}$ 22$_{\rm 4,18}$--22$_{\rm 3,19}$ EE  & 162.4107& & & 407\\
\vspace{0.25cm} 
CH$_{\rm 3}$OCH$_{\rm 3}$ 22$_{\rm 4,18}$--22$_{\rm 3,19}$ AA  & 162.4120& & & 153\\

CH$_{\rm 3}$OCH$_{\rm 3}$ 8$_{\rm 1,8}$--7$_{\rm 0,7}$ EA  & 162.5290 &  \multirow{4}{*}{15} & \multirow{4}{*}{33} & 36 & \multirow{4}{*}{8} & \multirow{4}{*}{--} & \multirow{4}{*}{--} & \multirow{4}{*}{--} & \multirow{4}{*}{140 (10)}\\
CH$_{\rm 3}$OCH$_{\rm 3}$ 8$_{\rm 1,8}$--7$_{\rm 0,7}$ AE  & 162.5290 &  & & 55\\
CH$_{\rm 3}$OCH$_{\rm 3}$ 8$_{\rm 1,8}$--7$_{\rm 0,7}$ EE  & 162.5296 & & & 145\\
\vspace{0.25cm} 
CH$_{\rm 3}$OCH$_{\rm 3}$ 8$_{\rm 1,8}$--7$_{\rm 0,7}$ AA  & 162.5302 & & & 91\\

CH$_{\rm 3}$OCH$_{\rm 3}$ 13$_{\rm 1,12}$--12$_{\rm 2,11}$ AA  & 202.4906 &  \multirow{4}{*}{12} & \multirow{4}{*}{86} & 77 & \multirow{4}{*}{18} & \multirow{4}{*}{--} & \multirow{4}{*}{--} & \multirow{4}{*}{--} & \multirow{4}{*}{180 (20)}\\
CH$_{\rm 3}$OCH$_{\rm 3}$ 13$_{\rm 1,12}$--12$_{\rm 2,11}$ EE  & 202.4916 &  & & 123\\
CH$_{\rm 3}$OCH$_{\rm 3}$ 13$_{\rm 1,12}$--12$_{\rm 2,11}$ AE  & 202.4926 & & & 46\\
\vspace{0.25cm} 
CH$_{\rm 3}$OCH$_{\rm 3}$ 13$_{\rm 1,12}$--12$_{\rm 2,11}$ EA  & 202.4926 & & & 31\\

CH$_{\rm 3}$OCH$_{\rm 3}$ 8$_{\rm 4,4}$--8$_{\rm 3,5}$ EE $^c$  & 204.5520 &  \multirow{3}{*}{12} & 55 & 75 & \multirow{3}{*}{18} & \multirow{3}{*}{--} & \multirow{3}{*}{--} & \multirow{3}{*}{--} & \multirow{3}{*}{150 (20)}\\
CH$_{\rm 3}$OCH$_{\rm 3}$ 8$_{\rm 4,5}$--8$_{\rm 3,5}$ EA  $^c$& 204.5525 &  & 55& 17\\
\vspace{0.25cm} 
CH$_{\rm 3}$OCH$_{\rm 3}$ 11$_{\rm 4,8}$--11$_{\rm 3,9}$ EE  $^c$ & 204.5526 &  & 83& 151\\

CH$_{\rm 3}$OCH$_{\rm 3}$ 12$_{\rm 0,12}$--11$_{\rm 1,11}$ AA  & 212.7559&  \multirow{4}{*}{12} & \multirow{4}{*}{70} & 91 & \multirow{4}{*}{11} & \multirow{4}{*}{--} & \multirow{4}{*}{--} & \multirow{4}{*}{--} & \multirow{4}{*}{210 (20)}\\
CH$_{\rm 3}$OCH$_{\rm 3}$ 12$_{\rm 0,12}$--11$_{\rm 1,11}$ AA  & 212.7561&    & & 243\\
CH$_{\rm 3}$OCH$_{\rm 3}$ 12$_{\rm 0,12}$--11$_{\rm 1,11}$ AA  & 212.7562&  & & 61\\
\vspace{0.25cm} 
CH$_{\rm 3}$OCH$_{\rm 3}$ 12$_{\rm 0,12}$--11$_{\rm 1,11}$ AA  & 212.7562&  & & 30\\

CH$_{\rm 3}$OCH$_{\rm 3}$ 12$_{\rm 1,12}$--11$_{\rm 0,11}$ EA  & 225.5988&  \multirow{4}{*}{11} & \multirow{4}{*}{70} & 62 & \multirow{4}{*}{18} & \multirow{4}{*}{--} & \multirow{4}{*}{--} & \multirow{4}{*}{--} & \multirow{4}{*}{320 (20)}\\
CH$_{\rm 3}$OCH$_{\rm 3}$ 12$_{\rm 1,12}$--11$_{\rm 0,11}$ AE  & 225.5988&    & & 93\\
CH$_{\rm 3}$OCH$_{\rm 3}$ 12$_{\rm 1,12}$--11$_{\rm 0,11}$ EE  & 225.5991&  & & 247\\
\vspace{0.25cm} 
CH$_{\rm 3}$OCH$_{\rm 3}$ 12$_{\rm 1,12}$--11$_{\rm 0,11}$ AA  & 225.5995&  & & 154\\

CH$_{\rm 3}$OCH$_{\rm 3}$ 9$_{\rm 2,8}$--8$_{\rm 1,7}$ EA  & 237.6188 &  \multirow{4}{*}{10} & \multirow{4}{*}{46} & 23 & \multirow{4}{*}{9} & \multirow{4}{*}{--} & \multirow{4}{*}{--} & \multirow{4}{*}{--} & \multirow{4}{*}{370 (20)}\\
CH$_{\rm 3}$OCH$_{\rm 3}$ 9$_{\rm 2,8}$--8$_{\rm 1,7}$ AE  & 237.6188 &   & & 12\\
CH$_{\rm 3}$OCH$_{\rm 3}$ 9$_{\rm 2,8}$--8$_{\rm 1,7}$ EE  & 237.6209 &   & & 93\\
\vspace{0.25cm} 
CH$_{\rm 3}$OCH$_{\rm 3}$ 9$_{\rm 2,8}$--8$_{\rm 1,7}$ AA  & 237.6230 &   & & 35\\

CH$_{\rm 3}$OCH$_{\rm 3}$ 13$_{\rm 1,13}$--12$_{\rm 0,12}$ AE  & 241.9462 &  \multirow{4}{*}{10} & \multirow{4}{*}{81} & 34 & \multirow{4}{*}{12} & \multirow{4}{*}{--} & \multirow{4}{*}{--} & \multirow{4}{*}{--} & \multirow{4}{*}{330 (20)}\\
CH$_{\rm 3}$OCH$_{\rm 3}$ 13$_{\rm 1,13}$--12$_{\rm 0,12}$ EA  & 241.9462 &    & & 69\\
CH$_{\rm 3}$OCH$_{\rm 3}$ 13$_{\rm 1,13}$--12$_{\rm 0,12}$ EE  & 241.9465 &    & & 274\\
\vspace{0.25cm} 
CH$_{\rm 3}$OCH$_{\rm 3}$ 13$_{\rm 1,13}$--12$_{\rm 0,12}$ AA  & 241.9468 &    & & 103\\

CH$_{\rm 3}$OCH$_{\rm 3}$ 15$_{\rm 1,14}$--14$_{\rm 2,13}$ AA  & 249.9238 &  \multirow{4}{*}{10} & \multirow{4}{*}{113} & 105 & \multirow{4}{*}{11} & \multirow{4}{*}{--} & \multirow{4}{*}{--} & \multirow{4}{*}{--} & \multirow{4}{*}{100 (10)}\\
CH$_{\rm 3}$OCH$_{\rm 3}$ 15$_{\rm 1,14}$--14$_{\rm 2,13}$ EE  & 249.9245 &    & & 168\\
CH$_{\rm 3}$OCH$_{\rm 3}$ 15$_{\rm 1,14}$--14$_{\rm 2,13}$ AE  & 249.9251 &   & & 63\\
\vspace{0.25cm} 
CH$_{\rm 3}$OCH$_{\rm 3}$ 15$_{\rm 1,14}$--14$_{\rm 2,13}$ EA  & 249.9251 &    & & 42\\

CH$_{\rm 3}$OCH$_{\rm 3}$ 21$_{\rm 5,16}$--21$_{\rm 4,17}$ AE  & 251.1408 &  \multirow{4}{*}{10} & \multirow{4}{*}{246} & 118 & \multirow{4}{*}{13} & \multirow{4}{*}{--} & \multirow{4}{*}{--} & \multirow{4}{*}{--} & \multirow{4}{*}{100 (20)}\\
CH$_{\rm 3}$OCH$_{\rm 3}$ 21$_{\rm 5,16}$--21$_{\rm 4,17}$ EA  & 251.1408 &    & & 79\\
CH$_{\rm 3}$OCH$_{\rm 3}$ 21$_{\rm 5,16}$--21$_{\rm 4,17}$ EE  & 251.1424 &  & & 316\\
CH$_{\rm 3}$OCH$_{\rm 3}$ 21$_{\rm 5,16}$--21$_{\rm 4,17}$ AA  & 251.1441 &    & & 197\\
\hline
\noalign{\vskip 2mm}
\end{tabular}}
\caption{List of transitions and line properties (in $T_{\rm MB}$ scale) of the CH$_{3}$OCH$_{3}$ emission detected towards SVS13-A. We report the frequency of each transition (GHz),
the telescope HPBW ($\arcsec$),
the excitation energies of the upper level $E_{\rm up}$ (K),
the S$\mu^{2}$ product (D$^{2}$), the line rms (mK),
the peak temperature (mK), the peak velocities (km s$^{-1}$), the line full width at half maximum (FWHM) (km s$^{-1}$)
and the velocity integrated line intensity $I_{\rm int}$ (mK km s$^{-1}$).}

\small {$^a$ Frequencies and spectroscopic parameters are extracted from the Cologne Database
 for Molecular Spectroscopy (CDMS\footnote{http://www.astro.uni-koeln.de/cdms/};\citealt{Muller2001},\citealt{Muller2005}) molecular database.
$^b$ Gaussian fit is not performed given the asymmetric line profiles.
$^c$The transition is not used for the further analysis.
} \\
\label{Table:DME}
\end{table*}

\addtocounter{table}{-1}
\begin{table*}
\resizebox{\textwidth}{!}{
\begin{tabular}{lccccccccc}

\hline
\multicolumn{1}{c}{Transition} &
\multicolumn{1}{c}{$\nu$$^a$} &
\multicolumn{1}{c}{$HPBW$} &
\multicolumn{1}{c}{$E_{\rm u}$$^a$} &
\multicolumn{1}{c}{$S\mu^2$$^a$} &
\multicolumn{1}{c}{rms} &
\multicolumn{1}{c}{$T_{\rm peak}$$^b$} &
\multicolumn{1}{c}{$V_{\rm peak}$$^b$} &
\multicolumn{1}{c}{$FWHM$$^b$} &
\multicolumn{1}{c}{$I_{\rm int}$$^b$} \\
\multicolumn{1}{c}{ } &
\multicolumn{1}{c}{(GHz)} &
\multicolumn{1}{c}{($\arcsec$)} &
\multicolumn{1}{c}{(K)} &
\multicolumn{1}{c}{(D$^2$)} &
\multicolumn{1}{c}{(mK)} &
\multicolumn{1}{c}{(mK)} &
\multicolumn{1}{c}{(km s$^{-1}$)} &
\multicolumn{1}{c}{(km s$^{-1}$)} &
\multicolumn{1}{c}{(mK km s$^{-1}$)} \\

\hline

CH$_{\rm 3}$OCH$_{\rm 3}$ 12$_{\rm 5,8}$--12$_{\rm 4,9}$ AE  & 262.8893 &  \multirow{3}{*}{9} & \multirow{3}{*}{106} & 58 & \multirow{3}{*}{15} & \multirow{3}{*}{--} & \multirow{4}{*}{--} & \multirow{3}{*}{--} & \multirow{3}{*}{110 (20)}\\
CH$_{\rm 3}$OCH$_{\rm 3}$ 12$_{\rm 5,8}$--12$_{\rm 4,9}$ EE  & 262.8902 &    & & 112\\
\vspace{0.25cm} 
CH$_{\rm 3}$OCH$_{\rm 3}$ 12$_{\rm 5,8}$--12$_{\rm 4,9}$ AA  & 262.8954 &    & & 97\\

CH$_{\rm 3}$OCH$_{\rm 3}$ 15$_{\rm 0,15}$--14$_{\rm 1,14}$ EA  $^c$ & 269.6088 &  \multirow{5}{*}{9} & \multirow{4}{*}{106} & 82 & \multirow{5}{*}{15} & \multirow{5}{*}{--} & \multirow{5}{*}{--} & \multirow{5}{*}{--} & \multirow{5}{*}{220 (20)}\\
CH$_{\rm 3}$OCH$_{\rm 3}$ 15$_{\rm 0,15}$--14$_{\rm 1,14}$ AE  $^c$ & 269.6088 &     & & 123\\
CH$_{\rm 3}$OCH$_{\rm 3}$ 15$_{\rm 0,15}$--14$_{\rm 1,14}$ EE $^c$ & 269.6088 &     & & 329\\
CH$_{\rm 3}$OCH$_{\rm 3}$ 15$_{\rm 0,15}$--14$_{\rm 1,14}$ AA  $^c$ & 269.6088 &     & & 205\\
\vspace{0.25cm} 
CH$_{\rm 3}$OCH$_{\rm 3}$ 22$_{\rm 3,19}$--21$_{\rm 4,18}$ EE  $^c$ & 269.6096&    &246 & 106\\

CH$_{\rm 3}$OCH$_{\rm 3}$ 15$_{\rm 1,15}$--14$_{\rm 0,14}$ AE  & 275.3817 &  \multirow{4}{*}{9} & \multirow{4}{*}{106} & 41 & \multirow{4}{*}{18} & \multirow{4}{*}{--} & \multirow{4}{*}{--} & \multirow{4}{*}{--} & \multirow{4}{*}{220 (20)}\\
CH$_{\rm 3}$OCH$_{\rm 3}$ 15$_{\rm 1,15}$--14$_{\rm 0,14}$ EA  & 275.3817 &    & & 82\\
CH$_{\rm 3}$OCH$_{\rm 3}$ 15$_{\rm 1,15}$--14$_{\rm 0,14}$ EE  & 275.3819 &  & & 330\\
CH$_{\rm 3}$OCH$_{\rm 3}$ 15$_{\rm 1,15}$--14$_{\rm 0,14}$ AA  & 275.3821 &   & & 124\\
\hline
\noalign{\vskip 2mm}
\end{tabular}}
\caption{{\it Continued.}}
\small {$^a$ Frequencies and spectroscopic parameters are extracted from the Cologne Database
 for Molecular Spectroscopy (CDMS\footnote{http://www.astro.uni-koeln.de/cdms/};\citealt{Muller2001},\citealt{Muller2005}) molecular database.
$^b$ Gaussian fit is not performed given the asymmetric line profiles.
$^c$The transition is not used for the further analysis.
} \\

\end{table*}

\begin{table*}
\resizebox{\textwidth}{!}{
\begin{tabular}{lccccccccc}
\hline
\multicolumn{1}{c}{Transition} &
\multicolumn{1}{c}{$\nu$$^{\rm a}$} &
\multicolumn{1}{c}{$HPBW$} &
\multicolumn{1}{c}{$E_{\rm u}$$^a$} &
\multicolumn{1}{c}{$S\mu^2$$^a$} &
\multicolumn{1}{c}{rms} &
\multicolumn{1}{c}{$T_{\rm peak}$$^b$} &
\multicolumn{1}{c}{$V_{\rm peak}$$^b$} &
\multicolumn{1}{c}{$FWHM$$^b$} &
\multicolumn{1}{c}{$I_{\rm int}$$^b$} \\
\multicolumn{1}{c}{ } &
\multicolumn{1}{c}{(GHz)} &
\multicolumn{1}{c}{($\arcsec$)} &
\multicolumn{1}{c}{(K)} &
\multicolumn{1}{c}{(D$^2$)} &
\multicolumn{1}{c}{(mK)} &
\multicolumn{1}{c}{(mK)} &
\multicolumn{1}{c}{(km s$^{-1}$)} &
\multicolumn{1}{c}{(km s$^{-1}$)} &
\multicolumn{1}{c}{(mK km s$^{-1}$)} \\
\hline


g-CH$_{\rm 3}$CH$_{\rm 2}$OH 8$_{\rm 3,6}$--8$_{\rm 2,6}$ &  221.8117 & 11 & 103 & 5 & 2 & 33 (02) & +8.6 (0.1) & 1.4 (0.1) & 48 (02) \\

g-CH$_{\rm 3}$CH$_{\rm 2}$OH 13$_{\rm 2,12}$--12$_{\rm 2,11}$ & 222.6771 & 11 & 137 & 20 & 5 & 36 (05) & +8.5 (0.1) & 1.5 (0.1) & 57 (05) \\

a-CH$_{\rm 3}$CH$_{\rm 2}$OH 14$_{\rm 0,14}$--13$_{\rm 1,13}$ & 230.9914 & 11 & 86 & 14 & 9 & 38 (09) & +8.5 (0.1) & 2.7 (0.3) & 110 (10) \\

a-CH$_{\rm 3}$CH$_{\rm 2}$OH 8$_{\rm 5,3}$--8$_{\rm 4,4}$ & 234.9842 & 10 & 62 & 7 & 8 & 21 (06) & +8.2 (0.2) & 2.5 (0.5) & 60 (10) \\

a-CH$_{\rm 3}$CH$_{\rm 2}$OH 7$_{\rm 3,4}$--6$_{\rm 2,5}$ & 254.3841 & 10 & 35 & 7 & 8 & 20 (08) & +8.3 (0.3) & 3.4 (0.7) & 70 (10) \\
\hline

\noalign{\vskip 2mm}

\end{tabular}}
\caption{List of transitions and line properties (in $T_{\rm MB}$ scale) of the CH$_{\rm 3}$CH$_{\rm 2}$OH emission detected towards SVS13-A.  We report the frequency of each transition (GHz),
the telescope HPBW ($\arcsec$),
the excitation energies of the upper level $E_{\rm up}$ (K),
the S$\mu^{2}$ product (D$^{2}$), the line rms (mK),
the peak temperature (mK), the peak velocities (km s$^{-1}$), the line full width at half maximum (FWHM) (km s$^{-1}$)
and the velocity integrated line intensity $I_{\rm int}$ (mK km s$^{-1}$).}
\small {$^a$ Frequencies and spectroscopic parameters are extracted from the Jet Propulsion Laboratory database \citep{Pickett1998}.
$^b$ The errors in brackets are the gaussian fit uncertainties.} \\
\label{Table:ethanol}
\end{table*}

\label{lastpage}

\end{document}